\documentclass{aa}  
\makeatletter

\renewcommand*\aa@pageof{, page \thepage{} of \pageref*{LastPage}}

\usepackage{natbib}
\bibpunct{(}{)}{;}{a}{}{,}%
\def\bibfont{\aa@bibliographyfont}%
\makeatother

\usepackage[T1]{fontenc}
\usepackage{amsmath}	
\usepackage{soul}
\usepackage{xspace}
\usepackage{subcaption}
\usepackage{gensymb}
\usepackage{bm}	
\usepackage{graphicx}
\usepackage{newtxtext}
\usepackage[smallerops,cmbraces]{newtxmath}
\usepackage{xcolor}
\usepackage{newtxtext,newtxmath}
\usepackage{rotating}
\usepackage{adjustbox}
\usepackage{multicol}

\usepackage{subcaption}

\usepackage[T1]{fontenc}
\definecolor{xlinkcolor}{cmyk}{1,1,0,0}

\usepackage[
  bookmarks=true,         
  pdfnewwindow=true,      
  colorlinks=true,        
  linkcolor=xlinkcolor,   
  citecolor=xlinkcolor,   
  filecolor=xlinkcolor,   
  urlcolor=xlinkcolor,    
  final=true,
]{hyperref}
\usepackage{soul}

\usepackage[normalem]{ulem} 

\usepackage{bm} 
\usepackage{textgreek} 

\usepackage[capitalize]{cleveref} 
\crefname{section}{Sect.}{Sects.}
\creflabelformat{equation}{#2#1#3} 
\crefname{enumi}{item}{items} 

\usepackage{siunitx} 
\DeclareSIUnit[number-unit-product = ]\percent{\char`\%} 

\usepackage{csquotes} 

\begin{document}


\title{Intra-Cluster Light as a Dynamical Clock for Galaxy Clusters: Insights from the MAGNETICUM, IllustrisTNG, Hydrangea and Horizon-AGN Simulations}
\titlerunning{ICL and Cluster Relaxation States}

\author{
    Lucas C. Kimmig\inst{\ref{inst:usm}}\thanks{E-mail: lkimmig@usm.lmu.de}
    \and
    Sarah Brough\inst{\ref{inst:unsw}}
    \and
    Klaus Dolag\inst{\ref{inst:usm},\ref{inst:mpa}}
    \and
    Rhea-Silvia Remus\inst{\ref{inst:usm}}
    \and
    Yannick M.~Bah\'e\inst{\ref{inst:epfl},\ref{inst:nottingham}}
    \and
    Garreth Martin\inst{\ref{inst:nottingham}}
    \and
    Annalisa Pillepich\inst{\ref{inst:mpia}}
    \and
    Nina Hatch\inst{\ref{inst:nottingham}}
    \and
    Mireia Montes\inst{\ref{inst:stsi}}
    \and
    Syeda Lammim Ahad\inst{\ref{inst:wat},\ref{inst:wat2}}
    \and 
    Callum Bellhouse\inst{\ref{inst:nottingham}}
    \and 
    Harley J. Brown\inst{\ref{inst:nottingham}}
    \and 
    Amaël Ellien\inst{\ref{inst:oca}}
    \and 
    Jesse B. Golden-Marx\inst{\ref{inst:nottingham}}
    \and 
    Anthony H. Gonzalez\inst{\ref{inst:florida}}
    \and 
    Enrica Iodice\inst{\ref{inst:inaf}}
    \and 
    Yolanda Jim\'enez-Teja\inst{\ref{inst:anda},\ref{inst:onac}}
    \and 
    Matthias Kluge\inst{\ref{inst:mpe}}
    \and 
    Johan H. Knapen\inst{\ref{inst:canaria},\ref{inst:laguna}}
    \and 
    J. Christopher Mihos\inst{\ref{inst:casewest}}
    \and 
    Rossella Ragusa\inst{\ref{inst:inaf}}
    \and 
    Marilena Spavone\inst{\ref{inst:inaf}}
    }
\authorrunning{L. C. Kimmig et al.}

\institute{
    Universit\"ats-Sternwarte, Fakult\"at f\"ur Physik, Ludwig-Maximilians-Universit\"at M\"unchen, Scheinerstr.\ 1, 81679 M\"unchen, Germany\label{inst:usm}
    \and
    School of Physics, University of New South Wales, NSW 2052, Australia\label{inst:unsw}
    \and
    Max Planck Institute for Astrophysics, D-85748 Garching, Germany\label{inst:mpa}
    \and
    Institute of Physics, Ecole Polytechnique F\'{e}d\'{e}rale de Lausanne (EPFL), Observatoire de Sauverny, 1290 Versoix, Switzerland\label{inst:epfl}
    \and
    School of Physics and Astronomy, University of Nottingham, University Park, Nottingham NG7 2RD, UK\label{inst:nottingham}
    \and
    Max-Planck-Institut f{\"u}r Astronomie, K{\"o}nigstuhl 17, 69117 Heidelberg, Germany\label{inst:mpia}
    \and 
    Institute of Space Sciences (ICE, CSIC), Campus UAB, Carrer de Can Magrans, s/n, 08193 Barcelona, Spain\label{inst:stsi}
    \and 
    Waterloo Centre for Astrophysics, University of Waterloo, Waterloo, Ontario N2L 3G1, Canada\label{inst:wat}
    \and
    Department of Physics and Astronomy, University of Waterloo, Waterloo, Ontario N2L 3G1, Canada\label{inst:wat2}
    \and 
    OCA, P.H.C Boulevard de l’Observatoire CS 34229, 06304 Nice Cedex 4, France\label{inst:oca}
    \and 
    Department of Astronomy, University of Florida, Gainesville, FL 32611, USA\label{inst:florida}
    \and 
    Instituto de Astrof\'isica de Andaluc\'ia--CSIC, Glorieta de la Astronom\'ia s/n, E--18008 Granada, Spain\label{inst:anda}
    \and 
    Observat\'orio Nacional, Rua General Jos\'e Cristino, 77 - Bairro Imperial de S\~ao Crist\'ov\~ao, Rio de Janeiro, 20921-400, Brazil\label{inst:onac}
    \and 
    INAF-Osservatorio Astronomico di Capodimonte, Salita Moiariello 16, 80131 Napoli, Italy\label{inst:inaf}
    \and 
    Max Planck Institute for Extraterrestrial Physics, Giessenbachstr. 1, 85748 Garching, Germany\label{inst:mpe}
    \and 
    Instituto de Astrof\'{i}sica de Canarias, V\'{i}a L\'{a}ctea S/N, E-38205 La Laguna, Spain\label{inst:canaria}
    \and
    Departamento de Astrof\'{i}sica, Universidad de La Laguna, E-38206 La Laguna, Spain\label{inst:laguna}
    \and 
    Department of Astronomy, Case Western Reserve University, 10900 Euclid Avenue, Cleveland, OH 44106, USA\label{inst:casewest}
    }

\date{Received XXX / Accepted YYY}

\abstract
{As the most massive nodes of the cosmic web, galaxy clusters represent the best probes of structure formation. Over time, they grow by accreting and disrupting satellite galaxies, adding those stars to the brightest cluster galaxy (BCG) and the intra-cluster light (ICL). However, the formation pathways of different galaxy clusters can vary significantly.}
{To inform upcoming large surveys, we aim to identify observables that can distinguish galaxy cluster formation pathways.}
{Using four different hydrodynamical simulations, Magneticum, TNG100 of IllustrisTNG, Horizon-AGN, and Hydrangea, we study how the fraction of stellar mass in the BCG and ICL ($f_\mathrm{ICL+BCG}$) relates to the galaxy cluster mass assembly history.}
{For all simulations, $f_\mathrm{ICL+BCG}$ is the best tracer for the time at which the cluster has accumulated 50\%~of its mass ($z_\mathrm{form}$), performing better than other typical dynamical tracers, such as the subhalo mass fraction, the halo mass, and the position offset of the cluster mass barycenter to the BCG. More relaxed clusters have higher $f_\mathrm{ICL+BCG}$, in rare cases up to 90\% of all stellar mass, while dynamically active clusters have lower fractions, down to 20\%, which we find to be independent of the exact implemented baryonic physics. We determine the average increase in $f_\mathrm{ICL+BCG}$ from stripping and mergers to be between 3-4\%~per Gyr. $f_\mathrm{ICL+BCG}$ itself is tightly traced by the stellar mass ratio between the BCG and both the second (M12) and fourth (M14) most massive cluster galaxy. The average galaxy cluster has assembled half of its halo mass by $z_\mathrm{form}=0.67$ (about~$6\,$Gyr ago), though individual histories vary significantly from $z_\mathrm{form}=0.06$~to~$z_\mathrm{form}=1.77$ (0.8~to~$10\,$Gyr ago).}
{As all four cosmological simulations consistently find that $f_\mathrm{ICL+BCG}$ is an excellent tracer of the cluster dynamical state, upcoming surveys can leverage measurements of $f_\mathrm{ICL+BCG}$ to statistically quantify the assembly of the most massive structures through cosmic time.}


\maketitle


\section{Introduction}

Galaxy clusters are the most massive bound, virialized structures in the observable Universe, representing less than one percent of its volume but on the order of 5-$10\%$ of the total mass \citep{libeskind:2018}. Understanding their formation is akin to understanding all of cosmic structure formation, as they represent the final stage in bottom-up $\Lambda$CDM cosmology \citep[e.g.,][]{chon:2015,seidel:2024}. As clusters assemble mass from the cosmic web, which is dominated by dark matter, they simultaneously also assemble luminous matter in the form of galaxies (which then become satellite galaxies), and build up a significant amount of stellar mass both in the brightest cluster galaxy (BCG) and in a diffuse stellar component commonly referred to as the intra-cluster light (ICL; \citealt{Montes2022}). A typical estimation is that each of the three components amounts to approximately one third of the total stellar mass hosted within a cluster \citep{Dolag2010}, but individual values between clusters for the ICL range from about $10\%$ \citep[e.g.,][]{Zibetti2005} to $35\%$ \citep[e.g.,][]{Gonzalez2007,Spavone2020,ragusa2023,kluge:2024,spavone:2024}. Similar fractions are also reported at higher redshifts of $z=0.3-0.4$, varying from about $10\%$ \citep[e.g.,][]{montes:2018} to $30\%$ \citep[e.g.,][]{Furnell2021}. Variations of ICL fractions measured directly via flux are even larger, from $3\%$ \citep{Jimenez-Teja2018} to $30\%$ \citep{Jimenez-Teja2021,jimenez-teja2023,jimenez-teja:2024}. Simulations find similarly diverse pictures, with the ICL containing more \citep{Dolag2010} or less \citep{brown:2024} stellar mass compared to the BCG, likely depending on varying definitions of where the BCG ends and the ICL starts, something which is still under debate \citep[see][for an overview on different methods]{brough24}. However, independent of the exact definition, the diffuse ICL accounts for a significant amount of the stellar mass in galaxy clusters, but is simultaneously the component which is most difficult to measure due to its low surface brightness nature \citep[e.g.,][]{murante:2004,Montes2022}.

From cosmological simulations several pathways are known that contribute to the formation and growth of the ICL \citep[e.g.,][]{brown:2024,bilata:woldeyes:2025}, including ex-situ origin of the stars as well as in-situ star formation \citep[e.g.,][]{Puchwein10,ahvazi:2024}. The ex-situ pathways include stars that are formed in other structures and then assembled through pre-processing \citep[e.g.,][]{mihos:2004,Mihos2005, Contini2014}, minor and major mergers \citep[e.g.,][]{Monaco2006, Murante2007,Contini2014}, the stripping of intermediate or massive satellite galaxies \citep[e.g.,][]{Rudick06,Rudick2009, Rudick2011, Contini2014, Contini2018,chun:2024}, as well as from disrupting dwarfs \citep[e.g.,][]{Conroy07, Giallongo2014}. It is becoming clear from simulations \citep[e.g.,][]{brown:2024}, models \citep[e.g.,][]{hao:2024} and observations \citep[e.g.,][]{Spavone2017b,Spavone2020,kluge:2024,zhang:2024,spavone:2024,golden-marx:2025} that the most dominant contribution to the stellar component of clusters is of an ex-situ nature and originates mostly from stellar stripping and mergers. Which of these is the more important is still debated, with evidence provided for stripping \citep[e.g.,][]{Contini2014} as well as for merging \citep[e.g.,][]{Murante2007, montenegro:2023}. It is further unclear if the dominant component of the ICL was formed within a single structure (i.e., galaxy) and then directly stripped within the cluster halo, or if, as found by \citet{joo:2024}, the majority had already been pre-processed by another structure (galaxy or group) which only afterwards then fell into the cluster \citep{mihos:2004,Mihos2017}. Additionally, the efficiency of the stripping of satellites to build up the ICL may be both halo mass \citep{Bahe2019,contini:2024b} and resolution dependent \citep{martin:2024}. However, in all cases it is clear that the ICL of galaxy clusters originates primarily from ex-situ accretion.

All other cluster stars are within the BCGs, which represent the most massive galaxies in the Universe. They predominantly grow through mergers, evidence for which is found both in models \citep[e.g.,][]{kochfar:2003,DeLucia2007} and from observations \citep[e.g.,][]{brough:2011,oliva-altamirano:2014,edwards:2024}. The cores of these BCGs are often extremely old \citep[e.g.,][]{oliva:2015,edwards:2020,santucci:2020}, and for early forming clusters their BCGs are typically the first galaxies to collapse on very short timescales \citep[e.g.,][]{rennehan:2020,remus:2022}. However, some of the observed BCGs also show ongoing star formation, perhaps most extreme in the Phoenix cluster \citep{reefe:2025}, but also on a lower level in our neighborhood \citep{donahue:2015,tremblay:2015}, or indications of episodes of star formation shown in populations of stars with younger ages \citep{mittal:2015,edwards:2024}, hinting at either recent ongoing merging events or even gas accretion or cooling from the hot halo. Nonetheless, compared to the outskirts and the ICL, the BCGs are generally older and more metal rich \citep[e.g.,][]{MT14,oliva:2015,Edwards2016,montes:2018,edwards:2020}, which suggests that the BCGs form earlier and the ICL was accreted more recently \citep{Montes2021}.

While these overall trends are well established, separating the ICL component from the BCG is an as of yet open topic. Generally, it has been established that there are two kinematically distinct stellar components in galaxy clusters \citep{Dolag2010,bender:2015,longobardi:2015,remus17,marini:2024}, which can be used to separate the central galaxy of a cluster from the surrounding ICL. Similarly, the radial luminosity profiles of galaxy clusters typically show two or sometimes even more components, which would suggest different origins for the stars belonging to each \citep[][among others]{Iodice2016,Spavone2017b,Kluge2020,Spavone2020,Montes2021,ragusa2023,Ahad2023}. In principle, the transition between the innermost component and the others could be used to split the ICL and BCG. Furthermore, surface brightness cuts can be applied \citep[e.g.,][]{Feldmeier2004,Burke2015,montes:2018,Martinez-Lombilla2023a}, or also non-parametric measures \citep[e.g.,][]{Martinez-Lombilla2023a}. There are more methods \citep[e.g.,][]{ellien:2021}, as discussed in more detail by \citet{brough24}, but unfortunately, these methods do not agree on the separation between the ICL and the BCG even when applied to the same galaxy clusters \citep{remus17,remus:2022,brough24}. Indeed, difficulties in quantifying the ICL between different methods were already presented by \citet{Rudick2011}, who found increased ICL fractions when including the current velocity or past accretion history of individual particles as opposed to just surface brightness.

Nonetheless, both the BCG and ICL components are clearly dominated by ex-situ growth, building up through accretion events throughout a galaxy cluster's lifetime. Consequently, galaxy clusters are in a unique position, as they represent the extreme end of the highest ex-situ fractions \citep[e.g.,][]{rodriguez:2016,dubois:2016,pillepich18b,davison:2020,remus:2022,Martin2022}. This has a few interesting implications for determining the current dynamical state of galaxy clusters as well as predicting their past mass accretion history. 

Firstly, with little in-situ formation of stars, the total stellar mass instead grows along with the accreted dark matter that dominates the total mass budget. So long as there are no large variances in the structure being assembled from cluster to cluster, this implies a self-similar scaling relation for their total stellar mass to halo mass. Indeed, both observations \citep{budzynski14,kravtsov18} as well as simulations \citep{bahe17,pillepich18b} have found slopes near unity for the total stellar to halo mass. 

Secondly, given a self-similar budget of total stellar mass, this would imply that the distribution of this mass into the components of central galaxy (BCG), stellar halos (ICL) and satellite galaxies should depend on the individual formation pathway of the galaxy cluster, as stripping and merging processes all require time to transform the assembled satellite galaxies into the BCG and ICL component. Taking the crossing time as a proxy for the dynamical timescale, this results in satellite disruption on the order of $1-2$~Gyr \citep{jiang16,contreras22,haggar24}, though this may be halo mass dependent \citep{Bahe2019}.

By contrast, the time-scale of the accretion of new mass onto the galaxy cluster as a whole is on the order of $M/\dot{M}\approx 10$~Gyr \citep{mcbride09,jiang17}. This results in a picture where on average additional satellite galaxies are accreted more slowly than the mass in existing ones is disrupted. Consequently, the amount of mass in the cluster satellites should act as a dynamical clock, as already suggested by \citet{Dolag2010} based on galaxy clusters selected from the simulation of a local-Universe analogue. Additional credence for this idea was recently provided from models by \citet{Contini2023A,contini:2024}, showing that clusters with higher concentrations, i.e., dynamically older clusters, have larger fractions of ICL. Similar results were also found from The Three Hundred Simulations \citep{contreras24} and the Horizon Run-5 simulation \citep{yoo:2024}. 

Furthermore, there is evidence from observations that the distribution of mass between the ICL and BCG versus the satellites does not depend simply on the total mass of the galaxy cluster. Both \citet{Kluge2021} and \citet{ragusa2023} do not find any correlation between the observed fraction of the ICL and cluster mass, despite using different methods of ICL measurement, and both report a large scatter in the fractions. A trend is also observed between ICL fraction and the fraction of early type galaxies in nearby groups and clusters, with the ICL fraction being on average larger in those environments dominated by early-type galaxies \citep[e.g.,][]{Aguerri2006,DaRocha2008,RAGUSA2021,Ragusa2022FrASS...952810R, ragusa2023,Spavone2020,spavone:2024}. Even in more massive systems at higher redshifts, \cite{montes:2018} noted that the most relaxed cluster in their sample of 6 (at $0.3<z<0.55$) has a higher fraction of ICL compared to the other clusters. Intriguingly, \citet{Jimenez-Teja2018} find at $0.2<z<0.4$ evidence for higher ICL fractions in the bluer bands of merging clusters compared to redder bands, indicating that these trends may further depend on the exact wavelengths used to measure the fractions. Nevertheless, all observations hint at a connection between the fraction of ICL and the dynamical state of clusters, in agreement with the theoretical considerations.

Interpreting this link is made more difficult by the variety of definitions of the dynamical state. Established parameters include the center shift, meaning the distance between the center of mass and the point of the deepest potential s$=\Delta_r / R_\mathrm{200c}$ \citep[e.g.,][]{mann12,biffi16,cui17,roberts18,zenteno20,contreras22}, the total mass contained in subhalos $f_\mathrm{sub}$ \citep[e.g.,][]{neto07,jiang17,cui:2018}, or the virial ratio $\eta=(2T-E_\mathrm{s})/|W|$, given by the total kinetic $T$, potential $W$, and surface pressure energy $E_\mathrm{s}$ \citep[e.g.,][]{shaw06,cui17,haggar24}. This list has expanded significantly in recent years to include, among others, the mass difference between the BCG and the second most massive galaxy (M12) \citep[e.g.,][]{jones03,ragagnin17,raouf:2019,casas:2024}, the offset between the central galaxy and the luminosity center of the total structure s$_\mathrm{stars}$ \citep[e.g.,][]{raouf:2019}, the fraction of total mass contained in the eighth most massive substructure of the cluster $f_{8}$ \citep{kimmig23}, and the formation redshift z$_\mathrm{form}$, defined as the time when the galaxy cluster first reached $50\%$ of its final $z=0$ mass \citep[e.g.,][]{haggar20,contreras22}. For a more in depth overview see the work by \citet{haggar24}.

However, many of these definitions require multi-wavelength observations to measure (center shift is often defined in the X-ray \citep{biffi16,zenteno20}, while the magnitude gap explicitly depends on the chosen band), while others such as $f_\mathrm{sub}$ require gravitational lensing \citep[e.g.,][]{jauzac16} which may suffer from projection effects \citep{schwinn18,mao18,kimmig23}. Additionally, as shown by \citet{haggar24}, different parameters may be measuring different types of cluster dynamical state. Regardless, it is of great importance to better quantify and understand the dynamical states of galaxy clusters, as there are indications of links to the properties of their galaxy populations \citep{aldas23,brambila23,veliz24}. With several facilities capable of probing the low-surface brightness regime that are either ongoing, such as {\it Euclid} \citep{laureijs:2011,kluge:2024} or upcoming, for example Roman \citep{montes:2023} or Vera Rubin Observatory's Legacy Survey of Space and Time \citep[LSST, see][]{ivezic:2019,brough:2020}, as well as with improved methods \citep{canepa:2025}, it is of increasing importance to have a tool capable of inferring dynamical states for the expected large number of galaxy clusters. 

In this work we utilize three large hydrodynamical cosmological simulation suites, namely Magneticum Pathfinder \citep[][Dolag et al., in prep]{teklu15}, IllustrisTNG \citep{Nelson2019}, and Horizon-AGN \citep{dubois:2014}, as well as the cosmological hydrodynamical zoom-in simulation set of galaxy clusters Hydrangea \citep{bahe17,Barnes_et_al_2017}, to investigate the connection between the ICL/BCG and the dynamical age of galaxy clusters. Of these, we focus in particular on the Magneticum simulations, as they have a statistically meaningful number of galaxy clusters over the full mass range from $M_\mathrm{vir} = 10^{14}M_\odot$ to $M_\mathrm{vir} = 2\times10^{15}M_\odot$ (for a statistical error $\sigma\propto \sqrt{N}$ we have a signal-to-noise ratio of $N/\sqrt{N}\approx30$). However, to confirm that all predicted trends are independent of resolution or included physics, we compare with the other three simulations which have a higher particle resolution. This set of four simulations has already been shown by \citet{brough24} to produce similar fractions of ICL when observational techniques are run on self consistently-generated mock images.

In Sec.~\ref{sec:sims_methods} the four different simulations are introduced. Sec.~\ref{sec:results} discusses the evolution of the ICL+BCG fraction with time and their connection to the dynamical state of clusters. Furthermore, the average disruption rates of satellites is calculated. Finally, in Sec.~\ref{sec:disc} the results are summarized and discussed. Throughout this work, we assume the native cosmology of each of the simulations as described in Section~\ref{sec:sims_methods}.

\section{Simulations}\label{sec:sims_methods}

This study is performed using three different hydrodynamical cosmological simulation suites, Box2~hr of the Magneticum Pathfinder suite, TNG100 of the IllustrisTNG suite, Horizon-AGN, as well as the Hydrangea zoom-in simulation suite, which vary in their included physics, resolutions and box volumes. As we aim to quantify the connection between cluster mass assembly and the fraction of stellar mass in the BCG and ICL, we require a statistically significant number of clusters to cover a sufficient diversity in assembly histories. To this end, we focus in particular on the Magneticum Pathfinder simulations which provide the largest simulation volume and thus the highest number of clusters by $z=0$, reaching $886$~galaxy clusters of $M_\mathrm{200c}\geq10^{14} M_\odot$ for the Box2~hr simulation.

The other two uniform-volume simulations have a higher resolution but significantly fewer galaxy clusters, with~14 of $M_{200c}>10^{14}M_\odot$ for both TNG100 and Horizon-AGN. The Hydrangea zoom-in simulations meanwhile have~46~such galaxy clusters. The set of simulations used for this work has already been shown in the previous study by \citet{brough24} to provide galaxy clusters with comparable properties in terms of their ICL and BCG when applying the same criteria, providing reasonable ground for a further comparison. Following the approach introduced in that work, we will also calculate all quantities used in this study in the exact same way for all simulations, as described below. 

The clusters for Horizon-AGN and TNG100 are equivalent to those selected by \citet{brough24}. For Hydrangea we use all available 46~clusters, while for Magneticum Pathfinder Box2~hr we include most of the $886$~total clusters by $z=0$, removing a few clusters with assembly histories which break before $z=3$ to ensure the same number of structures are followed at every point in time. Our final sample thus consists of 14, 14, 46 and 868 galaxy clusters with $M_\mathrm{200c}\geq10^{14} M_\odot$ from Horizon-AGN, TNG100, Hydrangea and Magneticum Pathfinder Box2~hr, respectively. The details of these four cosmological hydrodynamical simulations are described in more detail by \cite{brough24}. Here, we provide only a brief summary and refer the reader to that work for more details, especially to the extensive Table~A1 comparing the different simulation properties. 

\begin{figure}
    \includegraphics[width=\columnwidth]{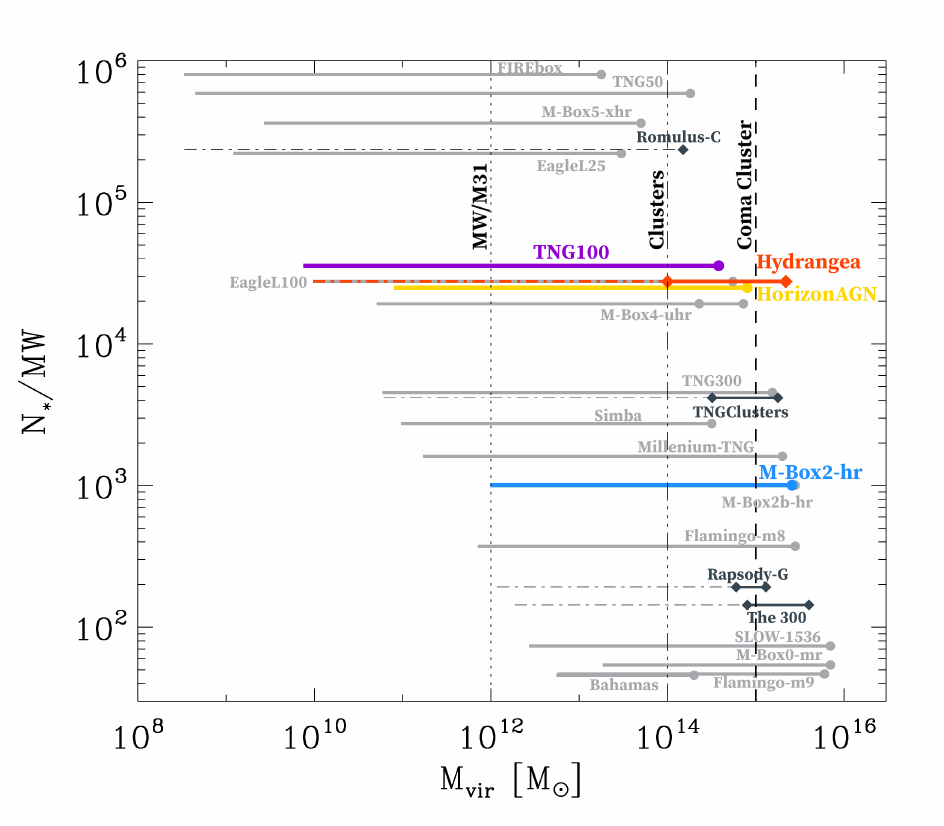}
    \caption{The number of stellar particles equating to the Milky Way-like (MW) stellar mass versus the halo mass covered by the simulation. The simulations used for this study are shown in color, with Magneticum Box2~hr shown in blue, TNG100 shown in purple, Horizon-AGN shown in gold, and Hydrangea shown in orange. Other commonly known simulations are shown in gray. 
    Full box cosmological simulations are in light gray, with $M_\mathrm{vir,max}(z=0)$ indicated by a filled circle while the line goes down to the resolution limit of 1000~dark matter particles. Zoom-in simulations are plotted in dark gray, going from the most massive halo in the set (diamond) down to the same 1000~dark matter particle limit (dashed thin line).
    Vertical dotted lines mark the MW halo mass, the $10^{14}M_\odot$~mass threshold used to define clusters, and the Coma cluster mass.
    Full box simulation suites shown are: Magneticum Pathfinder (Dolag et al., 2025), IllustrisTNG \citep{nelson:2019}, Horizon-AGN \citep{dubois:2014}, SLOW \citep{dolag:2023,seidel:2024}, EAGLE \citep{Schaye.2015}, Flamingo \citep{schaye:2023}, Millenium-TNG \citep{pakmor:2023}, and Bahamas \citep{mccarthy:2017}.
    Zoom-in simulation suites shown are: Hydrangea \citep{bahe17}, Romulus-C \citep{tremmel:2019}, TNG-Clusters \citep{nelson:2024}, Rapsody-G \citep{hahn:2017}, and The300 Project \citep{cui:2018}.
    }
    \label{fig:simres}
\end{figure}

For a comparison between the different simulations, Fig.~\ref{fig:simres} shows the number of stellar particles that the typical progenitor galaxy building up the ICL, which is about Milky-Way (MW) mass, $M^\mathrm{MW}_\mathrm{vir} = 1\times10^{12}M_\odot$ and $M^\mathrm{MW}_\mathrm{*} = 5\times10^{10}M_\odot$ \citep{Contini2014,brown:2024}, would have given the simulation stellar particle resolution, as a function of the range of halo masses covered by the simulations. The upper bound is given by the maximum mass of a halo at $z=0$, while the lower bound is given by the mass of~$1000$~DM particles. This clearly demonstrates the reason for using both the large Magneticum Box2~hr volume (blue) in combination with the smaller but higher resolution simulations of TNG100 (purple), Horizon-AGN (gold), and Hydrangea (orange), as the former is required for covering the galaxy clusters in a statistical manner, while the latter are better resolved but lack sample size for a statistical approach. 

\subsection{Magneticum Pathfinder}
For the majority of this work we focus on a set of clusters from the Magneticum Pathfinder\footnote{\url{www.magneticum.org}} hydrodynamical cosmological simulations, a suite of~6~different simulation volumes with three different resolution levels. For this work, we use in particular Box2~hr, the simulation that has a large volume of $(352\mathrm{cMpc}/h)^3$, with particle masses for dark matter of $m_\mathrm{dm}=9.8\times10^{8} M_{\odot}$ and for gas of $m_\mathrm{gas} = 2\times10^{8} M_{\odot}$. 
Because gas particles can spawn up to four generations of stellar particles, the mass resolution in the stellar component is a factor four higher, with masses of $m_\mathrm{*}\approx m_\mathrm{gas}/4\approx5\times10^{7} M_{\odot}$, allowing for galaxies of $M_\mathrm{*}\geq 10^{10} M_\odot$ to be sufficiently resolved with $200$~stellar particles. Box2~hr harbors a significant number of clusters by $z=0$, with $886$~galaxy clusters of $M_\mathrm{200c}\geq10^{14} M_\odot$. The properties of the galaxy clusters from this particular simulation have already been used to study the dynamics of the ICL and BCG components by \citet{remus17}, and their galaxy populations have been investigated and compared to observations by \citet{lotz19, lotz:2021}.

The softening lengths for the particle types are $\epsilon_\mathrm{dm} = \epsilon_\mathrm{gas} = 5.3~\mathrm{kpc}$ and $\epsilon_\mathrm{*} = 2.84~\mathrm{kpc}$, while the assumed cosmology is given by WMAP-7 \citep{komatsu11}, with $h=0.704$, $\Omega_m = 0.272$, $\Omega_b = 0.0451$, $\Omega_\lambda = 0.728$, $\sigma_8 = 0.809$ and $n_s = 0.963$. The simulation was performed with a modified version of \textsc{GADGET-2} \citep{springel05}, with upgrades covering the smoothed particle hydrodynamics (SPH), thermal conduction \citep{dolag05} and artificial viscosity \citep{dolag04}, among others \citep{donnert13,beck:2015}. The implementation of the baryonic physics is described in more detail by \citet{teklu15}. We identify galaxies and galaxy clusters using the structure finder \textsc{SUBFIND} \citep{Springel_et_al_2001,dolag09}.

\subsection{IllustrisTNG}
The Next Generation Illustris\footnote{\url{www.tng-project.org}} simulations (IllustrisTNG) are a suite of cosmological hydrodynamical simulations, of which we use the intermediate-sized medium-resolution volume called TNG100 \citep{Marinacci2018,Naiman2018,Nelson2018,pillepich18b,Springel2018}. The simulation has a volume of $(75\mathrm{cMpc}/h)^3$, and particle masses of $m_\mathrm{dm}=7.5\times10^{6} M_{\odot}$ for the dark matter and $m_\mathrm{*,gas} = 1.4\times10^{6}M_\odot$ for the stellar and gas masses. 
This results in galaxies of $M_\mathrm{*}\geq 10^{10} M_\odot$ to be highly resolved with around $7000$~stellar particles.
At $z=0$, the simulation includes $14$~galaxy clusters with masses of $M_\mathrm{200c}\geq10^{14} M_\odot$. Many previous works have considered both the total stellar content \citep[e.g.,][]{pillepich18b} as well as galaxy populations \citep[e.g.,][]{stevens:2018,donnari:2021} of the simulation galaxy clusters.

For the particle species, the softening lengths are $\epsilon_\mathrm{dm} = \epsilon_\mathrm{*} = 0.74~\mathrm{kpc}$, while the spatial resolution of the gas cell is adaptive and is better at higher densities. The assumed cosmology is given by the Planck results \citep{Planck2016}, with $h=0.6774$, $\Omega_m = 0.3089$, $\Omega_b = 0.0486$, $\Omega_\lambda = 0.6911$, $\sigma_8 = 0.8159$ and $n_s = 0.9667$.
The simulation was performed with the moving-mesh code \textsc{AREPO} \citep{Springel2010}, with updates as described by \citet{pillepich18a,pillepich18b}. 
Galaxies and galaxy clusters are identified via the structure finder \textsc{SUBFIND} \citep{Springel_et_al_2001,dolag09}.

\subsection{Horizon-AGN}
Horizon-AGN \citep{dubois:2014} is a cosmological hydrodynamical simulation with a volume of $(100\mathrm{cMpc}/h)^3$, a dark matter particle resolution of $m_\mathrm{dm}=8\times10^7M_\odot$, and a stellar particle resolution of $m_\mathrm{*}=2\times10^6M_\odot$. 
This results in galaxies of $M_\mathrm{*}\geq 10^{10} M_\odot$ being well resolved with approximately $5000$~stellar particles.
At $z=0$, the simulation includes $14$~galaxy clusters with masses above $M_\mathrm{200c}\geq10^{14} M_\odot$. Prior work has been done on the build up of the ICL \citep{brown:2024}, and the shape \citep{okabe:2018}, spin \citep{choi:2018} and star formation histories of cluster galaxies \citep{jeon:2022}.

The assumed cosmology is given by WMAP-7 \citep{komatsu11}, as for Magneticum (i.e., $h=0.704$, $\Omega_m = 0.272$, $\Omega_b = 0.045$, $\Omega_\lambda = 0.728$, $\sigma_8 = 0.81$ and $n_s = 0.967$).
The simulation was performed using the adaptive mesh refinement (AMR)-based Eulerian hydrodynamics code \textsc{RAMSES} \citep{Teyssier.2002}. Since the simulation was performed with a grid code, the initially-uniform $1024^3$ cell gas grid is refined according to a quasi-Lagrangian criterion, with the smallest cell sizes fixed at 1 physical kpc. For mode details on the technical implementations, see \citet{dubois:2014}.
We identify galaxies and clusters using the \textsc{AdaptaHOP} structure finder \citep{Tweed2009} performed separately on the stars and DM. Structures are identified on the basis of local particle density with no unbinding procedure preformed. Structures and substructures are assembled from the groups created by linking particles with densities greater than 178 and 80 times the total matter density for stars and DM respectively according to their closest local density maxima. \textsc{AdaptaHOP} then links structures based on saddle points in the density field.

\subsection{Hydrangea}
Hydrangea \citep[see also \citealt{Barnes_et_al_2017}]{bahe17} is a suite of 24 cosmological hydrodynamical zoom-in simulations of massive galaxy clusters using a variant of the EAGLE simulation model \citep{Schaye.2015}. The parent simulation has a volume of $(2252\,\mathrm{cMpc}/h)^3$. The particle resolutions of the Hydrangea clusters are $m_\mathrm{dm}= 9.7\times10^6M_\odot$ for dark matter, and $m_\mathrm{gas,*} \approx 1.8\times10^6M_\odot$, for gas and stars, respectively. 
This results in galaxies of $M_\mathrm{*}\geq 10^{10} M_\odot$ to be well resolved with $\gtrsim 6000$~stellar particles. Softening lengths are $\epsilon_\mathrm{dm} = \epsilon_\mathrm{gas} = \epsilon_\mathrm{*} = 0.7~\mathrm{kpc}$, while the assumed cosmology is given by the Planck results \citep{Planck2014}, with $h=0.6777$, $\Omega_m = 0.307$, $\Omega_b = 0.04825$, $\Omega_\lambda = 0.693$, $\sigma_8 = 0.8288$ and $n_s = 0.9611$. 

While the parent simulation contains $>\,10^5$ galaxy clusters, the Hydrangea set of zoom-in simulations contains $54$ galaxy clusters with masses above $M_\mathrm{200c}\geq10^{14} M_\odot$. We here use a subsample of~46 clusters, with masses of $14 \leq \log_{10}(M_\mathrm{200c}\,[\mathrm{M}_\odot]) < 15.4$, thus expanding on the set used by \citet{brough24}. Previous works have shown that the stellar mass function of cluster galaxies in Hydrangea agrees well with observations, both at $z \approx 0$ \citep{bahe17} and $z \sim 1$ \citep{Ahad_et_al_2021}. Although star formation in satellites is very efficiently quenched, DM stripping leads to a $\approx$1 dex positive offset of the
stellar-to-halo mass relation from the field \citep{bahe17}. As shown by \citet{Bahe2019}, only a few per cent of satellite galaxies accreted after $z \approx 2$ are fully disrupted. Nevertheless, the simulations predict a substantial ICL component that accounts for $\sim20\%$~of the total stellar mass even on group scales \citep[see also \citealt{brough24}]{Ahad2023}.

\begin{figure*}
    \includegraphics[width=\textwidth]{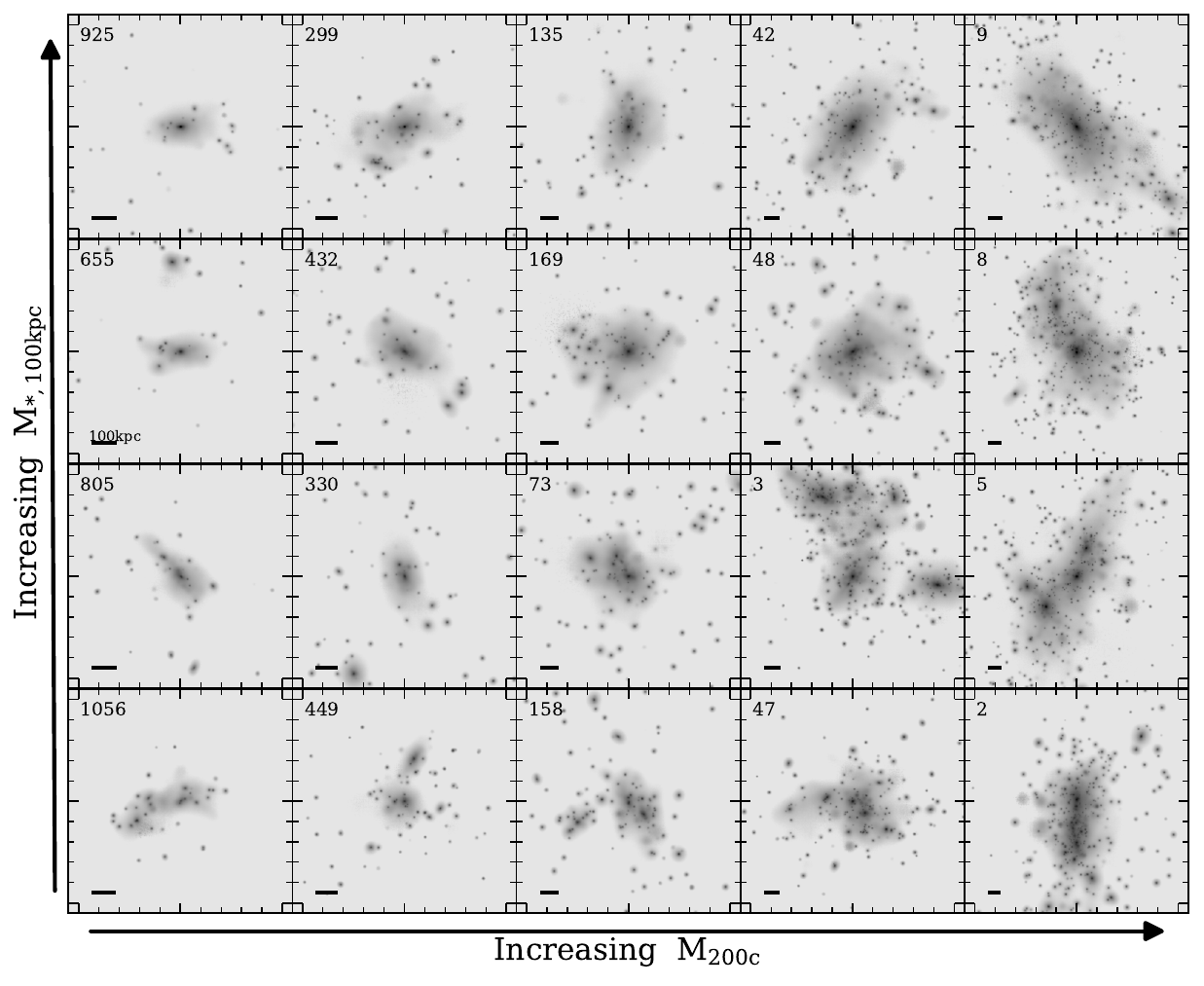}
    \caption{r-band mock images of example galaxy clusters from the Magneticum Box2~hr simulation in a random projection, with increasing mass from left to right and cluster~IDs indicated in the top left. Clusters in the left column have about $M_\mathrm{200c} = 1\times10^{14}M_\odot$, with clusters on the right having $M_\mathrm{200c} > 1\times10^{15}M_\odot$. From top to bottom we plot galaxy clusters with decreasing BCG stellar mass (within $100$~kpc, excising satellites) within a narrow bin of cluster halo mass. Each panel shows the full $R_\mathrm{200c}$ size of the cluster, with the black line in the bottom left of each panel marking the effective length of $100$~kpc while the axes ticks are in units of $0.2 r_\mathrm{200c}$. 
    }
    \label{fig:clusters}
\end{figure*}

The sub-grid parameters are the same as for the 50~Mpc `AGNdT9' simulation of the EAGLE suite, which differs from the `Reference' model as used for the 100~Mpc EAGLE simulation in the AGN heating temperature increment $\Delta T$ and viscosity parameter $C_\mathrm{visc}$ \citep{Schaye.2015}. The simulations were performed with a significantly modified version of the SPH code \textsc{GADGET-3} (last described by \citealt{springel05}). Major updates include the `Anarchy' pressure-entropy SPH scheme \citep{schaller:2015} as well as subgrid physics models for reionization, gas cooling, star formation, metal enrichment, and stellar and AGN feedback \citep[see also \citealt{crain:2015}]{Schaye.2015}. 
As for TNG100 and Magneticum, we identify galaxies and galaxy clusters via the structure finder \textsc{SUBFIND} \citep{Springel_et_al_2001,dolag09}.

\subsection{Property Definitions}

Throughout this work, we define our quantities within the radius enclosing 200~times the critical density, $R_\mathrm{200c}$. Halo mass is then given as the total mass in that radius, $M_\mathrm{200c}$. The total stellar mass $M_\mathrm{*,tot}$ is the sum of all stellar mass in $R_\mathrm{200c}$, while the ICL+BCG stellar mass $M_\mathrm{*,ICL+BCG}$ is the sum of all stellar mass within $R_\mathrm{200c}$ that is assigned to the central subhalo of the cluster, i.e., excluding mass in satellites. The ratio between the stellar mass in the ICL+BCG to total is given as $f_\mathrm{ICL+BCG}=M_\mathrm{*,ICL+BCG}/M_\mathrm{*,tot}$, and is by definition smaller than unity. Similarly, mass ratios to satellite galaxies are determined for satellites whose positions lie within $R_\mathrm{200c}$ of the cluster center as given by the BCG.

\section{Results}\label{sec:results}
As the main goal of this study is to understand to what extent the ICL and the BCG can be used as tracers for the evolution of galaxy clusters, we will first investigate the cluster populations with respect to their evolution before we connect them to the ICL and BCG properties. This also provides the basis for comparisons between the different simulations. Unless otherwise noted, we define the halo mass as $M_{200c}$, which is the total mass within a sphere of mean density of $200$~times the critical density. 

\begin{figure*}
    \includegraphics[width=\textwidth]{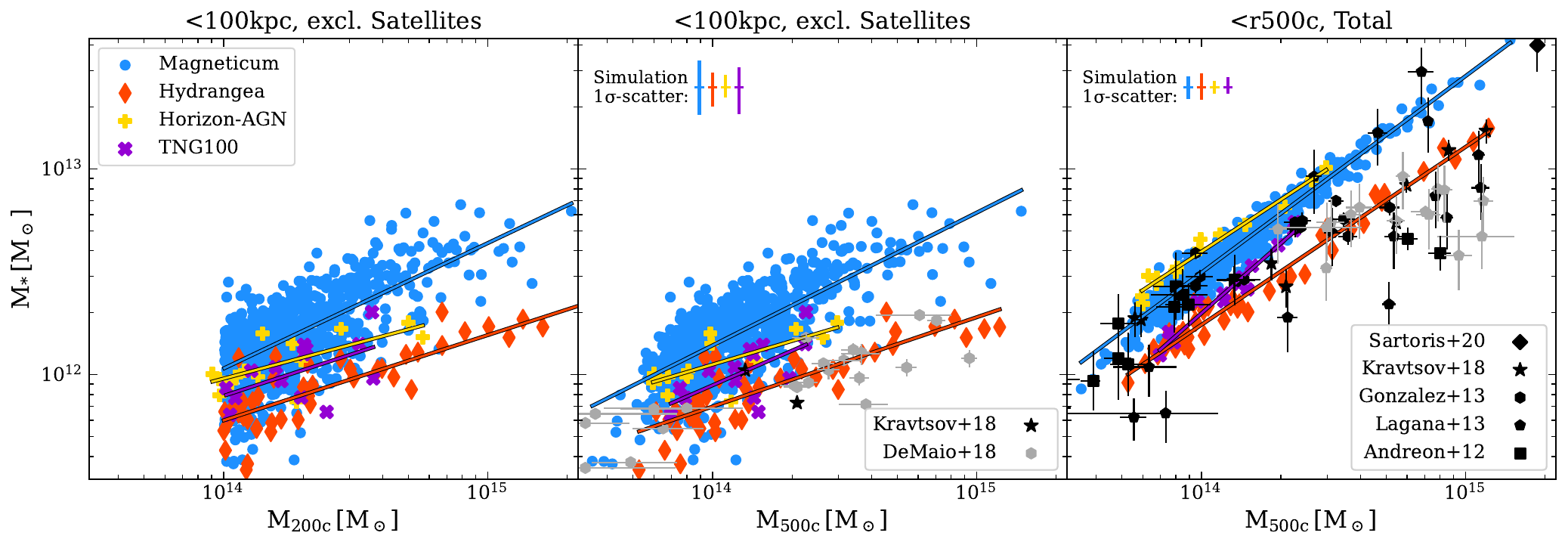}
    \caption{\textit{Left:} Stellar mass of the central BCG within $100$~kpc versus the total halo mass $M_\mathrm{200c}$ for the four simulations, scaled to a common $h$. Magneticum is plotted in blue, IllustrisTNG in purple, Hydrangea in orange and Horizon-AGN in gold. \textit{Center:} The same stellar mass plotted against the halo mass as given by $M_\mathrm{500c}$. Also plotted are observations by \citet{demaio18} and \citet{kravtsov18}. \textit{Right:} The total stellar mass (BCG, ICL and satellites) against $M_\mathrm{500c}$ for the simulations, as well as observations by \citet{andreon:2012,lagana:2013,gonzalez13,budzynski14,kravtsov18,sartoris:2020}.
    }
    \label{fig:smhm}
\end{figure*}

\subsection{Populations of Galaxy Clusters}\label{subsec:pop}
Galaxy clusters come in a variety of shapes and sizes, with the usual definition spanning over an order of magnitude in mass from $M_\mathrm{200c}\geq10^{14}M_\odot$ to $>10^{15}M_\odot$. Their differences result from the different assembly histories, with some assembling a majority of their mass already early in the Universe \citep[e.g.,][]{sorce:2020,remus:2023}, while others have only recently undergone major mergers, such as the Coma cluster \citep[e.g.,][]{burns:1994,sanders:2013,seidel:2024} or the Abell~2744 cluster \citep[e.g.,][]{owers:2011,kimmig23}. 

Fig.~\ref{fig:clusters} shows r-band mock images of 20~example galaxy clusters ranging from $M_\mathrm{200c}=1\times10^{14}M_\odot$ to $M_\mathrm{200c}\approx 1.5\times10^{15}M_\odot$. The clusters are ordered to have increasing $M_\mathrm{200c}$ from left to right, with clusters in the same column having the same $M_\mathrm{200c}$. The rows go from the highest to the lowest BCG mass (top to bottom), where, as motivated by \citet{brough24}, we use $M_\mathrm{*,100kpc}$, defined as the stellar mass within $100\mathrm{kpc}$, excluding satellites. For each cluster, we show the projected image of radius $R_\mathrm{200c}$ out to a depth of $3\cdot R_\mathrm{200c}$.
As can be seen immediately, the clusters show a larger amount of diffuse light with increasing halo mass (i.e., from left to right). However, it is also clear that in the galaxy clusters with the highest BCG masses (top row) have relatively less stellar mass (and therefore luminosity) in the satellites. This implies more recent merging activity for those clusters with lower BCG masses, which is indeed what we find. The galaxy clusters in the bottom rows commonly host two massive galaxies that would make the definition of a single BCG difficult. 

As structure merges into the growing cluster, it brings with it a majority of its mass in the form of dark matter, as well as baryons in the form of stars and gas. Depending on the nature of the orbit of the merging galaxies, not all of the stars may reach into the deepest parts of the galaxy cluster potential \citep{karademir:2019}, resulting in a range of stellar masses for the central BCG. This is also the origin for the scatter in the stellar mass-halo mass relation that we show in the left panel of Fig.~\ref{fig:smhm}, with a range of around~$0.7$~dex in $M_\mathrm{*,100kpc}$ at a given halo mass. Note that this scatter is entirely driven by the central BCG+ICL mass, as we exclude any contributions from satellites. For our set of example clusters in Fig.~\ref{fig:clusters}, they range in $M_\mathrm{*,100kpc}/M_\mathrm{200c}$ from $3.4$\%~for cluster~ID$\,925$ (top left) to $0.5$\%~for cluster~ID$\,158$ (bottom of the third column). Because the higher resolved simulations TNG100, Hydrangea and Horizon-AGN (purple, orange and gold) show a similar scatter in the left panel of Fig.~\ref{fig:smhm}, we can conclude this to be inherent in the way mass accretes onto the galaxy clusters. Finally, as also noted by \citet{brough24}, all simulations reproduce the underlying BCG-halo mass relation \citep[e.g.,][]{Brough2008,Lidman2012}, though there are some differences. Magneticum and Horizon-AGN tend toward higher stellar masses at equivalent halo mass relative to Hydrangea and TNG100. Crucially, however, the slopes of the $M_*-M_\mathrm{halo}$ relation are similar. 

To see that this scatter is not due to large fluctuations in the halo mass, which may be driven by recently infalling structure which resides far away from the BCG, we can instead plot $M_\mathrm{*,100kpc}$ against $M_\mathrm{500c}$ (middle panel of Fig.~\ref{fig:smhm}). We find that the range in stellar masses at a given $M_\mathrm{500c}$ is still significant, and comparable for Magneticum~Box2~hr to that found for TNG100. This scatter also matches the $\pm0.13\,$dex found by \citet{pillepich18b} for TNG300, the larger volume box of the IllustrisTNG suite, with the same galaxy formation model of TNG100. When comparing to individual galaxy cluster observations by \citet{demaio18} and \citet{kravtsov18}, we find that the simulations tend toward higher stellar masses compared to the observations. We note that we have rescaled all masses to a common $h$, and that observations at $z>0.2$ are plotted in light-gray, which may at least in part be a cause of the differences. The size of the offset may also be mass dependent, where observational measurements for larger clusters could be biased towards lower masses as compared to the simulations \citep{kravtsov18}.

In light of recent findings by \citet{popesso:2024}, who find Magneticum best reproduces the observed hot gas mass fractions in galaxy clusters, it is interesting to note that Magneticum lies at the highest stellar masses. This may imply the need for even stronger feedback, to neither overproduce the amount of stars nor the amount of hot gas in clusters. Alternatively, given that observations at high redshifts are finding significantly enhanced number densities of quenched galaxies compared to most cosmological simulations \citep{carnall:2023,valentino:2023}, this may instead indicate the need for an earlier onset of significant AGN feedback \citep{kimmig:2023q}.

Notably, the scatter found in the simulations in Fig.~\ref{fig:smhm} is also seen between the individual observations. To see there persists such a variety also for the total stellar content of galaxy clusters, we plot in the right panel of Fig.~\ref{fig:smhm} the total stellar mass in the cluster $M_\mathrm{*,tot}$ (BCG+ICL+Satellites) against $M_\mathrm{500c}$. We find a significant reduction in the scatter for all simulations, as well as closer agreement to observations of individual galaxy clusters by \citet{andreon:2012,lagana:2013,gonzalez13,kravtsov18,sartoris:2020}. As slopes of the $M_\mathrm{*,tot}$-$M_\mathrm{500c}$ relation, the simulations find~$0.96\pm0.02$, $0.86\pm0.05$, $0.84\pm0.07$~and~$1.13\pm$ for Magneticum, Hydrangea, Horizon-AGN and TNG100, respectively, where it is of note that \citet{pillepich18b} find a less steep $0.84$~for TNG300. These slopes are comparable to those found by observations, with $0.52\pm0.04$, $1.05\pm0.05$~and~$0.69\pm0.9$ found by \citet{gonzalez13}, \citet{budzynski14} and \citet{kravtsov18}. This could imply either that the larger relative stellar masses in the BCGs found in the simulations result from faster stripping of the satellites, or that the process of masking the satellites in observations also removes a significant amount of stars belonging to the diffuse ICL. Further investigation into this is needed in a more detailed comparison with observations, which will be the focus of a follow-up study to this work.

It is interesting to note, aside from TNG and \citet{budzynski14}, the slope of the $M_\mathrm{*,tot}$-$M_\mathrm{500c}$ relation lies below that of self-similarity (slope of~1). As we expect the great majority of the stellar mass at the cluster scale to be accreted \citep[e.g.,][]{pillepich18b,remus:2022}, this may indicate that satellites in smaller clusters of $M_\mathrm{500c}=10^{14}M_\odot$ possess ongoing star formation \citep[e.g.,][]{brown:2024}, while the more massive clusters are relatively more efficient in suppressing their satellite galaxies' star formation. This is consistent with the findings by \citet{lotz19}, that gas is stripped more quickly in more massive clusters. Alternatively, the infalling regions for lower mass clusters may on average include galaxies that were more efficient at converting gas into stars, so which are potentially earlier formed when the baryon conversion efficiency was higher \citep{boylan-kolchin:2023,turner:2025}, as compared to more massive clusters. 

We further note that there are considerable differences between individual observations of total stellar versus halo mass. Halo masses based on X-ray gas using hydrostatic assumptions suffer from hydrostatic mass bias \citep{nelson:2014} while weak-lensing measurements are impacted by weak lensing bias \citep{grandis:2024}. Stellar masses are impacted by uncertain mass-to-light ratios due to the not well known low-mass slope of the initial mass function, as well as a degeneracy between age and metallicity when stellar populations are inferred using broad-band colors \citep{swindle:2011,leauthaud:2012}.

\begin{figure}
    \includegraphics[width=\columnwidth]{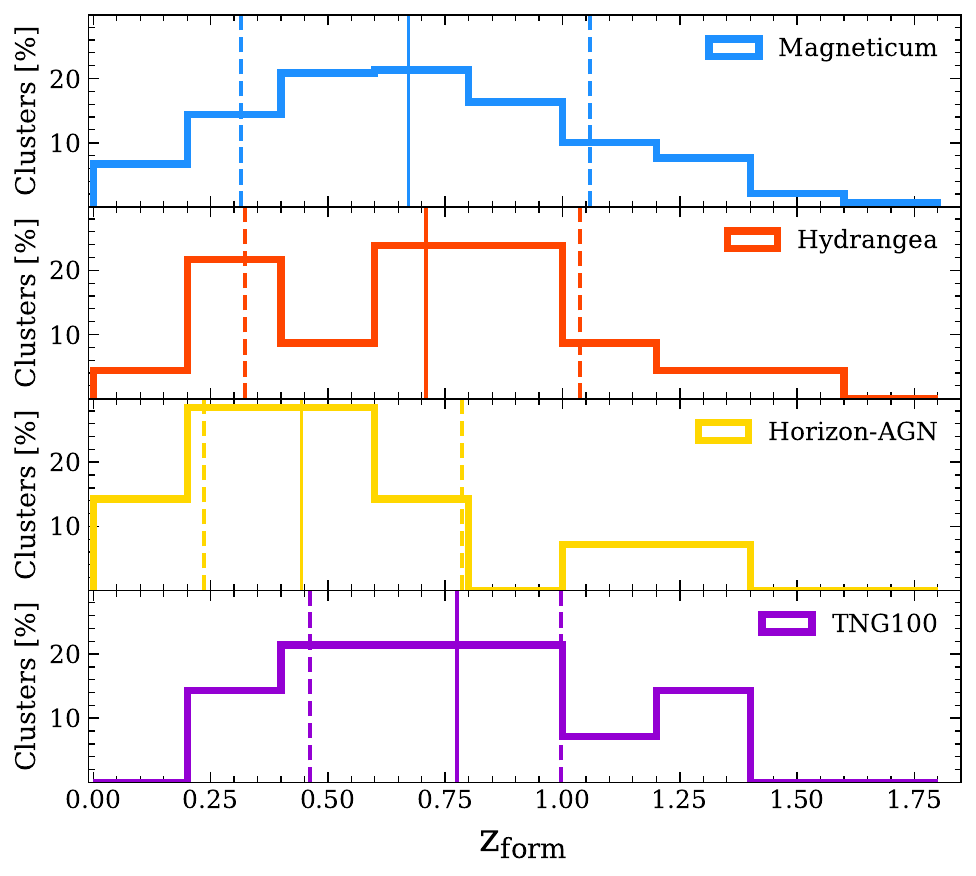}
    \caption{Histograms of the formation redshifts for all clusters from Magneticum Box2~hr (blue), Hydrangea (orange), Horizon-AGN (gold), and TNG100 (purple), going from top to bottom. Median (solid) and 1-$\sigma$~bounds (dashed) are indicates as vertical lines.
    }
    \label{fig:zformhist}
\end{figure}

\begin{figure*}
    \includegraphics[width=\textwidth]{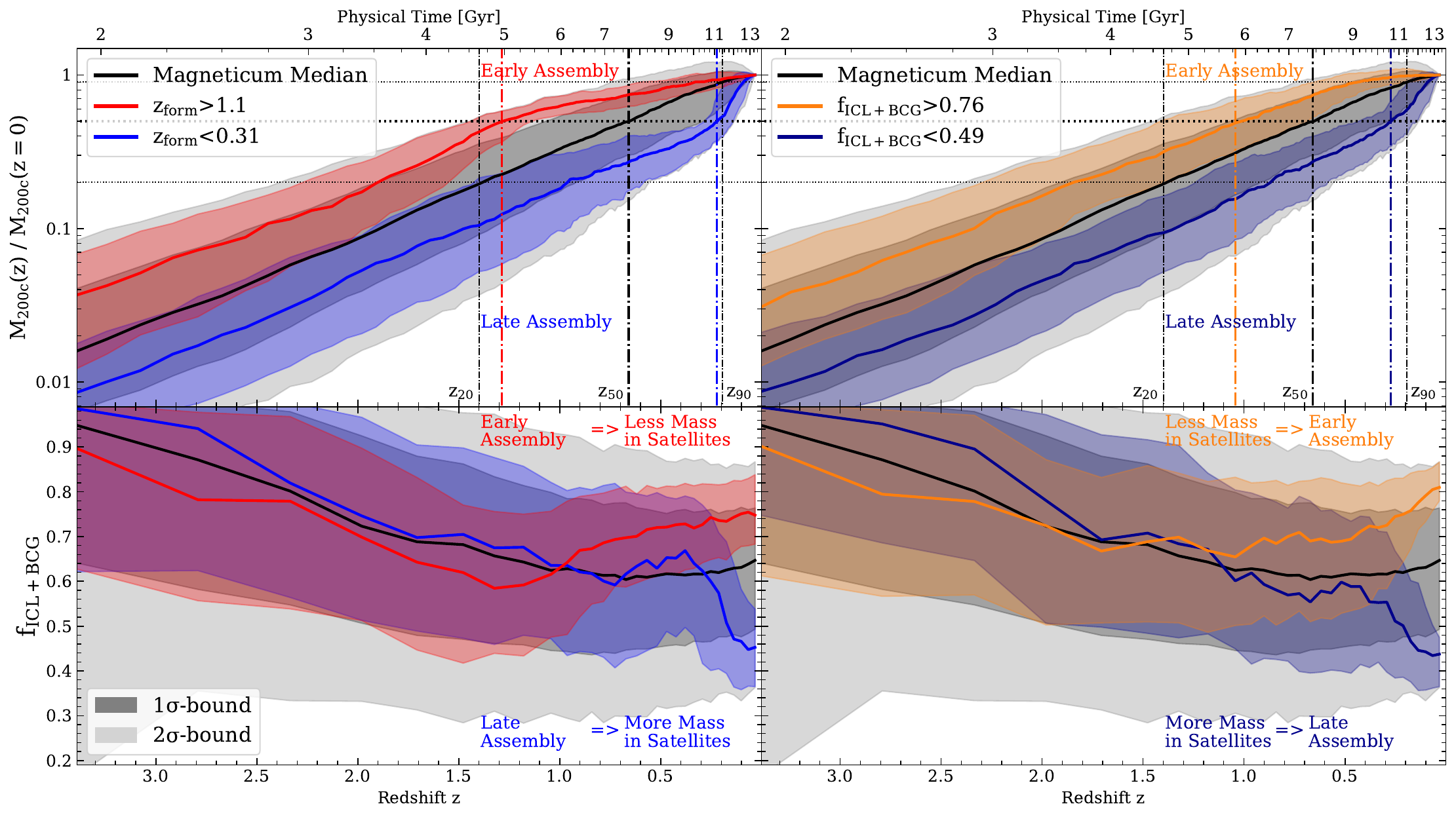}
    \caption{\textit{Upper row:} Evolution of the halo mass $M_\mathrm{200,c}$ with redshift $z$ for all clusters in the Magneticum~Box2~hr simulation, normalized to the final mass $M_\mathrm{200,c}(z=0)$. The black line shows the median, with the dark shaded area marking the $1\sigma$-bounds and the light shaded area marking the $2\sigma$-bounds. The colored lines and their $1\sigma$-bounds show the evolution of a subset of clusters split by their formation redshift (\textit{left}) or their $f_\mathrm{ICL+BCG}$ (\textit{right}) as given in the legend. Black dash-dotted vertical lines mark the redshifts at which the average cluster has accumulated 20\%, 50\%, and 90\% of its total mass (horizontal black lines), marked by $z_\mathrm{20}$, $z_\mathrm{50}$, and $z_\mathrm{90}$, respectively. Note that $z_\mathrm{50}$ is what we define as formation redshift of a cluster. 
    \textit{Lower row}: Evolution of $f_\mathrm{ICL+BCG}$ with redshift $z$. As in the upper panel, the black line marks the average evolution of all clusters from Magneticum, with the gray shaded areas showing the $1\sigma$- and $2\sigma$-bounds. The colored lines and regions show the median and $1\sigma$-bounds for the same subsets of galaxy clusters. 
    }
    \label{fig:tracks}
\end{figure*}

\subsection{Defining Cluster Dynamical State}\label{subsec:state}
Given the large scatter in BCG masses at a given halo mass, we now turn to consider how clusters grow. This means both their average total mass assembly, as well as how much of their stellar mass is already stripped/merged into the BCG or ICL relative to the total amount present. 

To do so, we first require a proxy of the mass assembly history. We chose here the formation redshift $z_\mathrm{form}$, which is defined as the redshift when the galaxy cluster first reached half of its final halo mass $M_\mathrm{200c}$ at $z=0$. A higher $z_\mathrm{form}$~means an earlier formation and thus higher degree of relaxation of the galaxy cluster relative to those with lower $z_\mathrm{form}$. This is a definition commonly used in the literature \citep[e.g.,][among others]{boylan:2009,power:2012}. We note that by using $z_\mathrm{form}$, we care particularly about the galaxy clusters dynamical evolution through cosmic time and the question of whether a given cluster lived in an early or late collapsing node, and care less about the most recent merging activity. Different definitions of the dynamical state or relaxedness of a galaxy cluster may be better traced by other parameters \citep{haggar24}, such as the X-ray center shift, given that the hydrodynamical gas component is stripped and relaxes on faster timescales than the dispersionless stars and dark matter. A more thorough look of which parameters are best suited across different timescales is needed, though it is beyond the scope of this study. 

Fig.~\ref{fig:zformhist} shows the distribution of the formation redshift $z_\mathrm{form}$ for all clusters from the Magneticum simulations (blue) and the other three simulations. 
For the large number of Magneticum Box2~hr clusters, we find that the average galaxy cluster assembled half of its mass by a redshift of $z_\mathrm{form}\approx0.67$, or just over~$6$~Gyr ago. This is in agreement with previous results for large cosmological dark matter only simulations, where \citet{power:2012} find average formation redshifts of $z_\mathrm{form}\approx0.7$ for clusters of about $M_\mathrm{vir}=10^{14}M_\odot$, and of $z_\mathrm{form}\approx0.45$ for clusters of about $M_\mathrm{vir}=10^{15}M_\odot$. Similar results are also reported for the Millennium and Millennium-II simulations by \citet{boylan:2009}, although these values are slightly older than what is predicted from the Press-Schechter Theorem \citep[see][]{power:2012}.

The range of formation redshifts found for clusters from Magneticum reaches from formation redshifts as early as $z_\mathrm{form}=1.8$ to as late as $z_\mathrm{form}=0.06$.
The other simulations used in this work also find a large spread in the formation redshift of their galaxy clusters, despite the lower number statistics of TNG100, Horizon-AGN and Hydrangea, as shown in Fig.~\ref{fig:zformhist}. The average formation redshift between the samples varies slightly, with the Horizon-AGN clusters being the youngest at $\langle z_\mathrm{form}^\mathrm{Horizon}\rangle=0.44$, both Magneticum and Hydrangea at similar formation times with $\langle z_\mathrm{form}^\mathrm{Magneticum}\rangle=0.67$ and $\langle z_\mathrm{form}^\mathrm{Hydrangea}\rangle=0.71$, while TNG100 has the earliest forming clusters at $\langle z_\mathrm{form}^\mathrm{TNG}\rangle=0.78$.

We note that both TNG100 and Horizon-AGN lack clusters with $z_\mathrm{form}>1.4$, as well as show a tighter spread. This is likely because they are simulations of about $100\,$Mpc boxlength and thus lack the largest modes of the initial power spectrum of the density fluctuations \citep[e.g.,][]{remus:2023}. As the Magneticum Box2~hr is a factor~$3$ larger in box-length, this equates to a factor of~$27$~in volume, and thus it includes not only significantly more galaxy clusters, but also more of the larger modes that can collapse earlier, leading to a larger variety in formation times. 

The Hydrangea clusters, meanwhile, are drawn from a very large cosmological parent box of $3200\,$pMpc boxlength and thus show a similar spread that includes both the earliest and latest forming galaxy clusters. It is interesting that their selection, requiring them to be in relative isolation (no more massive halo within $30\,$pMpc~or~$20R_\mathrm{200c}$, whatever is larger), does not seem to imprint in the formation redshifts, unlike what may be typically assumed for more isolated nodes in the cosmic web \citep[see also][for more details on this effect of early formation and starvation]{remus:2023,seidel:2024}. Overall, Fig.~\ref{fig:zformhist} demonstrates the large variety in formation histories for galaxy clusters, making it a non-trivial problem to distinguish those from observations.

\subsection{The Fraction of ICL+BCG and Cluster Dynamical State}\label{subsec:dyn}

To get a better understanding of how the mass of these galaxy clusters is assembled, we plot in the upper panels of Fig.~\ref{fig:tracks} the amount of halo mass $M_\mathrm{200c}(z)$ present in the galaxy clusters of the large Magneticum sample through time, relative to their final halo masses $M_\mathrm{200c}(z=0)$. The black line shows the median growth of all clusters, with the dark and light gray shades marking the $1\sigma$- and $2\sigma$-bounds. 

The overall spread in the histories increases the farther back we look, with the $1\sigma$-bounds reaching $0.5$~dex (or a factor of~3 in range) by $z=1$. On one hand, it is remarkable that the relative growth over the last nearly $8$~Gyr remains on average within a factor of~3 for a large sample of clusters that spans well over an order of magnitude in mass at $z=0$. On the other hand, by a redshift of $z=1.5$ the range of typical progenitors for a $M_\mathrm{200c}=10^{14}M_\odot$ cluster already goes from large galaxies $M_\mathrm{200c}\approx7\cdot10^{12}M_\odot$ up to intermediate groups $M_\mathrm{200c}\approx3\cdot10^{13}M_\odot$. Additionally, we note that only the clusters in the upper $2\sigma$-bound actually retain up to $10$\%~of their halo mass down to $z=3$, which would place them in the typical range of what is considered a protocluster with $M_\mathrm{200c}>1\cdot10^{13}M_\odot$  \citep{remus:2023}.

\begin{figure*}
    \includegraphics[width=\textwidth]{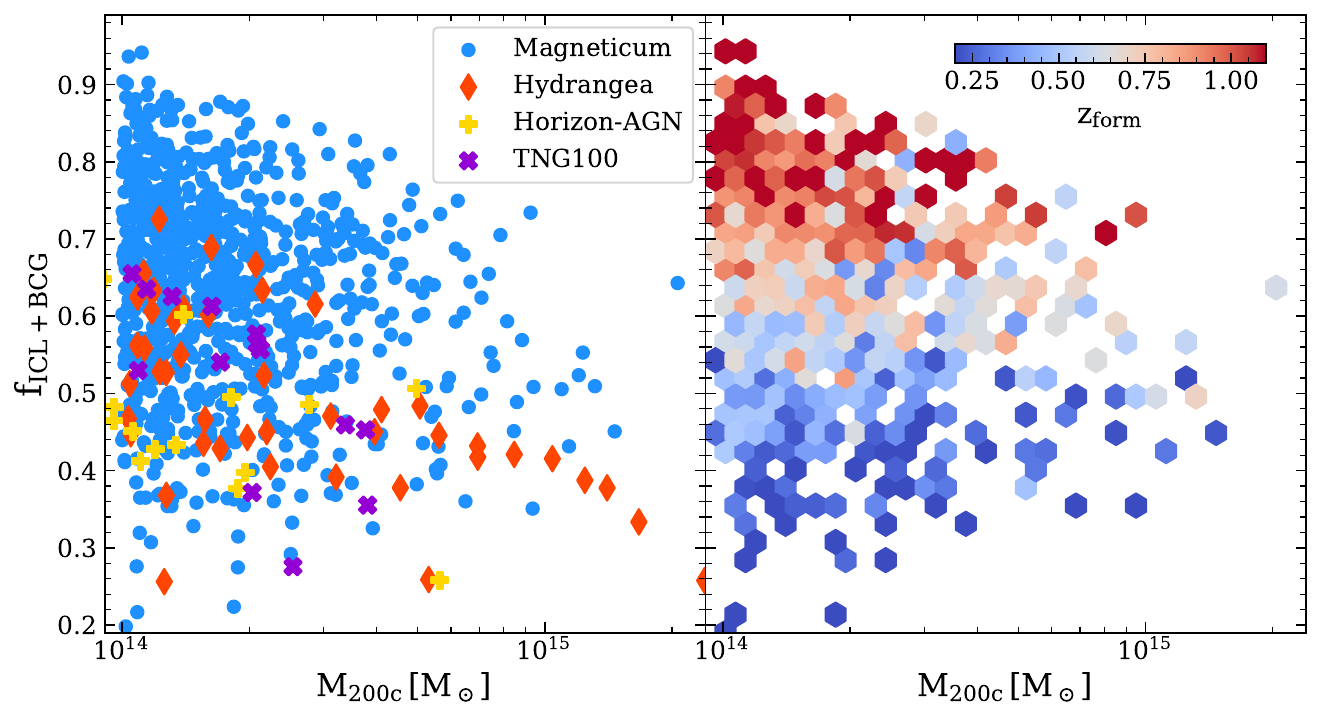}
    \caption{\textit{Left panel:} The fraction of stellar mass within $R_\mathrm{200c}$ that is not bound to satellites but instead in the BCG or ICL, $f_\mathrm{ICL+BCG}$, against the halo mass $M_\mathrm{200c}$, for the Magneticum (blue), IllustrisTNG (purple), Horizon-AGN (gold) and Hydrangea (orange) clusters.
    \textit{Right panel:} Same as the left panel but for the Magneticum galaxy clusters only, colored by the cluster formation redshift $z_\mathrm{form}$. Colors as indicated in the legend. 
    }
    \label{fig:iclbcg}
\end{figure*}

To emphasize this point even further, in the upper panel of Fig.~\ref{fig:tracks} we split the total sample (black) into two subsets, those with $z_\mathrm{form}$ in the upper/lower $1\sigma$-range of the total distribution (red/blue), with the exact cutoff shown in the legend. This results in around~$140$~clusters for each group. By construction, this results in the largest possible split between the formation redshifts of each group, with those assembling early (red) already reaching $50$\% of their final halo mass at $z\approx1.4$ (vertical red line), while the late assembling clusters do so only by $z\approx0.2$ (vertical blue line). Interestingly, even though the groups are split based on $z_\mathrm{form}$, they are still largely separate at $z>2$, indicating that clusters in either group inhabit fundamentally different assembling nodes through cosmic time. This raises the question if there is an observable that could be used to differentiate such starkly diverging histories.

Based on Fig.~\ref{fig:smhm}, we know that there is a range in the amount of mass already present in the BCG or ICL in comparison to the total stellar mass, where the range at a given halo mass is much more constrained. This is indicative of the mass transfer from satellite galaxies to the BCG or ICL as time passes, due to stripping, merging or other dynamical disruption events. 

Therefore, we ask the question whether the mass assembly history of a galaxy cluster can be traced through the fraction of stellar mass contained in the BCG and ICL relative to the total stellar mass, $f_\mathrm{ICL+BCG}$. We choose this definition instead of only the BCG or the ICL for two reasons. First, there is no clean definition of how to split the ICL from the BCG, as discussed in depth in the literature \citep[see][and discussion therein]{brough24}, with strong disagreement between the kinematic selection and the surface brightness selection methods \citep[see][for a direct comparison]{remus17}. Second, the distribution of stars through different stripping and merging events strongly depends on the orbit of the satellite that is disrupted or merged \citep{karademir:2019} with the BCG, and as such there are second order effects when considering just the BCG or just the ICL beyond the tracing of dynamical timescales. Finally, though there appear to be two kinematically distinct components in clusters \citep{remus17,marini:2024}, these need not also be physically separate.

\begin{figure}
    \includegraphics[width=0.95\columnwidth]{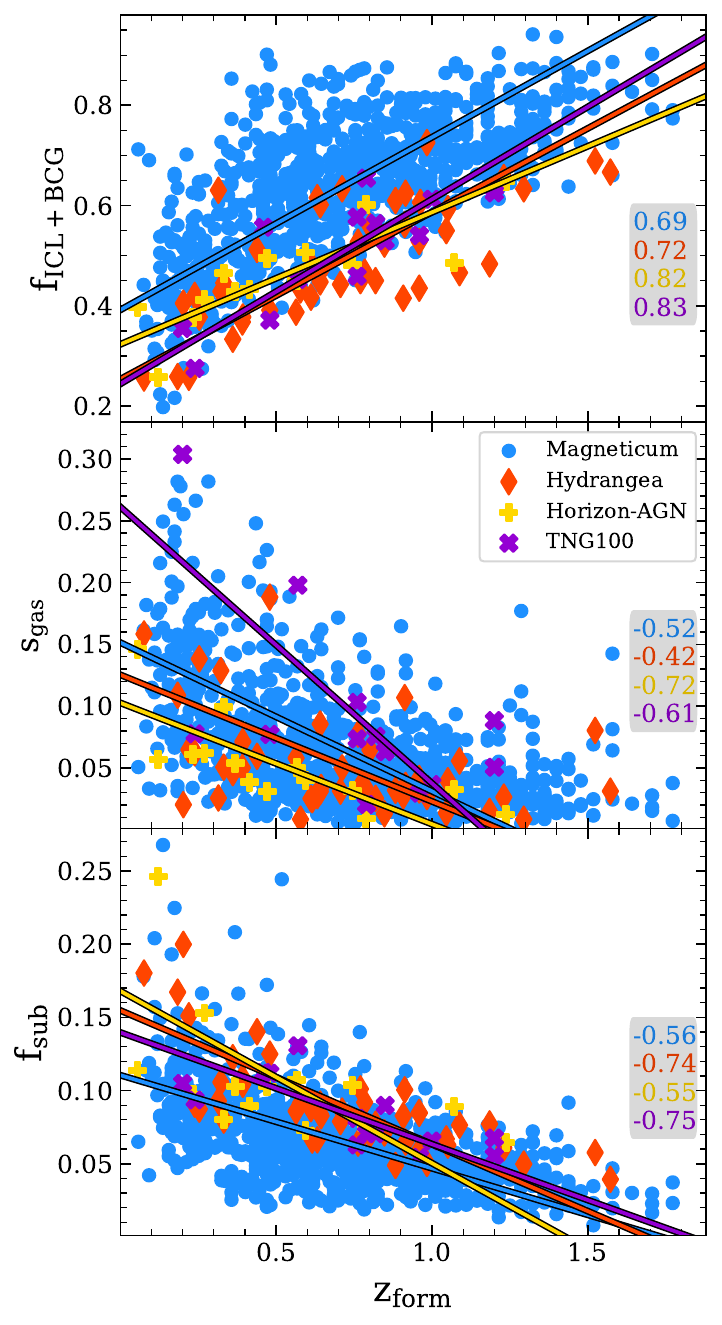}
    \caption{\textit{Upper panel:} $f_\mathrm{ICL+BCG}$ versus formation redshift $z_\mathrm{form}$, for the galaxy clusters from the Magneticum (blue), TNG100 (purple), Horizon-AGN (gold), and Hydrangea simulations (orange). The gray box displays the Pearson correlation coefficient for each simulation between $f_\mathrm{ICL+BCG}$ and $z_\mathrm{form}$. The closer the absolute value is to unity, the tighter the correlation. We further include best fit lines for each simulation determined via the least squares method (colored lines).  
    \textit{Middle panel:} The center shift between the $3$D~gas barycenter and the location of the BCG (normalized by $R_\mathrm{200c}$), plotted as a function of the formation redshift, with the colors the same as the upper panel.
    \textit{Lower panel:} The fraction of the total halo mass mass which is in all subhalos $f_\mathrm{sub}$ (so excluding the central and diffuse halo) versus formation redshift, with colors as in the upper two panels.
    }
    \label{fig:dyncor}
\end{figure}

We first consider how $f_\mathrm{ICL+BCG}$ evolves with time, plotted in the lower left panel of Fig.~\ref{fig:tracks}. The median of the Magneticum clusters is shown as a black solid line and the $1\sigma$ and $2\sigma$ bounds as dark and light shaded areas, respectively. Overall, $f_\mathrm{ICL+BCG}$ decreases with redshift up to $z\approx0.6$, after which it shows a slight increase again. This increase at later times is interesting when we consider the expectations from the assembly history of dark matter only simulations, where at later times the disruption of substructure occurs on faster timescales ($\approx1\,$Gyr) than the accretion of new matter \citep[around $10\,$Gyr according to][]{jiang17}. In this context, we thus expect an increase in $f_\mathrm{ICL+BCG}$ as more stars are disrupted than new galaxies are accreted. The lowest average value of $f_\mathrm{ICL+BCG}$ is reached at the time when the average cluster has assembled 50\%~of its mass. This implies that along the history of a galaxy cluster, the time when it has the highest fraction of stellar mass in satellites is when it reaches an intermediate group mass, which we discuss in more detail in Sec.~\ref{subsec:evo}. 

The lower left panel of Fig.~\ref{fig:tracks} also shows the evolution of $f_\mathrm{ICL+BCG}$ for the two subsets split by their $z_\mathrm{form}$, where we find a clear difference in their evolution. The early assembling clusters (red) reach their lowest $f_\mathrm{ICL+BCG}$ at $z=1.3$, which is comparable to their median $z_\mathrm{form}=1.25$. After this point they grow much slower, taking around $8$~Gyrs to assemble the remaining half of their final mass, nearly half of which is spent on the final $10$\%. Over this same time their $f_\mathrm{ICL+BCG}$ steadily increases up to $f_\mathrm{ICL+BCG}=74$\%~by $z=0$, as the clusters grow very slowly with sufficient time to strip and disrupt any accreted satellites. By contrast, the late assembling clusters' $f_\mathrm{ICL+BCG}$ (blue line in the bottom left panel of Fig.~\ref{fig:tracks}) behaves much like the total sample of clusters up until $z=0.3$, at which point the late assembling clusters sharply drop in $f_\mathrm{ICL+BCG}$, down to $f_\mathrm{ICL+BCG}=0.45$ by $z=0$. This coincides with the timeframe where they experience strong total mass growth, more than doubling their halo mass over a time of just around~$2\,$Gyrs. 

To confirm the implied anti-correlation between accretion events and $f_\mathrm{ICL+BCG}$, we also split the sample of clusters based on their $f_\mathrm{ICL+BCG}$ at $z=0$. The two groups of around~$140$~galaxy clusters with the highest/lowest $f_\mathrm{ICL+BCG}$ are shown in the right panels of Fig.~\ref{fig:tracks}. We find remarkably similar behavior to when we used the formation redshift to split the clusters, where a high/low $f_\mathrm{ICL+BCG}$ directly implies an earlier/later assembly. What is of particular note is how significantly their mass assembly histories differ, with the $1\sigma$-bounds of the two sets of clusters only overlapping at $z=2$. This means that a measurement of $f_\mathrm{ICL+BCG}$ allows us to differentiate the cluster's mass assembly history over the last~$10\,$~Gyrs. Generally, whenever another halo merges into the cluster, $f_\mathrm{ICL+BCG}$ drops as the number of satellite galaxies is increased significantly. After the merger, as long as the cluster stays undisturbed, it starts to disrupt and merge the satellite galaxies, thereby adding the mass to the ICL and the BCG while the cluster is relaxing its dynamical state. However, as we are considering the stellar component which lies in the deepest parts of the satellites' potential wells, this process is imprinted over long timescales of up to~$10\,$Gyrs.

\begin{figure*}
    \includegraphics[width=\textwidth]{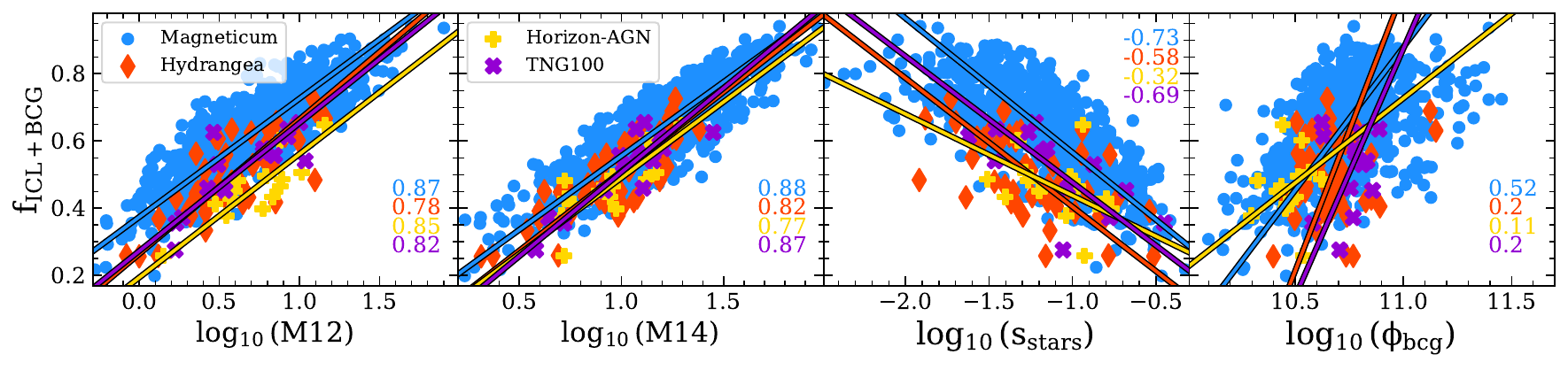}
    \caption{$f_\mathrm{ICL+BCG}$ versus different dynamical tracers, for all individual clusters from Magneticum Box2 (blue), TNG100 (purple), Horizon-AGN (gold), and Hydrangea (orange). The numbers in each panel indicate the Pearson correlation coefficient found for each relation and simulation (colors). We plot $f_\mathrm{ICL+BCG}$ against the stellar mass ratio of the BCG to the second (\textit{left panel}) and fourth (\textit{second panel}) most massive galaxy inside the cluster, counting the BCG. The \textit{third panel} shows $f_\mathrm{ICL+BCG}$ as a function of the offset between the stellar barycenter and the BCG position $\mathrm{s}_\mathrm{stars}$, while the \textit{right panel} plots the fraction against the stellar potential $\phi_\mathrm{stars}=M_\mathrm{*}/R_\mathrm{*,half}$ determined within $0.1\cdot R_\mathrm{200c}$. 
    }
    \label{fig:iclcor}
\end{figure*}

\setlength{\tabcolsep}{1.65pt}
\begin{table*}
\small
\begin{center}
\caption{Fit parameters from the orthogonal distance regression for the relations between various parameters for all four simulations, where $f=f_\mathrm{ICL+BCG}$. We find the 1-$\sigma$ error in predicting $z_\mathrm{form}$ from $f_\mathrm{ICL+BCG}$ to lie between $\pm0.16$ and $\pm0.28$, increasing with the sample size and considered mass range of the simulation. 
}
\label{tab:sims3}
\begin{tabular}{ll|cclcclcclcc}
   & \vline&	\multicolumn{2}{c}{Magneticum} &\vline& \multicolumn{2}{c}{Hydrangea} &\vline& \multicolumn{2}{c}{Horizon-AGN} &\vline& \multicolumn{2}{c}{TNG100} \\
   & \vline& A & B &\vline& A & B &\vline& A & B &\vline& A & B \\
   \hline\hline
   $z_\mathrm{form} = A\cdot f+B$ &\vline& $2.90\pm0.04$ & $-1.14\pm0.03$ &\vline& $3.00\pm0.12$ & $-0.78\pm0.06$ &\vline & $3.80\pm0.24$ & $-1.23\pm0.11$ &\vline & $2.72\pm 0.19$ & $-0.66\pm 0.10$ \\
   $z_\mathrm{form} = A\cdot \mathrm{s}_\mathrm{gas}+B$ &\vline& $-8.32\pm 0.06$ & $1.26\pm0.01$ &\vline& $-9.76\pm 0.23$ & $1.22 \pm 0.02$ &\vline& $-10.19\pm 0.28$ & $1.05\pm 0.02$ &\vline& $-4.47\pm0.39$ & $1.17\pm0.05$\\
   $z_\mathrm{form} = A\cdot f_\mathrm{sub} + B$ &\vline& $-15.7\pm 0.1$ & $1.7\pm 0.1$ &\vline& $-10.9\pm 0.2$ & $1.7\pm 0.1$ &\vline & $-8.5\pm 0.5$ & $1.4\pm 0.1$ &\vline& $-13.1\pm 0.3$ & $1.8\pm 0.1$\\
   \hline
   $f = A\cdot\log_{10}{(\mathrm{M}12)} + B$ &\vline& $0.35\pm0.02$ & $0.37\pm0.02$ &\vline & $0.40\pm0.12$ & $0.26\pm0.08$ &\vline& $0.37\pm0.16$ & $0.19\pm0.13$ &\vline& $0.38\pm 0.15$ & $0.27\pm 0.11$\\
   $f = A\cdot\log_{10}{(\mathrm{M}14)} + B$ &\vline& $0.44\pm0.02$ & $0.12\pm0.02$ &\vline & $0.47\pm0.11$ & $0.04\pm0.11$ &\vline& $0.45\pm0.15$ & $0.04\pm0.14$ &\vline& $0.50\pm 0.18$ & $0.01\pm 0.19$\\
   $f = A\cdot\log_{10}{(\mathrm{s}_\mathrm{stars})}+B$ &\vline& $-0.40\pm0.03$ & $0.17\pm0.04$ &\vline& $-0.38\pm0.18$ & $0.02\pm0.23$ &\vline& $-0.24\pm0.23$ & $0.19\pm0.26$ &\vline& $-0.38\pm 0.31$ & $0.10\pm 0.35$\\
   $f = A\cdot\log_{10}{(\phi_\mathrm{bcg})} + B$ &\vline& $0.9\pm0.1$ & $-8.5\pm0.9$ &\vline& $1.7\pm 1.0$ & $-17.3\pm 9.7$ &\vline& $0.5\pm 0.3$ & $-4.8\pm 3.1$ &\vline& $1.5\pm 0.6$ & $-15.4\pm 6.5$\\
\hline

\end{tabular}
\end{center}
\end{table*}

\subsection{Halo Mass (In-)Dependence}\label{subsec:halo}
To test whether there is a systematic trend with galaxy cluster mass, we plot $f_\mathrm{ICL+BCG}$ against halo mass $M_\mathrm{200c}$ in the left panel of Fig.~\ref{fig:iclbcg}. First, we find no clear correlation between the two parameters for any of the simulations. There is a slight tendency of the very most massive clusters with $M_\mathrm{200c}\approx 10^{15}M_\odot$ toward lower $f_\mathrm{ICL+BCG}$, matching the observed trend \citep{Montes2022}. This is because these clusters generally have formed more recently due to the bottom-up structure formation for a $\Lambda$CDM cosmology, and thus had less time to strip their satellites' stars. However, it is unclear if the simulation volumes probed here are simply too small to house the extreme outlier cases of very early forming galaxy clusters with masses in excess of $M_\mathrm{200c}> 10^{15}M_\odot$, which may then exhibit higher $f_\mathrm{ICL+BCG}$, or if those do not form rapidly enough to be seen by $z=0$.

We note that we find differences in the absolute values of $f_\mathrm{ICL+BCG}$ between the simulations, with a median $f_\mathrm{ICL+BCG}=0.65$,~$0.47$,~$0.46$ and~$0.55$ for Magneticum, Hydrangea, Horizon-AGN and TNG100. On one hand, this may be caused by differences in the implemented subgrid physics, where \citet{popesso:2024} find that Magneticum best reproduces observed cluster gas properties among a set of simulations including IllustrisTNG and EAGLE, but in return produce overly massive central galaxies. On the other hand, the lower resolution of Magneticum could result in a lack of low-mass satellites in galaxy clusters, as they are stripped/disrupted too quickly because of their low particle numbers, which would increase $f_\mathrm{ICL+BCG}$. Indeed, galaxies of $M_*<10^{10}M_\odot$ are the progenitors of around $20$\% of the ICL+BCG stellar mass \citep{brown:2024}. However, as we will show in the following, this does not change the quantitative correlations found between the observables we consider.

When we color the points by the galaxy cluster formation redshift $z_\mathrm{form}$ in the right panel of Fig.~\ref{fig:iclbcg}, we find a clear color gradient. At a given halo mass $M_\mathrm{200c}$, there is a direct correlation between $f_\mathrm{ICL+BCG}$ and $z_\mathrm{form}$, where galaxy clusters at the highest (lowest) $f_\mathrm{ICL+BCG}$ have formed the earliest (latest). This provides strong credence to the idea that $f_\mathrm{ICL+BCG}$ can indeed be used as a dynamical clock for the formation of galaxy clusters. We also see this for the different higher-resolved simulations, where, on average, the older galaxy cluster from both TNG100 and Hydrangea (Fig.~\ref{fig:zformhist}) have higher $f_\mathrm{ICL+BCG}$ than the younger Horizon-AGN clusters (left panel of Fig.~\ref{fig:iclbcg}).

We test this claim in the upper panel of Fig.~\ref{fig:dyncor}, plotting $f_\mathrm{ICL+BCG}$ at $z=0$ against the formation redshift $z_\mathrm{form}$ of each galaxy cluster from all four simulations. It is apparent that there is a positive correlation, one which is present across all simulations and is therefore stable against calibrations of the subgrid physics. A higher $f_\mathrm{ICL+BCG}$ consistently implies an earlier formation redshift. We quantify this by the Pearson correlation coefficient, finding significant values of $p=0.69$, $p=0.72$, $p=0.82$ and $p=0.83$ for Magneticum, Hydrangea, Horizon-AGN and TNG100, respectively. We have tested and find similar correlation strengths for both the Spearman ($p>0.68$) and Kendall rank coefficients ($p>0.5$), which is to be expected given that there is not strong indication away from linearity between $f_\mathrm{ICL+BCG}$ and $z_\mathrm{form}$.

The lower two panels of Fig.~\ref{fig:dyncor} show two other common tracers of the dynamical state of galaxy clusters known from the literature, namely the gas center shift $\mathrm{s}_\mathrm{gas}$ (middle panel) and the fraction of total mass (including DM) in subhalos $f_\mathrm{sub}$ (lower panel), for all simulations. The former is given by the distance offset between the location of the BCG (typically at the deepest point in the potential) and the barycenter of all gas within $R_\mathrm{200c}$. We normalize this offset then by $R_\mathrm{200c}$. Larger values thus indicate a strongly disturbed distribution of gas compared to the underlying dark matter halo, indicative of recent merging activity \citep[e.g.,][]{biffi16,zenteno20,kimmig23}.

As expected from previous works, both $\mathrm{s}_\mathrm{gas}$ and $f_\mathrm{sub}$ trace the formation redshift. However, we find for all simulations that $f_\mathrm{ICL+BCG}$ correlates tighter (or comparably well) with $z_\mathrm{form}$ compared to established dynamical state tracers, allowing for a better overall predictive power. We compile the best fit lines for each simulation for the formation redshift $z_\mathrm{form}$ as a function of $f_\mathrm{ICL+BCG}$, $\mathrm{s}_\mathrm{gas}$ and $f_\mathrm{sub}$ in Table~\ref{tab:sims3}. We fit via orthogonal distance regression, which minimizes the errors in both directions, and provide the fit parameters in Table~\ref{tab:sims3}. For a discussion on the predictive power of the fit compared to least-squares fitting, we refer to the Appendix~\ref{app:fitting}.

\subsection{Tracing ICL+BGC through Other Observables}

\begin{figure*}
    \includegraphics[width=0.99\textwidth]{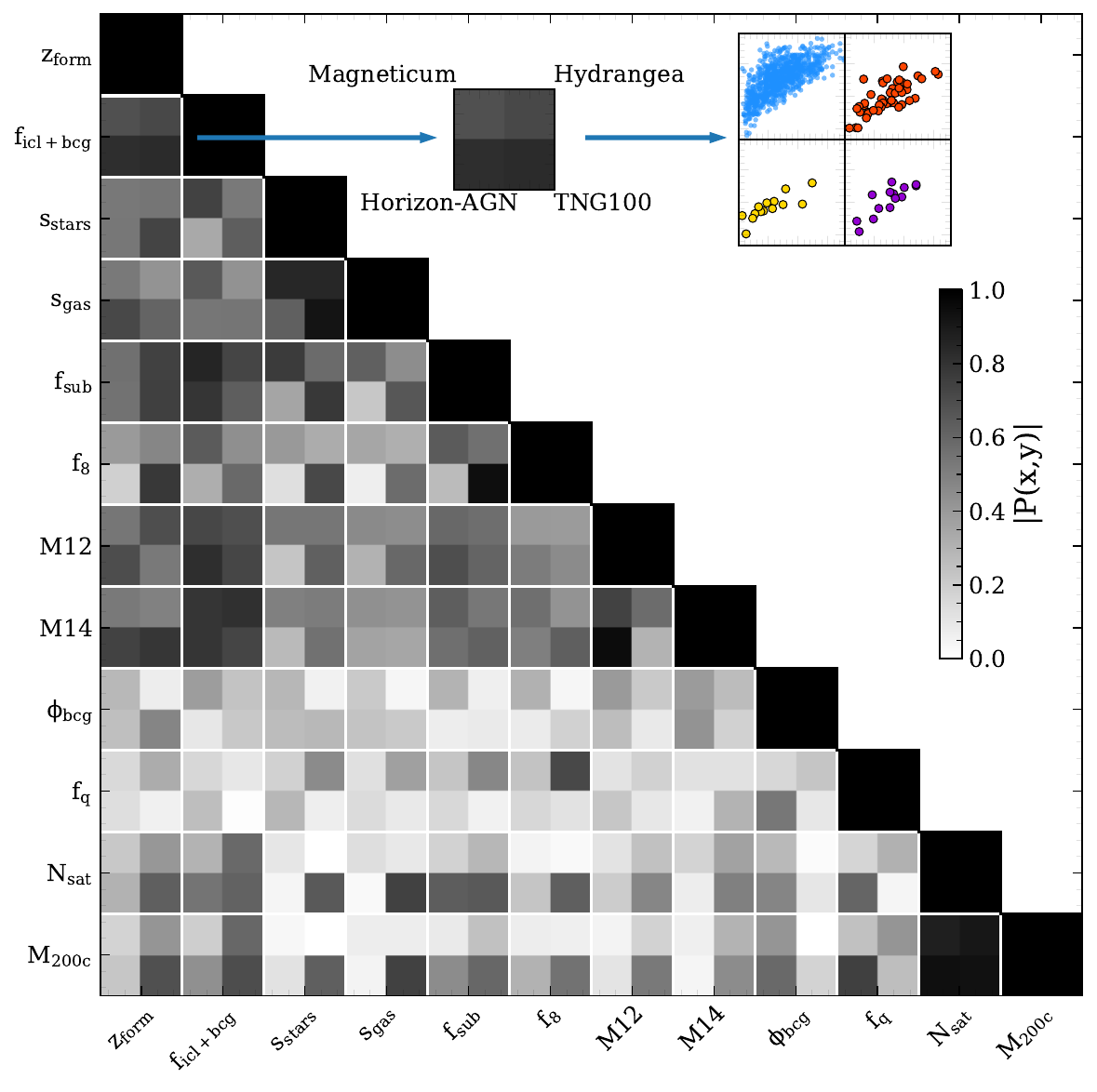}
    \caption{Full matrix of the correlation strength between different parameters, with darker colors indicating stronger correlations, for all four simulations as indicated at the respective tiles for each panel. From top to bottom and left to right are the same parameters, in the following order: formation redshift $z_\mathrm{form}$, the fraction of stellar mass within ICL plus BCG $f_\mathrm{ICL+BCG}$, the center shift between the stellar ($\mathrm{s}_\mathrm{stars}$) or gas barycenter ($\mathrm{s}_\mathrm{gas}$) and the BCG, the fraction of total mass within all substructures $f_\mathrm{sub}$ or within just the $8$th~most massive substructure $f_8$, the stellar mass ratio between the BCG and the second ($M12$) or fourth ($M14$) most massive galaxy in a cluster, the central stellar potential $\phi_\mathrm{BCG}$, the passive fraction $f_\mathrm{quench}$ of galaxies in the cluster with stellar masses above $1\times 10^{10}M_\odot$ as well as their total number $N_\mathrm{sat}$, and the total halo mass $M_\mathrm{200c}$. Note that black squares in the diagonal mark the self-correlation of the parameters and are thus all equal to one.
    }
    \label{fig:cormat}
\end{figure*}

As the ICL is difficult to observe even with the upcoming facilities, one question that remains is in how well the fraction of stars in the ICL plus BCG can be approximated by measuring other observables. In Fig.~\ref{fig:iclcor} we show four parameters that are generally easier to obtain than the stellar mass in the BCG and ICL itself. From left to right, we show the correlations between $f_\mathrm{ICL+BCG}$ and 1) the stellar mass ratio between the BCG and the second most massive galaxy within $R_\mathrm{200c}$ of the cluster, M$12\equiv M_\mathrm{*,BCG}/M_\mathrm{*,2^{nd}}$, also termed the fossilness parameter \citep{tremaine:1977,loh:2006,ragagnin17}; 2) the stellar mass ratio between the BCG and the fourth most massive galaxy in the cluster, M$14$ \citep{golden-marx:2018}; 3) the normalized offset $\mathrm{s}_\mathrm{stars}=\Delta_{r}/R_\mathrm{200c}$ between the barycenter of stellar mass of the cluster within $R_\mathrm{200c}$ and the position of the BCG, also known as the center shift \citep{mann12,biffi16,contreras22}; and 4) the quantity $\phi_\mathrm{BCG}=M_\mathrm{*,BCG}/R_\mathrm{*,1/2,BCG}$ in units of $M_\odot/\mathrm{kpc}$, an approximation for the central stellar potential depth following \citet{bluck:2023}, where we determine the stellar mass $M_\mathrm{*,BCG}$ and stellar half-mass radius $R_\mathrm{*,1/2,BCG}$ for the BCG within $0.1\cdot R_\mathrm{200c}$.

Across all four simulations (Magneticum in blue, TNG100 in purple, Horizon-AGN in gold, and Hydrangea in orange) we consistently find that the correlation between $f_\mathrm{ICL+BCG}$ and the two stellar mass ratios (M$12$ and M$14$) is tightest, with remarkably little scatter. The Pearson correlation coefficients are shown in the figure, with all simulations finding $p\geq0.7$. Interestingly, while there appears to be an offset between the simulations in the relation between $f_\mathrm{ICL+BCG}$ and M$12$, there is much less for the correlation between $f_\mathrm{ICL+BCG}$ and M$14$. This could imply that the fourth most massive galaxy is a more consistent representative of the overall stripping that occurred in the cluster, where the second most massive galaxy may be less affected. 
Nonetheless, we find that $f_\mathrm{ICL+BCG}$ may be well determined by both M$12$ and M$14$ for all simulations, with very similar slopes in the best fit lines for all, the values of which we provide in Table~\ref{tab:sims3}. When considering the relations of $f_\mathrm{ICL+BCG}$ to $\mathrm{s}_\mathrm{stars}$ and in particular to $\phi_\mathrm{BCG}$, we find significantly increased scatter. For the latter, aside from Magneticum, the simulations find little to no correlation. 

This means that good measurements of satellite stellar masses as well as the BCG (including, however, identifying cluster membership) are sufficient to tightly determine the overall fraction of stellar mass in the ICL and BCG $f_\mathrm{ICL+BCG}$. That fraction, in turn, well captures the mass assembly history of a galaxy cluster. Most significantly, we find this correlation to be stable against changes in the subgrid physics as well as the choice of structure finder. However, we note that when we test the partial correlation between $z_\mathrm{form}$ and both M$12$ and M$14$, we find that the underlying driver is their correlation with $f_\mathrm{ICL+BCG}$. Consequently, using the mass ratios to predict a cluster's $z_\mathrm{form}$ will generally add scatter compared to directly using $f_\mathrm{ICL+BCG}$. This point is further discussed in Appendix~\ref{app:fitting}.

We then broaden the spectrum of galaxy cluster parameters to also include a) the center shift of the gaseous component $\mathrm{s}_\mathrm{gas}$, that is the offset between the BCG and the barycenter of the gas, closely mimicking the X-ray determined center shift (see Appendix~\ref{app:center}); b) the fraction of total mass contained in the $8$th most massive substructure $f_8$ \citep[see][for more details]{kimmig23}, and both the total number (i.e., cluster richness) as well as passive fraction $f_\mathrm{quench}$ of cluster galaxies with $M_{*}>10^{10}M_\odot$ \citep[e.g.,][]{Aguerri2006}. The full matrix of correlations between all these different tracers of the dynamical state of a cluster is shown in Fig.~\ref{fig:cormat}.

Fig.~\ref{fig:cormat} is structured as follows: for a given parameter we determine the Pearson correlation strength with another parameter, doing this for every simulation separately to fill a two-by-two grid, with Magneticum in the upper left, Hydrangea in the upper right, Horizon-AGN in the lower left and finally TNG100 in the lower right. We color the cell based on the strength, going from white ($p=0$, no correlation) to black ($p=1$, one-to-one correlation). The upper inset of Fig.~\ref{fig:cormat} illustrates this process for the case of a correlation between $f_\mathrm{ICL+BCG}$ versus $z_\mathrm{form}$. Note that the full black squares in the top diagonal mark the relation between a parameter and itself, and are thus all fully black ($p=1$). 

The first column in Fig.~\ref{fig:cormat} shows the correlation strength between the formation redshift $z_\mathrm{form}$ with all other parameters, for each of the simulations. We find that the overall strongest correlation exists for $f_\mathrm{ICL+BCG}$, aside from Hydrangea where $f_\mathrm{sub}$ performs negligibly better ($p=0.74$ versus $p=0.72$). Additionally, all simulations find that the mass ratios M$12$ and M$14$ are good tracers of $z_\mathrm{form}$, which is explained by their strong correlation with $f_\mathrm{ICL+BCG}$ (see the second column of Fig.~\ref{fig:cormat} or Fig.~\ref{fig:iclcor}). $f_\mathrm{ICL+BCG}$ remains the true, tighter underlying correlation, however, as we discuss in Appendix~\ref{app:fitting}.

For all simulations, the number of satellite galaxies with stellar masses above $10^{10}M_\odot$, $N_\mathrm{>10^{10}}$ correlates significantly with the total halo mass $M_\mathrm{200c}$, as expected. What is interesting is that the simulations do not find a significant correlation between the halo mass $M_\mathrm{200c}$ and the formation redshift $z_\mathrm{form}$, aside from TNG100. For that simulation, however, $M_\mathrm{200c}$ also shows a notable correlation with $f_\mathrm{ICL+BCG}$, which may be the more fundamental link given that this is what the other three simulations find. Indeed, when we consider partial correlations, we find also for TNG100 that the correlation between $z_\mathrm{form}$ and $M_\mathrm{200c}$ practically vanishes when accounting for $f_\mathrm{ICL+BCG}$, as we discuss in Appendix~\ref{app:fitting}.

Thus, we conclude that the formation time of galaxy clusters, defined as the time when a cluster has assembled half of its present-day mass, is best traced through the amount of disruption suffered from the satellite component. This is best quantified by how much of the total stellar mass is bound to the full potential of the cluster, i.e., the BCG and the ICL, or in second order through the stellar mass ratios between the BCG and the most massive companion galaxies.

\subsection{Evolution Through Cosmic Time}\label{subsec:evo}

\begin{figure}
    \includegraphics[width=\columnwidth]{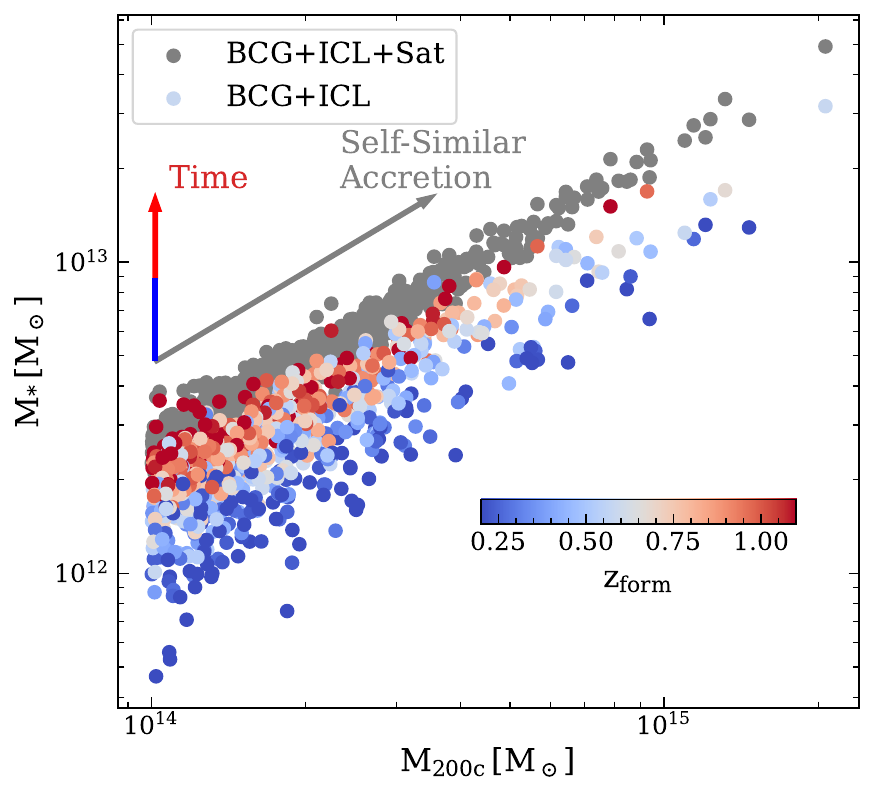}
    \caption{Stellar mass-halo mass relation for the Magneticum Box2~hr clusters. The gray points show the total stellar mass within $R_\mathrm{200c}$, including the BCG, ICL and satellites. The colored symbols show the stellar mass of only the ICL and the BCG together, without including satellite galaxies. These points are colored according to the formation redshift of the cluster from older and more relaxed galaxy clusters (red) to younger, recently assembled clusters (blue).
    }
    \label{fig:smhm_time}
\end{figure}

Finally, we want to understand the origin of the correlation between the stellar mass contained in the ICL plus BCG, and the formation redshift of galaxy clusters. As we are interested in the statistical behavior now, we will restrict the analysis in this last subsection to the Magneticum Box2~hr simulation only. To that end, we plot again the stellar-mass halo-mass relation in Fig.~\ref{fig:smhm_time}, but this time for all stars in $R_\mathrm{200c}$ (gray), as well as the subset of those stars which are not bound in satellites, and so belong to the BCG or ICL (colored). As already seen for all simulations in the right panel of Fig.~\ref{fig:smhm}, the total stellar mass is tightly increasing with halo mass, following a slope very close to 1.

On the other hand, the colored points which exclude the mass in satellites show a much larger scatter, with a slightly flatter slope. There is a clear color gradient driving this scatter, however, which is given by the formation redshift of the galaxy cluster. Clusters that have nearly all of their stellar mass in the BCG or ICL are consistently older and more relaxed, while clusters with lower BCG+ICL masses are more recently formed. As expected from Fig.~\ref{fig:iclbcg}, this is independent of halo mass. Consequently, if a cluster has time to disrupt and merge their satellite galaxies without being disturbed by additional merging, they will do so consistently across a full order of magnitude in halo mass. This demonstrates why $f_\mathrm{ICL+BCG}$ is such an effective clock for tracing the formation time, and mirrors prior results in the EAGLE simulations on the group mass scale by \citet{matthee:2017}, as well as recent work with TNG300 when comparing to both the concentration of the halo and the formation time \citep{montenegro-taborda:2025}.

\begin{figure*}
    \includegraphics[width=\textwidth]{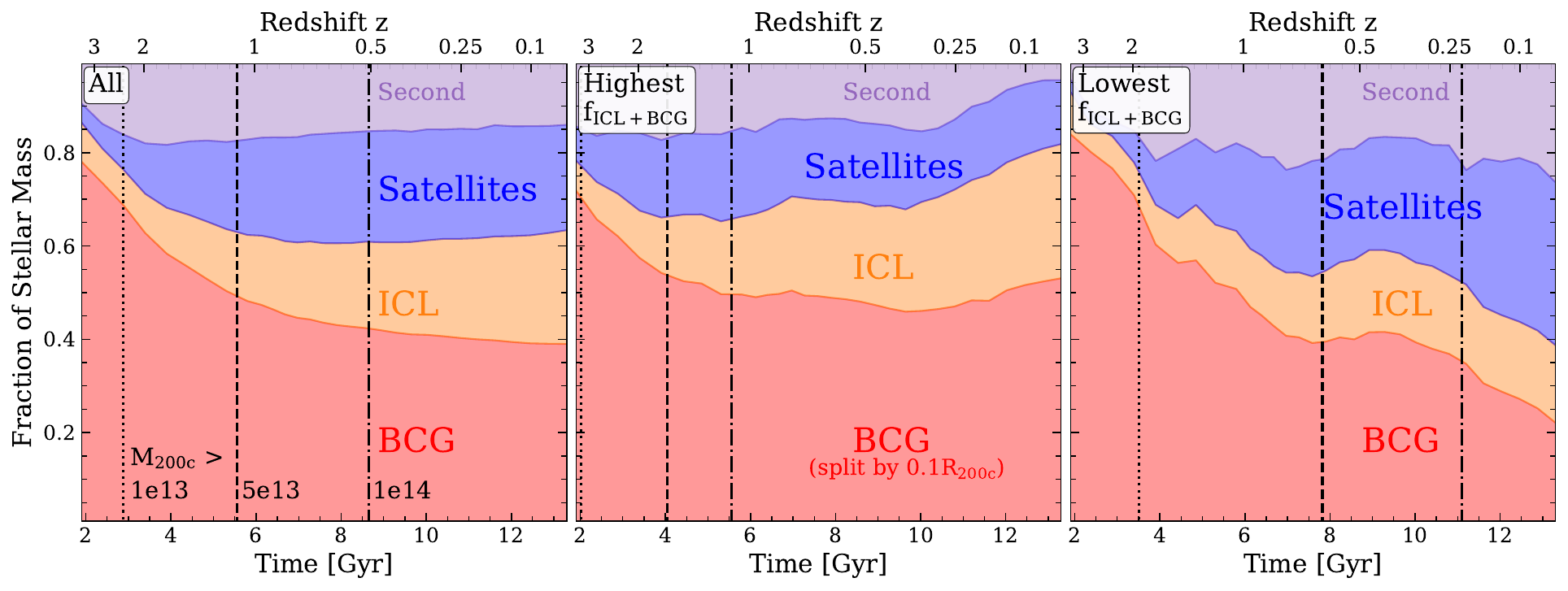}
    \caption{The fraction of the total stellar mass contained in the different components of the Magneticum Box2~hr galaxy clusters versus time, determined for all clusters (\textit{left panel}) and for the~50~clusters with the highest (\textit{central panel}) and the~50~clusters with the lowest $f_\mathrm{ICL+BCG}$ for their halo mass (\textit{right panel}). The fraction of stellar mass contained within the BCG is shown in red, while the fraction in the ICL, the second most massive satellite, and all other satellites are plotted in gold, violet and blue, respectively. The black vertical lines mark the times when the median mass of each group of galaxy clusters is $M_\mathrm{200,cri}>1\times10^{13}M_\odot$ (dotted), $M_\mathrm{200,cri}>5\times10^{13}M_\odot$ (dashed), and $M_\mathrm{200,cri}>1\times10^{14}M_\odot$ (dash-dotted). Note that here we split the BCG and ICL at $0.1\times R_\mathrm{200,cri}$. For different definitions of the split between the ICL and the BCG, see Fig.~\ref{fig:diff_def} in the Appendix.
    }
    \label{fig:evo}
\end{figure*}

We use this fact to split our sample of~868~galaxy clusters into two further groups. In ten equally log-spaced halo mass bins going from $M_\mathrm{200c}=10^{14}M_\odot$ to $M_\mathrm{200c}=5\times10^{14}M_\odot$ we select the~$5$~galaxy cluster with the highest/lowest $f_\mathrm{ICL+BCG}$. Using these two samples of~$50$~clusters each, which represent the extreme ends in $f_\mathrm{ICL+BCG}$ at a given halo mass, we trace the evolution in the distribution of stellar mass between the BCG (red), ICL (gold), second most massive satellite (violet) and the remaining satellites (blue) in Fig.~\ref{fig:evo}. We choose here to define the split between BCG and ICL at $0.1\times R_\mathrm{200c}$, which allows the split to vary with the scale of the halo over time. For further definitions including a fixed aperture cut, we refer to Fig.~\ref{fig:diff_def} in the appendix, and simply note here that the qualitative results do not change strongly depending on the definition of the split.

First, we consider in the left panel of Fig.~\ref{fig:evo} the evolution in the components for all~$868$~clusters, and find that while the progenitors of the clusters were in the group mass regime ($z\approx2$, given by the dotted vertical black line) the central galaxy dominated the overall stellar component, accounting for anywhere from $40$\% to $70$\%. As they grow from groups to clusters by $z=0.5$ (dash-dotted line), the fraction of total stellar mass contained in satellites (second most massive plus others) increases, even while the amount of ICL increases too. This is because the BCG does not grow as quickly in stellar mass as satellites are accreted onto the cluster over this time. Upon reaching galaxy cluster masses, the BCG accounts for just $40$\% of total stellar mass, with the fraction in ICL rising to $20$\%. The second most massive galaxy at this point contains $15$\%~of the galaxy cluster stellar mass, which is down from the group regime where it accounted for up to $20$\%. Consequently, as the overall contribution of satellites has increased this stellar mass is now distributed across a greater number of galaxies as opposed to being dominated by a single massive satellite. Finally, by $z=0$, the fraction in ICL has continued rising to $25$\%, this time drawing from both the BCG as well as the satellites, following a rate of growth comparable to that found by \citet{Rudick2011}. This is reasonable as it becomes more difficult for less massive infalling satellites to penetrate deeply into the growing galaxy cluster potential, and they are instead stripped already farther out \citep{lotz19}. 

When we consider the subset of galaxy clusters with the highest $f_\mathrm{ICL+BCG}$, we find generally similar behavior. The crucial difference is that these processes occur much earlier, as their progenitors are already groups at $z=3$ and would thus fulfill the definition of what is typically considered a protocluster \citep{remus22}. By $z=1$, they are already galaxy clusters, though with some differences to what we found for the total sample. They are more dominated by their BCG, which contains around half of the stellar mass (versus $40$\%), at the cost of both the ICL at $15$\% (versus $20$\%) and the satellites at $35$\% (versus $40$\%). This means at the moment these early forming clusters reach halo masses $M_\mathrm{200c}>10^{14}M_\odot$, they are already more centrally concentrated compared to the average clusters. Down to $z=0$ their BCGs actually \textit{increase} in stellar mass fraction (up to $53$\%), as does their ICL (up to $30$\%). This is because they strip, disrupt or merge with a significant portion of their satellites, given the nearly~$6\,$Gyr time they have to do so, and by $z=0$ their second most massive satellite barely accounts for $5$\% of the total stellar mass. 

Finally, for the sample of clusters with the lowest $f_\mathrm{ICL+BCG}$, we note first that although they reach group masses at a similar time to the total sample (at $z=1.9$ vs $z=2.3$), they stagnate significantly afterwards. By $z=0.5$, when the average galaxy cluster progenitor has reached halo masses $M_\mathrm{200c}>10^{14}M_\odot$, the progenitors of clusters with the lowest $f_\mathrm{ICL+BCG}$ are barely above massive groups. Indeed, it takes another nearly $2\,$Gyr~of additional time until they reach galaxy cluster mass. At that time their BCG accounts for just $35$\% of the total stellar mass, while their ICL is about $15$\%. Indeed, their satellites already dominate the overall stellar mass at nearly half, which only increases toward $z=0$ and ends at around $60$\% of their total stellar mass bound in satellites. The second most massive satellite is at a comparable stellar mass fraction to the BCG, at $25$\%. We therefore see here how differing mass assembly histories also result in a different \textit{distribution} of mass within the final $z=0$ structure.

\begin{figure*}
    \includegraphics[width=\textwidth]{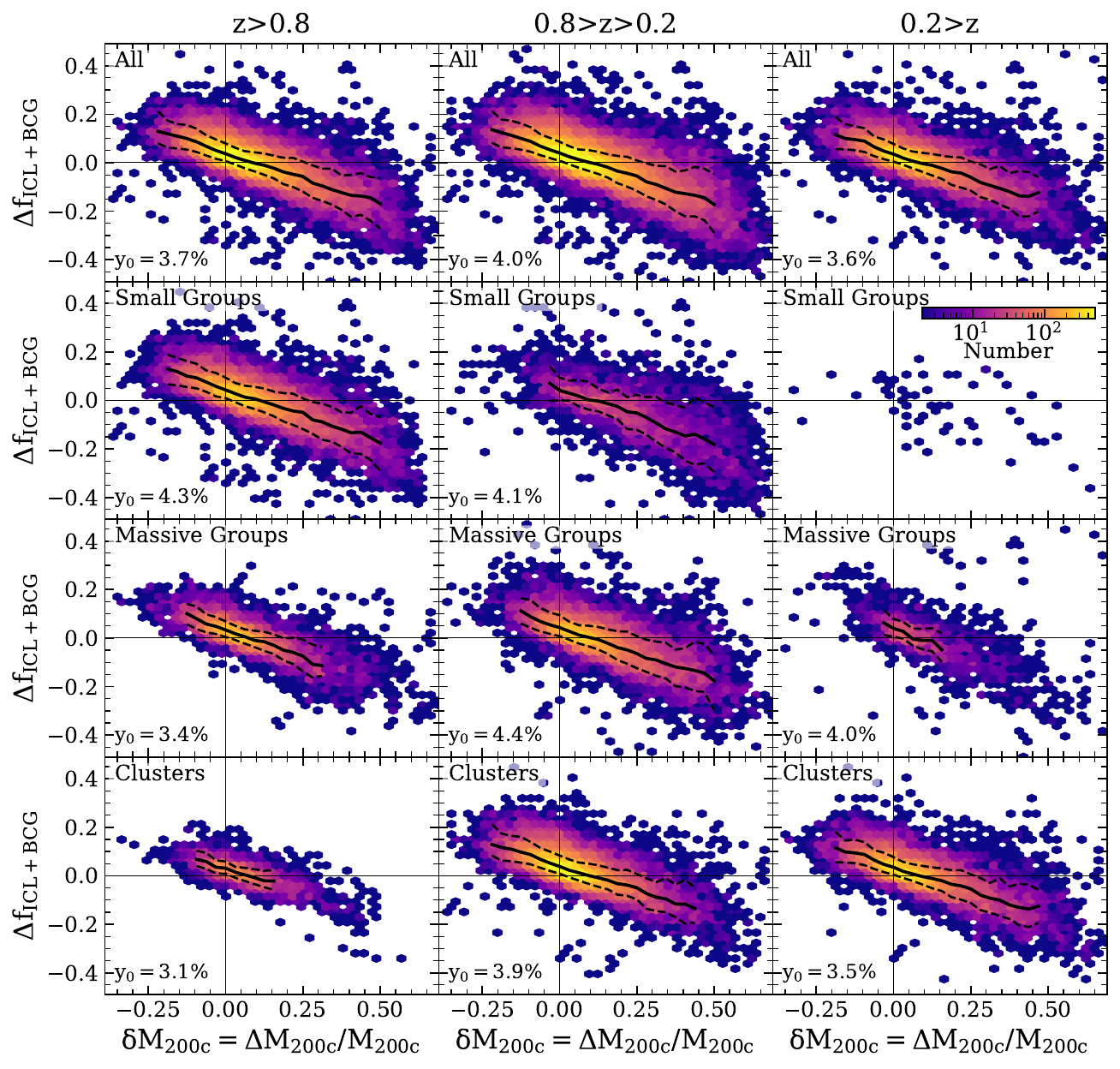}
    \caption{The absolute change in f$_\mathrm{ICL+BCG}$ versus the relative change in mass $\delta M_\mathrm{200c}$ as determined over a period of $T=1\,$Gyr for the Magneticum clusters. Columns from left to right split the sample into redshift bins, while the rows separate the clusters based on their current mass, going from all clusters in the top row to small and massive groups and finally clusters on the bottom row. The median line and $1\sigma$-bounds are plotted in black, while the thin vertical and horizontal black lines show the location of zero change in either f$_\mathrm{ICL+BCG}$ or cluster mass. $y_{0}$ indicated in each panel is the median change in f$_\mathrm{ICL+BCG}$ for halos where the total mass did not change, $\delta M_\mathrm{200c}=0$.
    }
    \label{fig:shredding}
\end{figure*}

\subsection{Rate of Satellite Shredding}\label{subsec:shred}

Thus, one interesting question which arises is how fast a cluster can disrupt its satellites to add their mass to the ICL+BCG. To this end, we consider how the halo mass and $f_\mathrm{ICL+BCG}$ change together over a characteristic time $T$. We choose $T=1\,$Gyr as it is approximately equal to the typical crossing time of our galaxy clusters \citep{jiang16,contreras22,haggar24}, though we have tested that our conclusions do not strongly depend on varying the timescale between $T=0.5\,$Gyr to $T=2\,$Gyr. Defining t$1$ as the starting point and t$2$ as the end of the window, then 

\begin{align}
    &\delta M_\mathrm{200c} \equiv \frac{\Delta M_\mathrm{200c}}{M_\mathrm{200c,t2}} = \frac{M_\mathrm{200c,t2}-M_\mathrm{200c,t1}}{M_\mathrm{200c,t2} },\\
    &\Delta f_\mathrm{ICL+BCG}\equiv f_\mathrm{ICL+BCG,t2}-f_\mathrm{ICL+BCG,t1}.
\end{align}

In Fig.~\ref{fig:shredding} we plot $\Delta f_\mathrm{ICL+BCG}$ against $\delta M_\mathrm{200c}$ based on the assembly histories of our~$868$~galaxy clusters. The columns show the relation at $z>0.8$ (left), $0.8>z>0.2$ (center) and $z<0.2$ (right), based on the redshift at the end of the time window~t$2$. We further split the sample based on the halo mass at the time~t$2$, plotting all clusters' assembly histories (top row), or only those clusters where the main progenitor has a halo mass of a small group $1\times10^{13}M_\odot\leq M_\mathrm{200c}<5\times10^{13}M_\odot$ (second row), massive group $5\times10^{13}M_\odot\leq M_\mathrm{200c}<1\times10^{14}M_\odot$ (third row), or is already a cluster $1\times10^{14}M_\odot\leq M_\mathrm{200c}$ (bottom row). We plot the median and $1\sigma$-bounds (solid and dashed thick black lines), as well as the horizontal and vertical lines corresponding to a net change of zero in either mass or $f_\mathrm{ICL+BCG}$.

Beginning with the top row, we find a negative dependence of $\Delta f_\mathrm{ICL+BCG}$ on $\delta M_\mathrm{200c}$. When the halo mass decreases, $\delta M_\mathrm{200c}<0$, then $\Delta f_\mathrm{ICL+BCG}$ is most positive, which means that $f_\mathrm{ICL+BCG}$ most strongly increases. This may be the result of either a fly-by or a merger. For the former, a structure had previously entered the halo, bringing with it also stellar mass, and has then exited again. Because both halo and satellite stellar mass are removed, $\delta M_\mathrm{200c}<0$ and $\Delta f_\mathrm{ICL+BCG}>0$. For the latter, this is a consequence of the definition of $M_\mathrm{200c}$. As a structure begins to merge, increasingly more mass of the infalling satellite lies within the radial definition of $R_\mathrm{200c}$, until a peak $M_\mathrm{200c}$ is reached that is approximately the sum of both structures total mass. However, if it is massive enough, the infalling structure does not instantly merge, instead continuing on its orbit until it reaches the apocenter. At this point, a portion of its mass may again lie outside of $R_\mathrm{200c}$, thus decreasing $M_\mathrm{200c}$ (and increasing $f_\mathrm{ICL+BCG}$ as satellite stellar mass is moved outside of the cluster extent). 

When the halo mass does not change over $T=1\,$Gyr, $\delta M_\mathrm{200c}=0$, then $\Delta f_\mathrm{ICL+BCG}\approx 4$\%. Thus, if left alone these systems increase the fraction of the total stellar mass contained within the BCG and ICL over time, consistent with the picture of stripping and merging. Finally, when $\delta M_\mathrm{200c}>0$, so when the halo grows, then $\Delta f_\mathrm{ICL+BCG}$ is either close to zero or negative. During large accretion events involving significant total mass, the fraction of the stellar mass in satellites therefore also increases (so $f_\mathrm{ICL+BCG}$ decreases). 

What is remarkable is that the median line of this behavior remains consistent both across time (left to right columns) as well as between structures of different masses (top to bottom rows). This may indicate that there is an average response between the distribution of the stellar mass and the growth (or lack thereof) of the halo mass of the structure. Generally, as can be seen from the color indicating the number of systems, most lie in the low right quadrant where they are experiencing slight growth in halo mass while $f_\mathrm{ICL+BCG}$ decreases as they continue to accrete more satellites over time. 

We can define the average change in $f_\mathrm{ICL+BCG}$ when a structure does not grow ($\delta M_\mathrm{200c}=0$) as the ``shredding rate'', and indicate this value in the panels of Fig.~\ref{fig:shredding}. We find that for galaxy groups $\Delta f_\mathrm{ICL+BCG}\approx 4\%$, while galaxy clusters lie slightly lower at $3$\% to $4$\% per Gyr. Therefore, as long as there is no new material being accreted, the stellar mass in the ICL+BCG will increase by around $4\%$ of the galaxy cluster total stellar mass. Slower disruption rates of galaxies in clusters compared to groups were also found by \citet{Bahe2019}, consistent with our findings here. We note that there is a relatively large scatter for individual clusters, as the actual rate of course depends on various factors. If the cluster is already strongly dominated by the BCG and ICL, then the remaining galaxies may be on more circular orbits that are stripped less quickly, while a cluster post major merger may suddenly and significantly increase its $f_\mathrm{ICL+BCG}$ when the two constituent BCGs merge. Indeed, Fig.~\ref{fig:shredding} shows that there are also clusters at $z<0.2$ which increase $f_\mathrm{ICL+BCG}$ by $20$\% while $\delta M_\mathrm{200c}=0$, which likely has to do with the orbits, masses, compactness and velocity of the satellite population. 

\section{Summary and Conclusions}
\label{sec:disc}
To inform the upcoming large surveys of galaxy clusters, which will provide a plethora of data on the largest forming structures throughout time, we have analyzed here the build up of the BCG and the ICL through four separate simulations. We have studied the connection between the fraction of the stellar mass in the ICL+BCG of clusters, $f_\mathrm{ICL+BCG}$, and their formation histories, following up on initial ideas presented by \citet{brough24}, and show that this fraction acts as a dynamical clock for galaxy clusters. We primarily employed Box2~hr of the Magneticum Pathfinder hydrodynamical simulation suite (Dolag et al., in prep), a simulation with a box volume of $(352\mathrm{cMpc}/h)^3$ and $886$~galaxy clusters with halo masses $M_\mathrm{200c}>10^{14}M_\odot$. We complemented this statistical sample with a smaller sample of higher resolved galaxy clusters from the TNG100 \citep{pillepich18b} and Horizon-AGN \citep{dubois:2014} simulations (14 clusters each), as well as from the Hydrangea suite of zoom-in cluster simulations \citep[][46 clusters]{bahe17}, to verify that the results found for Magneticum are independent of the resolution, numerical details, or subgrid physics prescriptions. 

Defining the formation redshift $z_\mathrm{form}$ of galaxy clusters as the redshift when the cluster first assembled $50$\%~of its $z=0$ halo mass, we find that clusters in Magenticum Box2~hr have on average assembled only recently with $\langle z_\mathrm{form}^\mathrm{Box2}\rangle=0.67$, in agreement with previous studies \citep[e.g.,][]{boylan:2009,power:2012}, but that some clusters already assembled as early as $z_\mathrm{form}=1.8$. This is consistent with the non-linearity in the growth of individual clusters compared to their statistical average and is matched nicely by linear growth theory (see also \citealt{remus:2023} and Kimmig et al., in prep.).

Here, it is of special interest to note that there is a difference in cluster formation times, with the (on average) latest forming clusters in Horizon-AGN at $\langle z_\mathrm{form}^\mathrm{Horizon}\rangle=0.44$, Magneticum and Hydrangea are similar at $\langle z_\mathrm{form}^\mathrm{Magneticum}\rangle=0.67$ and $\langle z_\mathrm{form}^\mathrm{Hydrangea}\rangle=0.71$, while TNG100 has the earliest forming clusters at $\langle z_\mathrm{form}^\mathrm{TNG}\rangle=0.78$. However, all simulations cover a similar range in formation redshifts (Fig.~\ref{fig:zformhist}), so that we can meaningfully compare the higher resolved simulations of TNG100, Horizon-AGN and Hydrangea to the larger sample of clusters in Magneticum Box2~hr. 

We find that the formation time of a galaxy cluster $z_\mathrm{form}$ is tightly traced by $f_\mathrm{ICL+BCG}$, so the fraction of total stellar mass within the ICL+BCG (Fig.~\ref{fig:dyncor}). We show this follows because the stripping/merging of stellar mass bound in satellite galaxies into the central potential of a galaxy cluster is proportional to how long a cluster has been left undisturbed (Fig.~\ref{fig:shredding}). More relaxed clusters have higher fractions of $f_\mathrm{ICL+BCG}$, with values up to $90$\%, while recent accretion can (temporarily) push the fraction down to as low as 20\% (Fig.~\ref{fig:iclbcg}).  

This process is also the origin for the scatter in the BCG stellar mass-halo mass relation (Fig.~\ref{fig:smhm}). Galaxy clusters at the upper end in stellar mass have formed much earlier and are thus both more compact and had more time to disrupt and strip their satellite populations (Fig.~\ref{fig:smhm_time}), supporting observational works that find this scatter to correlate with dynamical tracers such as M$14$ \citep{golden-marx:2018,golden-marx:2025}. If left undisturbed, that is to say without growing their halo mass, galaxy clusters will gradually disrupt their satellite population, stripping most or all of their mass even if a small core may survive. We calculate an average rate for this ``shredding'' at between $3$\%~and~$4$\% of the total stellar mass in the cluster per Gyr (Fig.~\ref{fig:shredding}). 

We considered two other common tracers of the dynamical state, namely the fraction of total mass (including DM) in the galaxy cluster's subhalos, $f_\mathrm{sub}$, and the gas center shift s$_\mathrm{gas}$, defined as the offset between the 3D~gas barycenter and the BCG position. Across all four simulations, $f_\mathrm{ICL+BCG}$ best traces $z_\mathrm{form}$ (Fig.~\ref{fig:dyncor}). This is likely because different dynamical tracers trace different dynamical times, as shown by \citet{haggar24}. In the future, a more detailed investigation in the timescales involved for the gas (X-ray centershift) and dark matter stripping ($f_\mathrm{sub}$) is necessary to understand when to use which tracer, to inform the new and upcoming measurements from multiple instruments that will significantly improve the set of observed galaxy clusters, such as {\it Euclid} \citep{laureijs:2011,kluge:2024}, Roman \citep{montes:2023} and Rubin/LSST \citep{ivezic:2019,brough:2020}. However, that is beyond the scope of this work. Instead, we provide here fitting functions to predict the formation redshift $z_\mathrm{form}$ of observed galaxy clusters from measurements of $f_\mathrm{ICL+BCG}$ (Table~\ref{tab:sims3}).

We show that $f_\mathrm{ICL+BCG}$, which is difficult to measure because of the low-surface brightness of the outer ICL, can best be approximated by the stellar mass ratio between the BCG and the second or fourth most massive galaxy, M$12$ or M$14$ (Fig.~\ref{fig:iclcor}), in agreement with \citet{golden-marx:2025}. Furthermore, all four simulations agree remarkably well on the presented relations, even though they are tuned to reproduce different relations, with Magneticum in particular focused on the hot gas component \citep{popesso:2024}. This is likely because the processes of stripping and merging for the stellar component of galaxy clusters is driven by gravity, which is treated equivalently between the simulations. We provide fitting functions from each simulation to estimate $f_\mathrm{ICL+BCG}$ based on M$12$, M$14$, the stellar barycenter offset $\mathrm{s}_\mathrm{stars}$ and the ratio of BCG stellar mass to half mass radius $\phi_\mathrm{stars}$ (Table~\ref{tab:sims3}). Finally, we determine that, although M$12$ and M$14$ well trace $z_\mathrm{form}$ for all simulations, the underlying driver of the correlation is $f_\mathrm{ICL+BCG}$ (Fig.~\ref{fig:cormat}). Accurate measurements of the low-surface brightness ICL of galaxy clusters are thus necessary to best predict galaxy cluster assembly. 

We conclude that the assembly histories of galaxy clusters are varied and complex, but that their present-day properties hold the key to disentangling their evolution. With the increasing number of massive galaxies observed by \textit{JWST}, finding a means to robustly connect the largest cosmic structures through time is becoming increasingly necessary. This study finds that the distribution of the stellar component of these clusters, quantified either via the fraction of the total stellar mass in the ICL+BCG $f_\mathrm{ICL+BCG}$ or alternatively the stellar mass ratios of the BCG to the second or fourth most massive galaxy in the cluster, M$12$ or M$14$, is a crucial tool to predict whether a cluster originates from an early or late collapsing node of the cosmic web. Therefore, despite the difficulty of measuring this low-surface brightness feature of galaxy clusters, it is of great importance as the field of galaxy cluster research moves into the next phase with statistically significant samples from our observed Universe.

\begin{acknowledgements}
LCK acknowledges support by the Deutsche Forschungsgemeinschaft (DFG, German Research Foundation) under project nr. 516355818. SB acknowledges funding support from the Australian Research Council through a Discovery Project DP190101943. KD and LCK acknowledge support for the COMPLEX project from the European Research Council (ERC) under the European Union’s Horizon 2020 research and innovation program grant agreement ERC-2019-AdG 882679. LCK, KD and RSR acknowledge support by DFG under Germany's Excellence Strategy -- EXC-2096 -- 3900783311. YMB acknowledges support from UK Research and Innovation through a Future Leaders Fellowship (grant agreement MR/X035166/1) and financial support from the Swiss National Science Foundation (SNSF) under project 200021\_213076. GM and NH are supported by STFC consolidated grant ST/X000982/1.

MM acknowledges support from grant RYC2022-036949-I financed by the MICIU/AEI/10.13039/501100011033 and by ESF+ and program Unidad de Excelencia Mar\'{i}a de Maeztu CEX2020-001058-M. HJB gratefully acknowledges support from the UK Science and Technology Facilities Council (STFC) under grant ST/Y509437/1. AE acknowledges funding by the CNES post-doctoral fellowship program. Y.J-T. acknowledges financial support from the State Agency for Research of the Spanish MCIU through Center of Excellence Severo Ochoa award to the Instituto de Astrofísica de Andalucía CEX2021-001131-S funded by CIN/AEI/10.13039/501100011033, and from the grant PID2022-136598NB-C32 Estallidos and project ref. AST22-00001-Subp-15 funded from the EU-NextGenerationEU. R.R. acknowledges financial support through grants PRIN-MIUR 2020SKSTHZ and through INAF-WEAVE StePS founds. MS acknowledges the support by the Italian Ministry for Education University and Research (MIUR) grant PRIN 2022 2022383WFT ``SUNRISE", CUP C53D23000850006 and by VST funds. Co-funded by the European Union (MSCA Doctoral Network EDUCADO, GA 101119830 and Widening Participation, ExGal-Twin, GA 101158446). JHK acknowledges Spanish grant PID2022-136505NB-I00 funded by MCIN/AEI/10.13039/501100011033 and EU, ERDF. This work used the DiRAC@Durham facility managed by the Institute for Computational Cosmology on behalf of the STFC DiRAC HPC Facility (www.dirac.ac.uk). The equipment was funded by BEIS capital funding via STFC capital grants ST/K00042X/1, ST/P002293/1 and ST/R002371/1, Durham University and STFC operations grant ST/S003908/1. DiRAC is part of the National e-Infrastructure.

The {\it Magneticum} simulations were performed at the Leibniz-Rechenzentrum with CPU time assigned to the Project {\it pr83li}. We are especially grateful for the support by M. Petkova through the Computational Center for Particle and Astrophysics (C2PAP).

The Horizon-AGN simulations were undertaken using the HPC resources of CINES under the allocations 2013047012, 2014047012 and 2015047012 made by GENCI and were partially supported by grant Spin(e) (ANR-13-BS05-0005, \url{http://cosmicorigin.org}). This work has made use of the Horizon cluster on which the simulation was post-processed, hosted by the Institut d’Astrophysique de Paris. Part of the analysis of the simulations was carried out on the DiRAC Facility jointly funded by BIS and STFC. We warmly thank S. Rouberol for running it smoothly.

The IllustrisTNG simulations were undertaken with compute time awarded by the Gauss Centre for Supercomputing (GCS) under GCS Large-Scale Projects GCS-ILLU and GCS-DWAR on the GCS share of the supercomputer Hazel Hen at the High Performance Computing Center Stuttgart (HLRS), as well as on the machines of the Max Planck Computing and Data Facility (MPCDF) in Garching, Germany.

This research was supported by the International Space Science Institute (ISSI) in Bern, through ISSI International Team project number 23-577.
\end{acknowledgements}

\section*{Data Availability}
The Magneticum Pathfinder simulation data are available at https://c2papcosmosim.uc.lrz.de/ \citep{ragagnin17}, with larger data sets on request.

The Hydrangea simulation data are publicly available at https://ftp.strw.leidenuniv.nl/bahe/Hydrangea/.

The Horizon-AGN data used in this work can be obtained upon request from https://www.horizon-simulation.org/data.html.

The IllustrisTNG data used in this work are publicly available at http://www.tng-project.org.



\bibliographystyle{aa}
\bibliography{icl}

\begin{thebibliography}{195}
\expandafter\ifx\csname natexlab\endcsname\relax\def\natexlab#1{#1}\fi

\bibitem[{{Aguerri} {et~al.}(2006){Aguerri}, {Castro-Rodr{\'\i}guez}, {Napolitano}, {Arnaboldi}, \& {Gerhard}}]{Aguerri2006}
{Aguerri}, J.~A.~L., {Castro-Rodr{\'\i}guez}, N., {Napolitano}, N., {Arnaboldi}, M., \& {Gerhard}, O. 2006, \aap, 457, 771

\bibitem[{{Ahad} {et~al.}(2023){Ahad}, {Bah{\'e}}, \& {Hoekstra}}]{Ahad2023}
{Ahad}, S.~L., {Bah{\'e}}, Y.~M., \& {Hoekstra}, H. 2023, \mnras, 518, 3685

\bibitem[{{Ahad} {et~al.}(2021){Ahad}, {Bah{\'e}}, {Hoekstra}, {van der Burg}, \& {Muzzin}}]{Ahad_et_al_2021}
{Ahad}, S.~L., {Bah{\'e}}, Y.~M., {Hoekstra}, H., {van der Burg}, R. F.~J., \& {Muzzin}, A. 2021, \mnras, 504, 1999

\bibitem[{{Ahvazi} {et~al.}(2024){Ahvazi}, {Sales}, {Navarro}, {Benson}, {Boselli}, \& {D'Souza}}]{ahvazi:2024}
{Ahvazi}, N., {Sales}, L.~V., {Navarro}, J.~F., {et~al.} 2024, The Open Journal of Astrophysics, 7, 111

\bibitem[{{Ald{\'a}s} {et~al.}(2023){Ald{\'a}s}, {Zenteno}, {G{\'o}mez}, {Hernandez-Lang}, {Carrasco}, {Vega-Mart{\'\i}nez}, \& {Nilo Castell{\'o}n}}]{aldas23}
{Ald{\'a}s}, F., {Zenteno}, A., {G{\'o}mez}, F.~A., {et~al.} 2023, \mnras, 525, 1769

\bibitem[{{Andreon}(2012)}]{andreon:2012}
{Andreon}, S. 2012, \aap, 548, A83

\bibitem[{{Bah{\'e}} {et~al.}(2017){Bah{\'e}}, {Barnes}, {Dalla Vecchia}, {Kay}, {White}, {McCarthy}, {Schaye}, {Bower}, {Crain}, {Theuns}, {Jenkins}, {McGee}, {Schaller}, {Thomas}, \& {Trayford}}]{bahe17}
{Bah{\'e}}, Y.~M., {Barnes}, D.~J., {Dalla Vecchia}, C., {et~al.} 2017, \mnras, 470, 4186

\bibitem[{{Bah{\'e}} {et~al.}(2019){Bah{\'e}}, {Schaye}, {Barnes}, {Dalla Vecchia}, {Kay}, {Bower}, {Hoekstra}, {McGee}, \& {Theuns}}]{Bahe2019}
{Bah{\'e}}, Y.~M., {Schaye}, J., {Barnes}, D.~J., {et~al.} 2019, \mnras, 485, 2287

\bibitem[{{Barnes} {et~al.}(2017){Barnes}, {Kay}, {Bah{\'e}}, {Dalla Vecchia}, {McCarthy}, {Schaye}, {Bower}, {Jenkins}, {Thomas}, {Schaller}, {Crain}, {Theuns}, \& {White}}]{Barnes_et_al_2017}
{Barnes}, D.~J., {Kay}, S.~T., {Bah{\'e}}, Y.~M., {et~al.} 2017, \mnras, 471, 1088

\bibitem[{{Beck} {et~al.}(2016){Beck}, {Murante}, {Arth}, {Remus}, {Teklu}, {Donnert}, {Planelles}, {Beck}, {F{\"o}rster}, {Imgrund}, {Dolag}, \& {Borgani}}]{beck:2015}
{Beck}, A.~M., {Murante}, G., {Arth}, A., {et~al.} 2016, \mnras, 455, 2110

\bibitem[{{Bender} {et~al.}(2015){Bender}, {Kormendy}, {Cornell}, \& {Fisher}}]{bender:2015}
{Bender}, R., {Kormendy}, J., {Cornell}, M.~E., \& {Fisher}, D.~B. 2015, \apj, 807, 56

\bibitem[{{Biffi} {et~al.}(2016){Biffi}, {Borgani}, {Murante}, {Rasia}, {Planelles}, {Granato}, {Ragone-Figueroa}, {Beck}, {Gaspari}, \& {Dolag}}]{biffi16}
{Biffi}, V., {Borgani}, S., {Murante}, G., {et~al.} 2016, \apj, 827, 112

\bibitem[{{Bilata-Woldeyes} {et~al.}(2025){Bilata-Woldeyes}, {Perea}, \& {Solanes}}]{bilata:woldeyes:2025}
{Bilata-Woldeyes}, B., {Perea}, J.~D., \& {Solanes}, J.~M. 2025, arXiv e-prints, arXiv:2502.14461

\bibitem[{{Bluck} {et~al.}(2023){Bluck}, {Piotrowska}, \& {Maiolino}}]{bluck:2023}
{Bluck}, A. F.~L., {Piotrowska}, J.~M., \& {Maiolino}, R. 2023, \apj, 944, 108

\bibitem[{{Boylan-Kolchin}(2023)}]{boylan-kolchin:2023}
{Boylan-Kolchin}, M. 2023, Nature Astronomy, 7, 731

\bibitem[{{Boylan-Kolchin} {et~al.}(2009){Boylan-Kolchin}, {Springel}, {White}, {Jenkins}, \& {Lemson}}]{boylan:2009}
{Boylan-Kolchin}, M., {Springel}, V., {White}, S. D.~M., {Jenkins}, A., \& {Lemson}, G. 2009, \mnras, 398, 1150

\bibitem[{{Brambila} {et~al.}(2023){Brambila}, {Lopes}, {Ribeiro}, \& {Cortesi}}]{brambila23}
{Brambila}, D., {Lopes}, P. A.~A., {Ribeiro}, A. L.~B., \& {Cortesi}, A. 2023, \mnras, 523, 785

\bibitem[{{Brough} {et~al.}(2024){Brough}, {Ahad}, {Bah{\'e}}, {Ellien}, {Gonzalez}, {Jim{\'e}nez-Teja}, {Kimmig}, {Martin}, {Mart{\'\i}nez-Lombilla}, {Montes}, {Pillepich}, {Ragusa}, {Remus}, {Collins}, {Knapen}, \& {Mihos}}]{brough24}
{Brough}, S., {Ahad}, S.~L., {Bah{\'e}}, Y.~M., {et~al.} 2024, \mnras, 528, 771

\bibitem[{{Brough} {et~al.}(2020){Brough}, {Collins}, {Demarco}, {Ferguson}, {Galaz}, {Holwerda}, {Martinez-Lombilla}, {Mihos}, \& {Montes}}]{brough:2020}
{Brough}, S., {Collins}, C., {Demarco}, R., {et~al.} 2020, arXiv e-prints, arXiv:2001.11067

\bibitem[{{Brough} {et~al.}(2008){Brough}, {Couch}, {Collins}, {Jarrett}, {Burke}, \& {Mann}}]{Brough2008}
{Brough}, S., {Couch}, W.~J., {Collins}, C.~A., {et~al.} 2008, \mnras, 385, L103

\bibitem[{{Brough} {et~al.}(2011){Brough}, {Tran}, {Sharp}, {von der Linden}, \& {Couch}}]{brough:2011}
{Brough}, S., {Tran}, K.~V., {Sharp}, R.~G., {von der Linden}, A., \& {Couch}, W.~J. 2011, \mnras, 414, L80

\bibitem[{{Brown} {et~al.}(2024){Brown}, {Martin}, {Pearce}, {Hatch}, {Bah{\'e}}, \& {Dubois}}]{brown:2024}
{Brown}, H.~J., {Martin}, G., {Pearce}, F.~R., {et~al.} 2024, \mnras, 534, 431

\bibitem[{{Budzynski} {et~al.}(2014){Budzynski}, {Koposov}, {McCarthy}, \& {Belokurov}}]{budzynski14}
{Budzynski}, J.~M., {Koposov}, S.~E., {McCarthy}, I.~G., \& {Belokurov}, V. 2014, \mnras, 437, 1362

\bibitem[{{Burke} {et~al.}(2015){Burke}, {Hilton}, \& {Collins}}]{Burke2015}
{Burke}, C., {Hilton}, M., \& {Collins}, C. 2015, \mnras, 449, 2353

\bibitem[{{Burns} {et~al.}(1994){Burns}, {Roettiger}, {Ledlow}, \& {Klypin}}]{burns:1994}
{Burns}, J.~O., {Roettiger}, K., {Ledlow}, M., \& {Klypin}, A. 1994, \apjl, 427, L87

\bibitem[{{Canepa} {et~al.}(2025){Canepa}, {Brough}, {Lanusse}, {Montes}, \& {Hatch}}]{canepa:2025}
{Canepa}, L., {Brough}, S., {Lanusse}, F., {Montes}, M., \& {Hatch}, N. 2025, \apj, 980, 245

\bibitem[{{Carnall} {et~al.}(2023){Carnall}, {McLeod}, {McLure}, {Dunlop}, {Begley}, {Cullen}, {Donnan}, {Hamadouche}, {Jewell}, {Jones}, {Pollock}, \& {Wild}}]{carnall:2023}
{Carnall}, A.~C., {McLeod}, D.~J., {McLure}, R.~J., {et~al.} 2023, \mnras, 520, 3974

\bibitem[{{Casas} {et~al.}(2024){Casas}, {Putnam}, {Mantz}, {Allen}, \& {Somboonpanyakul}}]{casas:2024}
{Casas}, M.~C., {Putnam}, K., {Mantz}, A.~B., {Allen}, S.~W., \& {Somboonpanyakul}, T. 2024, \apj, 967, 14

\bibitem[{{Choi} {et~al.}(2018){Choi}, {Yi}, {Dubois}, {Kimm}, {Devriendt}, \& {Pichon}}]{choi:2018}
{Choi}, H., {Yi}, S.~K., {Dubois}, Y., {et~al.} 2018, \apj, 856, 114

\bibitem[{{Chon} {et~al.}(2015){Chon}, {B{\"o}hringer}, \& {Zaroubi}}]{chon:2015}
{Chon}, G., {B{\"o}hringer}, H., \& {Zaroubi}, S. 2015, \aap, 575, L14

\bibitem[{{Chun} {et~al.}(2024){Chun}, {Shin}, {Ko}, {Smith}, \& {Yoo}}]{chun:2024}
{Chun}, K., {Shin}, J., {Ko}, J., {Smith}, R., \& {Yoo}, J. 2024, \apj, 969, 142

\bibitem[{{Conroy} {et~al.}(2007){Conroy}, {Wechsler}, \& {Kravtsov}}]{Conroy07}
{Conroy}, C., {Wechsler}, R.~H., \& {Kravtsov}, A.~V. 2007, \apj, 668, 826

\bibitem[{{Contini} {et~al.}(2014){Contini}, {De Lucia}, {Villalobos}, \& {Borgani}}]{Contini2014}
{Contini}, E., {De Lucia}, G., {Villalobos}, {\'A}., \& {Borgani}, S. 2014, \mnras, 437, 3787

\bibitem[{{Contini} {et~al.}(2023){Contini}, {Jeon}, {Rhee}, {Han}, \& {Yi}}]{Contini2023A}
{Contini}, E., {Jeon}, S., {Rhee}, J., {Han}, S., \& {Yi}, S.~K. 2023, arXiv e-prints, arXiv:2310.03263

\bibitem[{{Contini} {et~al.}(2024{\natexlab{a}}){Contini}, {Rhee}, {Han}, {Jeon}, \& {Yi}}]{contini:2024b}
{Contini}, E., {Rhee}, J., {Han}, S., {Jeon}, S., \& {Yi}, S.~K. 2024{\natexlab{a}}, \aj, 167, 7

\bibitem[{{Contini} {et~al.}(2024{\natexlab{b}}){Contini}, {Yi}, \& {Jeon}}]{contini:2024}
{Contini}, E., {Yi}, S.~K., \& {Jeon}, S. 2024{\natexlab{b}}, arXiv e-prints, arXiv:2404.01560

\bibitem[{{Contini} {et~al.}(2018){Contini}, {Yi}, \& {Kang}}]{Contini2018}
{Contini}, E., {Yi}, S.~K., \& {Kang}, X. 2018, \mnras, 479, 932

\bibitem[{{Contreras-Santos} {et~al.}(2024){Contreras-Santos}, {Knebe}, {Cui}, {Alonso Asensio}, {Dalla Vecchia}, {Ca{\~n}as}, {Haggar}, {Mostoghiu Paun}, {Pearce}, \& {Rasia}}]{contreras24}
{Contreras-Santos}, A., {Knebe}, A., {Cui}, W., {et~al.} 2024, \aap, 683, A59

\bibitem[{{Contreras-Santos} {et~al.}(2022){Contreras-Santos}, {Knebe}, {Pearce}, {Haggar}, {Gray}, {Cui}, {Yepes}, {De Petris}, {De Luca}, {Power}, {Mostoghiu}, {Nuza}, \& {Hoeft}}]{contreras22}
{Contreras-Santos}, A., {Knebe}, A., {Pearce}, F., {et~al.} 2022, \mnras, 511, 2897

\bibitem[{{Crain} {et~al.}(2015){Crain}, {Schaye}, {Bower}, {Furlong}, {Schaller}, {Theuns}, {Dalla Vecchia}, {Frenk}, {McCarthy}, {Helly}, {Jenkins}, {Rosas-Guevara}, {White}, \& {Trayford}}]{crain:2015}
{Crain}, R.~A., {Schaye}, J., {Bower}, R.~G., {et~al.} 2015, \mnras, 450, 1937

\bibitem[{{Cui} {et~al.}(2018){Cui}, {Knebe}, {Yepes}, {Pearce}, {Power}, {Dave}, {Arth}, {Borgani}, {Dolag}, {Elahi}, {Mostoghiu}, {Murante}, {Rasia}, {Stoppacher}, {Vega-Ferrero}, {Wang}, {Yang}, {Benson}, {Cora}, {Croton}, {Sinha}, {Stevens}, {Vega-Mart{\'\i}nez}, {Arthur}, {Baldi}, {Ca{\~n}as}, {Cialone}, {Cunnama}, {De Petris}, {Durando}, {Ettori}, {Gottl{\"o}ber}, {Nuza}, {Old}, {Pilipenko}, {Sorce}, \& {Welker}}]{cui:2018}
{Cui}, W., {Knebe}, A., {Yepes}, G., {et~al.} 2018, \mnras, 480, 2898

\bibitem[{{Cui} {et~al.}(2017){Cui}, {Power}, {Borgani}, {Knebe}, {Lewis}, {Murante}, \& {Poole}}]{cui17}
{Cui}, W., {Power}, C., {Borgani}, S., {et~al.} 2017, \mnras, 464, 2502

\bibitem[{{Da Rocha} {et~al.}(2008){Da Rocha}, {Ziegler}, \& {Mendes de Oliveira}}]{DaRocha2008}
{Da Rocha}, C., {Ziegler}, B.~L., \& {Mendes de Oliveira}, C. 2008, \mnras, 388, 1433

\bibitem[{{Davison} {et~al.}(2020){Davison}, {Norris}, {Pfeffer}, {Davies}, \& {Crain}}]{davison:2020}
{Davison}, T.~A., {Norris}, M.~A., {Pfeffer}, J.~L., {Davies}, J.~J., \& {Crain}, R.~A. 2020, \mnras, 497, 81

\bibitem[{{De Lucia} \& {Blaizot}(2007)}]{DeLucia2007}
{De Lucia}, G. \& {Blaizot}, J. 2007, \mnras, 375, 2

\bibitem[{{DeMaio} {et~al.}(2018){DeMaio}, {Gonzalez}, {Zabludoff}, {Zaritsky}, {Connor}, {Donahue}, \& {Mulchaey}}]{demaio18}
{DeMaio}, T., {Gonzalez}, A.~H., {Zabludoff}, A., {et~al.} 2018, \mnras, 474, 3009

\bibitem[{{Dolag} {et~al.}(2009){Dolag}, {Borgani}, {Murante}, \& {Springel}}]{dolag09}
{Dolag}, K., {Borgani}, S., {Murante}, G., \& {Springel}, V. 2009, \mnras, 399, 497

\bibitem[{{Dolag} {et~al.}(2004){Dolag}, {Jubelgas}, {Springel}, {Borgani}, \& {Rasia}}]{dolag04}
{Dolag}, K., {Jubelgas}, M., {Springel}, V., {Borgani}, S., \& {Rasia}, E. 2004, \apjl, 606, L97

\bibitem[{{Dolag} {et~al.}(2010){Dolag}, {Murante}, \& {Borgani}}]{Dolag2010}
{Dolag}, K., {Murante}, G., \& {Borgani}, S. 2010, \mnras, 405, 1544

\bibitem[{{Dolag} {et~al.}(2023){Dolag}, {Sorce}, {Pilipenko}, {Hern{\'a}ndez-Mart{\'\i}nez}, {Valentini}, {Gottl{\"o}ber}, {Aghanim}, \& {Khabibullin}}]{dolag:2023}
{Dolag}, K., {Sorce}, J.~G., {Pilipenko}, S., {et~al.} 2023, \aap, 677, A169

\bibitem[{{Dolag} {et~al.}(2005){Dolag}, {Vazza}, {Brunetti}, \& {Tormen}}]{dolag05}
{Dolag}, K., {Vazza}, F., {Brunetti}, G., \& {Tormen}, G. 2005, \mnras, 364, 753

\bibitem[{{Donahue} {et~al.}(2015){Donahue}, {Connor}, {Fogarty}, {Li}, {Voit}, {Postman}, {Koekemoer}, {Moustakas}, {Bradley}, \& {Ford}}]{donahue:2015}
{Donahue}, M., {Connor}, T., {Fogarty}, K., {et~al.} 2015, \apj, 805, 177

\bibitem[{{Donnari} {et~al.}(2021){Donnari}, {Pillepich}, {Joshi}, {Nelson}, {Genel}, {Marinacci}, {Rodriguez-Gomez}, {Pakmor}, {Torrey}, {Vogelsberger}, \& {Hernquist}}]{donnari:2021}
{Donnari}, M., {Pillepich}, A., {Joshi}, G.~D., {et~al.} 2021, \mnras, 500, 4004

\bibitem[{{Donnert} {et~al.}(2013){Donnert}, {Dolag}, {Brunetti}, \& {Cassano}}]{donnert13}
{Donnert}, J., {Dolag}, K., {Brunetti}, G., \& {Cassano}, R. 2013, \mnras, 429, 3564

\bibitem[{{Dubois} {et~al.}(2016){Dubois}, {Peirani}, {Pichon}, {Devriendt}, {Gavazzi}, {Welker}, \& {Volonteri}}]{dubois:2016}
{Dubois}, Y., {Peirani}, S., {Pichon}, C., {et~al.} 2016, \mnras, 463, 3948

\bibitem[{{Dubois} {et~al.}(2014){Dubois}, {Pichon}, {Welker}, {Le Borgne}, {Devriendt}, {Laigle}, {Codis}, {Pogosyan}, {Arnouts}, {Benabed}, {Bertin}, {Blaizot}, {Bouchet}, {Cardoso}, {Colombi}, {de Lapparent}, {Desjacques}, {Gavazzi}, {Kassin}, {Kimm}, {McCracken}, {Milliard}, {Peirani}, {Prunet}, {Rouberol}, {Silk}, {Slyz}, {Sousbie}, {Teyssier}, {Tresse}, {Treyer}, {Vibert}, \& {Volonteri}}]{dubois:2014}
{Dubois}, Y., {Pichon}, C., {Welker}, C., {et~al.} 2014, \mnras, 444, 1453

\bibitem[{{Edwards} {et~al.}(2016){Edwards}, {Alpert}, {Trierweiler}, {Abraham}, \& {Beizer}}]{Edwards2016}
{Edwards}, L.~O.~V., {Alpert}, H.~S., {Trierweiler}, I.~L., {Abraham}, T., \& {Beizer}, V.~G. 2016, \mnras, 461, 230

\bibitem[{{Edwards} {et~al.}(2024){Edwards}, {Hamel}, {Shy}, {Hernandez}, {Holguin Luna}, {Higgins}, {Chawla}, {Gavidia}, \& {Cole}}]{edwards:2024}
{Edwards}, L. O.~V., {Hamel}, K. A.~S.~J., {Shy}, J.~C., {et~al.} 2024, \mnras, 530, 3924

\bibitem[{{Edwards} {et~al.}(2020){Edwards}, {Salinas}, {Stanley}, {Holguin West}, {Trierweiler}, {Alpert}, {Coelho}, {Koppaka}, {Tremblay}, {Martel}, \& {Li}}]{edwards:2020}
{Edwards}, L. O.~V., {Salinas}, M., {Stanley}, S., {et~al.} 2020, \mnras, 491, 2617

\bibitem[{{Ellien} {et~al.}(2021){Ellien}, {Slezak}, {Martinet}, {Durret}, {Adami}, {Gavazzi}, {Raba{\c{c}}a}, {Da Rocha}, \& {Epit{\'a}cio Pereira}}]{ellien:2021}
{Ellien}, A., {Slezak}, E., {Martinet}, N., {et~al.} 2021, \aap, 649, A38

\bibitem[{Feldmeier {et~al.}(2004)Feldmeier, Mihos, Morrison, Harding, Kaib, \& Dubinski}]{Feldmeier2004}
Feldmeier, J.~J., Mihos, J.~C., Morrison, H.~L., {et~al.} 2004, \apj, 609, 617

\bibitem[{{Fu} {et~al.}(2024){Fu}, {Shankar}, {Ayromlou}, {Koutsouridou}, {Cattaneo}, {Bertemes}, {Bellstedt}, {Mart{\'\i}n-Navarro}, {Leja}, {Allevato}, {Bernardi}, {Boco}, {Dimauro}, {Gruppioni}, {Lapi}, {Menci}, {Rodr{\'\i}guez}, {Puglisi}, \& {Alonso-Tetilla}}]{hao:2024}
{Fu}, H., {Shankar}, F., {Ayromlou}, M., {et~al.} 2024, \mnras, 532, 177

\bibitem[{Furnell {et~al.}(2021)Furnell, Collins, Kelvin, Baldry, James, Manolopoulou, Mann, Giles, Bermeo, Hilton, Wilkinson, Romer, Vergara, Bhargava, Stott, Mayers, \& Viana}]{Furnell2021}
Furnell, K.~E., Collins, C.~A., Kelvin, L.~S., {et~al.} 2021, \mnras, 502, 2419

\bibitem[{{Giallongo} {et~al.}(2014){Giallongo}, {Menci}, {Grazian}, {Gallozzi}, {Castellano}, {Fiore}, {Fontana}, {Pentericci}, {Boutsia}, {Paris}, {Speziali}, \& {Testa}}]{Giallongo2014}
{Giallongo}, E., {Menci}, N., {Grazian}, A., {et~al.} 2014, \apj, 781, 24

\bibitem[{{Golden-Marx} \& {Miller}(2018)}]{golden-marx:2018}
{Golden-Marx}, J.~B. \& {Miller}, C.~J. 2018, \apj, 860, 2

\bibitem[{{Golden-Marx} {et~al.}(2025){Golden-Marx}, {Zhang}, {Ogando}, {Yanny}, {da Silva Pereira}, {Hilton}, {Aguena}, {Allam}, {Andrade-Oliveira}, {Bacon}, {Brooks}, {Carnero Rosell}, {Carretero}, {Cheng}, {da Costa}, {De Vicente}, {Desai}, {Doel}, {Everett}, {Ferrero}, {Frieman}, {Garc{\'\i}a-Bellido}, {Gatti}, {Giannini}, {Gruen}, {Gruendl}, {Gutierrez}, {Hinton}, {Hollowood}, {Honscheid}, {James}, {Kuehn}, {Lee}, {Mena-Fern{\'a}ndez}, {Menanteau}, {Miquel}, {Mohr}, {Palmese}, {Pieres}, {Plazas Malag{\'o}n}, {Samuroff}, {Sanchez}, {Schubnell}, {Sevilla-Noarbe}, {Smith}, {Suchyta}, {Tarle}, {Vikram}, {Walker}, {Weaverdyck}, \& {Wiseman}}]{golden-marx:2025}
{Golden-Marx}, J.~B., {Zhang}, Y., {Ogando}, R.~L.~C., {et~al.} 2025, \mnras, 538, 622

\bibitem[{{Gonzalez} {et~al.}(2013){Gonzalez}, {Sivanandam}, {Zabludoff}, \& {Zaritsky}}]{gonzalez13}
{Gonzalez}, A.~H., {Sivanandam}, S., {Zabludoff}, A.~I., \& {Zaritsky}, D. 2013, \apj, 778, 14

\bibitem[{{Gonzalez} {et~al.}(2007){Gonzalez}, {Zaritsky}, \& {Zabludoff}}]{Gonzalez2007}
{Gonzalez}, A.~H., {Zaritsky}, D., \& {Zabludoff}, A.~I. 2007, ApJ, 666, 147

\bibitem[{{Grandis} {et~al.}(2024){Grandis}, {Ghirardini}, {Bocquet}, {Garrel}, {Mohr}, {Liu}, {Kluge}, {Kimmig}, {Reiprich}, {Alarcon}, {Amon}, {Artis}, {Bahar}, {Balzer}, {Bechtol}, {Becker}, {Bernstein}, {Bulbul}, {Campos}, {Carnero Rosell}, {Carrasco Kind}, {Cawthon}, {Chang}, {Chen}, {Chiu}, {Choi}, {Clerc}, {Comparat}, {Cordero}, {Davis}, {Derose}, {Diehl}, {Dodelson}, {Doux}, {Drlica-Wagner}, {Eckert}, {Elvin-Poole}, {Everett}, {Ferte}, {Gatti}, {Giannini}, {Giles}, {Gruen}, {Gruendl}, {Harrison}, {Hartley}, {Herner}, {Huff}, {Kleinebreil}, {Kuropatkin}, {Leget}, {Maccrann}, {Mccullough}, {Merloni}, {Myles}, {Nandra}, {Navarro-Alsina}, {Okabe}, {Pacaud}, {Pandey}, {Prat}, {Predehl}, {Ramos}, {Raveri}, {Rollins}, {Roodman}, {Ross}, {Rykoff}, {Sanchez}, {Sanders}, {Schrabback}, {Secco}, {Seppi}, {Sevilla-Noarbe}, {Sheldon}, {Shin}, {Troxel}, {Tutusaus}, {Varga}, {Wu}, {Yanny}, {Yin}, {Zhang}, {Zhang}, {Alves}, {Bhargava}, {Brooks}, {Burke}, {Carretero}, {Costanzi}, {da Costa}, {Pereira}, {De Vicente},
  {Desai}, {Doel}, {Ferrero}, {Flaugher}, {Friedel}, {Frieman}, {Garc{\'\i}a-Bellido}, {Gutierrez}, {Hinton}, {Hollowood}, {Honscheid}, {James}, {Jeffrey}, {Lahav}, {Lee}, {Marshall}, {Menanteau}, {Ogando}, {Pieres}, {Plazas Malag{\'o}n}, {Romer}, {Sanchez}, {Schubnell}, {Smith}, {Suchyta}, {Swanson}, {Tarle}, {Weaverdyck}, \& {Weller}}]{grandis:2024}
{Grandis}, S., {Ghirardini}, V., {Bocquet}, S., {et~al.} 2024, \aap, 687, A178

\bibitem[{{Haggar} {et~al.}(2024){Haggar}, {De Luca}, {De Petris}, {Sazonova}, {Taylor}, {Knebe}, {Gray}, {Pearce}, {Contreras-Santos}, {Cui}, {Kuchner}, {Mostoghiu Paun}, \& {Power}}]{haggar24}
{Haggar}, R., {De Luca}, F., {De Petris}, M., {et~al.} 2024, \mnras, 532, 1031

\bibitem[{{Haggar} {et~al.}(2020){Haggar}, {Gray}, {Pearce}, {Knebe}, {Cui}, {Mostoghiu}, \& {Yepes}}]{haggar20}
{Haggar}, R., {Gray}, M.~E., {Pearce}, F.~R., {et~al.} 2020, \mnras, 492, 6074

\bibitem[{{Hahn} {et~al.}(2017){Hahn}, {Martizzi}, {Wu}, {Evrard}, {Teyssier}, \& {Wechsler}}]{hahn:2017}
{Hahn}, O., {Martizzi}, D., {Wu}, H.-Y., {et~al.} 2017, \mnras, 470, 166

\bibitem[{{Iodice} {et~al.}(2016){Iodice}, {Capaccioli}, {Grado}, {Limatola}, {Spavone}, {Napolitano}, {Paolillo}, {Peletier}, {Cantiello}, {Lisker}, {Wittmann}, {Venhola}, {Hilker}, {D'Abrusco}, {Pota}, \& {Schipani}}]{Iodice2016}
{Iodice}, E., {Capaccioli}, M., {Grado}, A., {et~al.} 2016, ApJ, 820, 42

\bibitem[{{Ito} {et~al.}(2024){Ito}, {Valentino}, {Brammer}, {Faisst}, {Gillman}, {G{\'o}mez-Guijarro}, {Gould}, {Heintz}, {Ilbert}, {Jespersen}, {Kokorev}, {Kubo}, {Magdis}, {McPartland}, {Onodera}, {Rizzo}, {Tanaka}, {Toft}, {Vijayan}, {Weaver}, {Whitaker}, \& {Wright}}]{ito:2024}
{Ito}, K., {Valentino}, F., {Brammer}, G., {et~al.} 2024, \apj, 964, 192

\bibitem[{{Ivezi{\'c}} {et~al.}(2019){Ivezi{\'c}}, {Kahn}, {Tyson}, {Abel}, {Acosta}, {Allsman}, {Alonso}, {AlSayyad}, {Anderson}, {Andrew}, {Angel}, {Angeli}, {Ansari}, {Antilogus}, {Araujo}, {Armstrong}, {Arndt}, {Astier}, {Aubourg}, {Auza}, {Axelrod}, {Bard}, {Barr}, {Barrau}, {Bartlett}, {Bauer}, {Bauman}, {Baumont}, {Bechtol}, {Bechtol}, {Becker}, {Becla}, {Beldica}, {Bellavia}, {Bianco}, {Biswas}, {Blanc}, {Blazek}, {Blandford}, {Bloom}, {Bogart}, {Bond}, {Booth}, {Borgland}, {Borne}, {Bosch}, {Boutigny}, {Brackett}, {Bradshaw}, {Brandt}, {Brown}, {Bullock}, {Burchat}, {Burke}, {Cagnoli}, {Calabrese}, {Callahan}, {Callen}, {Carlin}, {Carlson}, {Chandrasekharan}, {Charles-Emerson}, {Chesley}, {Cheu}, {Chiang}, {Chiang}, {Chirino}, {Chow}, {Ciardi}, {Claver}, {Cohen-Tanugi}, {Cockrum}, {Coles}, {Connolly}, {Cook}, {Cooray}, {Covey}, {Cribbs}, {Cui}, {Cutri}, {Daly}, {Daniel}, {Daruich}, {Daubard}, {Daues}, {Dawson}, {Delgado}, {Dellapenna}, {de Peyster}, {de Val-Borro}, {Digel}, {Doherty}, {Dubois},
  {Dubois-Felsmann}, {Durech}, {Economou}, {Eifler}, {Eracleous}, {Emmons}, {Fausti Neto}, {Ferguson}, {Figueroa}, {Fisher-Levine}, {Focke}, {Foss}, {Frank}, {Freemon}, {Gangler}, {Gawiser}, {Geary}, {Gee}, {Geha}, {Gessner}, {Gibson}, {Gilmore}, {Glanzman}, {Glick}, {Goldina}, {Goldstein}, {Goodenow}, {Graham}, {Gressler}, {Gris}, {Guy}, {Guyonnet}, {Haller}, {Harris}, {Hascall}, {Haupt}, {Hernandez}, {Herrmann}, {Hileman}, {Hoblitt}, {Hodgson}, {Hogan}, {Howard}, {Huang}, {Huffer}, {Ingraham}, {Innes}, {Jacoby}, {Jain}, {Jammes}, {Jee}, {Jenness}, {Jernigan}, {Jevremovi{\'c}}, {Johns}, {Johnson}, {Johnson}, {Jones}, {Juramy-Gilles}, {Juri{\'c}}, {Kalirai}, {Kallivayalil}, {Kalmbach}, {Kantor}, {Karst}, {Kasliwal}, {Kelly}, {Kessler}, {Kinnison}, {Kirkby}, {Knox}, {Kotov}, {Krabbendam}, {Krughoff}, {Kub{\'a}nek}, {Kuczewski}, {Kulkarni}, {Ku}, {Kurita}, {Lage}, {Lambert}, {Lange}, {Langton}, {Le Guillou}, {Levine}, {Liang}, {Lim}, {Lintott}, {Long}, {Lopez}, {Lotz}, {Lupton}, {Lust}, {MacArthur}, {Mahabal},
  {Mandelbaum}, {Markiewicz}, {Marsh}, {Marshall}, {Marshall}, {May}, {McKercher}, {McQueen}, {Meyers}, {Migliore}, {Miller}, \& {Mills}}]{ivezic:2019}
{Ivezi{\'c}}, {\v{Z}}., {Kahn}, S.~M., {Tyson}, J.~A., {et~al.} 2019, \apj, 873, 111

\bibitem[{{Jauzac} {et~al.}(2016){Jauzac}, {Eckert}, {Schwinn}, {Harvey}, {Baugh}, {Robertson}, {Bose}, {Massey}, {Owers}, {Ebeling}, {Shan}, {Jullo}, {Kneib}, {Richard}, {Atek}, {Cl{\'e}ment}, {Egami}, {Israel}, {Knowles}, {Limousin}, {Natarajan}, {Rexroth}, {Taylor}, \& {Tchernin}}]{jauzac16}
{Jauzac}, M., {Eckert}, D., {Schwinn}, J., {et~al.} 2016, \mnras, 463, 3876

\bibitem[{{Jeon} {et~al.}(2022){Jeon}, {Yi}, {Dubois}, {Chung}, {Devriendt}, {Han}, {Jackson}, {Kimm}, {Pichon}, \& {Rhee}}]{jeon:2022}
{Jeon}, S., {Yi}, S.~K., {Dubois}, Y., {et~al.} 2022, \apj, 941, 5

\bibitem[{{Jiang} \& {van den Bosch}(2016)}]{jiang16}
{Jiang}, F. \& {van den Bosch}, F.~C. 2016, \mnras, 458, 2848

\bibitem[{{Jiang} \& {van den Bosch}(2017)}]{jiang17}
{Jiang}, F. \& {van den Bosch}, F.~C. 2017, \mnras, 472, 657

\bibitem[{{Jim{\'e}nez-Teja} {et~al.}(2018){Jim{\'e}nez-Teja}, {Dupke}, {Ben{\'{\i}}tez}, {Koekemoer}, {Zitrin}, {Umetsu}, {Ziegler}, {Frye}, {Ford}, {Bouwens}, {Bradley}, {Broadhurst}, {Coe}, {Donahue}, {Graves}, {Grillo}, {Infante}, {Jouvel}, {Kelson}, {Lahav}, {Lazkoz}, {Lemze}, {Maoz}, {Medezinski}, {Melchior}, {Meneghetti}, {Mercurio}, {Merten}, {Molino}, {Moustakas}, {Nonino}, {Ogaz}, {Riess}, {Rosati}, {Sayers}, {Seitz}, \& {Zheng}}]{Jimenez-Teja2018}
{Jim{\'e}nez-Teja}, Y., {Dupke}, R., {Ben{\'{\i}}tez}, N., {et~al.} 2018, \apj, 857, 79

\bibitem[{{Jim{\'e}nez-Teja} {et~al.}(2024){Jim{\'e}nez-Teja}, {Dupke}, {Lopes}, \& {Dimauro}}]{jimenez-teja:2024}
{Jim{\'e}nez-Teja}, Y., {Dupke}, R.~A., {Lopes}, P. A.~A., \& {Dimauro}, P. 2024, \apjl, 960, L7

\bibitem[{{Jim{\'e}nez-Teja} {et~al.}(2023){Jim{\'e}nez-Teja}, {Dupke}, {Lopes}, \& {V{\'\i}lchez}}]{jimenez-teja2023}
{Jim{\'e}nez-Teja}, Y., {Dupke}, R.~A., {Lopes}, P.~A.~A., \& {V{\'\i}lchez}, J.~M. 2023, \aap, 676, A39

\bibitem[{{Jim{\'e}nez-Teja} {et~al.}(2021){Jim{\'e}nez-Teja}, {V{\'\i}lchez}, {Dupke}, {Lopes}, {de Oliveira}, \& {Coe}}]{Jimenez-Teja2021}
{Jim{\'e}nez-Teja}, Y., {V{\'\i}lchez}, J.~M., {Dupke}, R.~A., {et~al.} 2021, \apj, 922, 268

\bibitem[{{Jones} {et~al.}(2003){Jones}, {Ponman}, {Horton}, {Babul}, {Ebeling}, \& {Burke}}]{jones03}
{Jones}, L.~R., {Ponman}, T.~J., {Horton}, A., {et~al.} 2003, \mnras, 343, 627

\bibitem[{{Joo} \& {Jee}(2023)}]{joo:2023_icl}
{Joo}, H. \& {Jee}, M.~J. 2023, \nat, 613, 37

\bibitem[{{Joo} {et~al.}(2024){Joo}, {Jee}, {Kim}, {Lee}, {Ko}, {Park}, {Shin}, {Snaith}, {Pichon}, {Gibson}, \& {Kim}}]{joo:2024}
{Joo}, H., {Jee}, M.~J., {Kim}, J., {et~al.} 2024, arXiv e-prints, arXiv:2411.08117

\bibitem[{{Karademir} {et~al.}(2019){Karademir}, {Remus}, {Burkert}, {Dolag}, {Hoffmann}, {Moster}, {Steinwandel}, \& {Zhang}}]{karademir:2019}
{Karademir}, G.~S., {Remus}, R.-S., {Burkert}, A., {et~al.} 2019, \mnras, 487, 318

\bibitem[{{Khochfar} \& {Burkert}(2003)}]{kochfar:2003}
{Khochfar}, S. \& {Burkert}, A. 2003, \apjl, 597, L117

\bibitem[{{Kimmig} {et~al.}(2023){Kimmig}, {Remus}, {Dolag}, \& {Biffi}}]{kimmig23}
{Kimmig}, L.~C., {Remus}, R.-S., {Dolag}, K., \& {Biffi}, V. 2023, \apj, 949, 92

\bibitem[{{Kimmig} {et~al.}(2025){Kimmig}, {Remus}, {Seidel}, {Valenzuela}, {Dolag}, \& {Burkert}}]{kimmig:2023q}
{Kimmig}, L.~C., {Remus}, R.-S., {Seidel}, B., {et~al.} 2025, \apj, 979, 15

\bibitem[{Kluge {et~al.}(2021)Kluge, Bender, Riffeser, Goessl, Hopp, Schmidt, \& Ries}]{Kluge2021}
Kluge, M., Bender, R., Riffeser, A., {et~al.} 2021, \apjs, 252, 27

\bibitem[{{Kluge} {et~al.}(2024){Kluge}, {Hatch}, {Montes}, {Golden-Marx}, {Gonzalez}, {Cuillandre}, {Bolzonella}, {Lan{\c{c}}on}, {Laureijs}, {Saifollahi}, {Schirmer}, {Stone}, {Boselli}, {Cantiello}, {Sorce}, {Marleau}, {Duc}, {Sola}, {Urbano}, {Ahad}, {Bah{\'e}}, {Bamford}, {Bellhouse}, {Buitrago}, {Dimauro}, {Durret}, {Ellien}, {Jimenez-Teja}, {Slezak}, {Aghanim}, {Altieri}, {Andreon}, {Auricchio}, {Baldi}, {Balestra}, {Bardelli}, {Bender}, {Bonino}, {Branchini}, {Brescia}, {Brinchmann}, {Camera}, {Candini}, {Capobianco}, {Carbone}, {Carretero}, {Casas}, {Castellano}, {Cavuoti}, {Cimatti}, {Congedo}, {Conselice}, {Conversi}, {Copin}, {Courbin}, {Courtois}, {Cropper}, {Da Silva}, {Degaudenzi}, {Dinis}, {Duncan}, {Dupac}, {Dusini}, {Farina}, {Farrens}, {Ferriol}, {Fosalba}, {Frailis}, {Franceschi}, {Fumana}, {Galeotta}, {Garilli}, {Gillard}, {Gillis}, {Giocoli}, {G{\'o}mez-Alvarez}, {Granett}, {Grazian}, {Grupp}, {Guzzo}, {Haugan}, {Hoar}, {Hoekstra}, {Holmes}, {Hook}, {Hormuth}, {Hornstrup}, {Hudelot},
  {Jahnke}, {Keih{\"a}nen}, {Kermiche}, {Kiessling}, {Kitching}, {Kohley}, {Kubik}, {K{\"u}mmel}, {Kunz}, {Kurki-Suonio}, {Lahav}, {Ligori}, {Lilje}, {Lindholm}, {Lloro}, {Maiorano}, {Mansutti}, {Marggraf}, {Markovic}, {Martinet}, {Marulli}, {Massey}, {Maurogordato}, {McCracken}, {Medinaceli}, {Mei}, {Melchior}, {Mellier}, {Meneghetti}, {Merlin}, {Meylan}, {Moresco}, {Moscardini}, {Munari}, {Nichol}, {Niemi}, {Nightingale}, {Padilla}, {Paltani}, {Pasian}, {Pedersen}, {Percival}, {Pettorino}, {Pires}, {Polenta}, {Poncet}, {Popa}, {Pozzetti}, {Racca}, {Raison}, {Rebolo}, {Renzi}, {Rhodes}, {Riccio}, {Rix}, {Romelli}, {Roncarelli}, {Rossetti}, {Saglia}, {Sapone}, {Sartoris}, {Sauvage}, {Scaramella}, {Schneider}, {Schrabback}, {Secroun}, {Seidel}, {Seiffert}, {Serrano}, {Sirignano}, {Sirri}, {Skottfelt}, {Stanco}, {Tallada-Cresp{\'\i}}, {Taylor}, {Teplitz}, {Tereno}, {Toledo-Moreo}, {Torradeflot}, {Tutusaus}, {Valentijn}, {Valenziano}, {Vassallo}, {Verdoes Kleijn}, {Veropalumbo}, {Wang}, {Weller}, {Williams},
  {Zamorani}, {Zucca}, {Biviano}, {Burigana}, {De Lucia}, {George}, {Scottez}, {Simon}, {Mora}, {Mart{\'\i}n-Fleitas}, {Ruppin}, \& {Scott}}]{kluge:2024}
{Kluge}, M., {Hatch}, N.~A., {Montes}, M., {et~al.} 2024, arXiv e-prints, arXiv:2405.13503

\bibitem[{{Kluge} {et~al.}(2020){Kluge}, {Neureiter}, {Riffeser}, {Bender}, {Goessl}, {Hopp}, {Schmidt}, {Ries}, \& {Brosch}}]{Kluge2020}
{Kluge}, M., {Neureiter}, B., {Riffeser}, A., {et~al.} 2020, \apjs, 247, 43

\bibitem[{{Komatsu} {et~al.}(2011){Komatsu}, {Smith}, {Dunkley}, {Bennett}, {Gold}, {Hinshaw}, {Jarosik}, {Larson}, {Nolta}, {Page}, {Spergel}, {Halpern}, {Hill}, {Kogut}, {Limon}, {Meyer}, {Odegard}, {Tucker}, {Weiland}, {Wollack}, \& {Wright}}]{komatsu11}
{Komatsu}, E., {Smith}, K.~M., {Dunkley}, J., {et~al.} 2011, \apjs, 192, 18

\bibitem[{{Kravtsov} {et~al.}(2018){Kravtsov}, {Vikhlinin}, \& {Meshcheryakov}}]{kravtsov18}
{Kravtsov}, A.~V., {Vikhlinin}, A.~A., \& {Meshcheryakov}, A.~V. 2018, Astronomy Letters, 44, 8

\bibitem[{{Lagan{\'a}} {et~al.}(2013){Lagan{\'a}}, {Martinet}, {Durret}, {Lima Neto}, {Maughan}, \& {Zhang}}]{lagana:2013}
{Lagan{\'a}}, T.~F., {Martinet}, N., {Durret}, F., {et~al.} 2013, \aap, 555, A66

\bibitem[{{Laureijs} {et~al.}(2011){Laureijs}, {Amiaux}, {Arduini}, {Augu{\`e}res}, {Brinchmann}, {Cole}, {Cropper}, {Dabin}, {Duvet}, {Ealet}, {Garilli}, {Gondoin}, {Guzzo}, {Hoar}, {Hoekstra}, {Holmes}, {Kitching}, {Maciaszek}, {Mellier}, {Pasian}, {Percival}, {Rhodes}, {Saavedra Criado}, {Sauvage}, {Scaramella}, {Valenziano}, {Warren}, {Bender}, {Castander}, {Cimatti}, {Le F{\`e}vre}, {Kurki-Suonio}, {Levi}, {Lilje}, {Meylan}, {Nichol}, {Pedersen}, {Popa}, {Rebolo Lopez}, {Rix}, {Rottgering}, {Zeilinger}, {Grupp}, {Hudelot}, {Massey}, {Meneghetti}, {Miller}, {Paltani}, {Paulin-Henriksson}, {Pires}, {Saxton}, {Schrabback}, {Seidel}, {Walsh}, {Aghanim}, {Amendola}, {Bartlett}, {Baccigalupi}, {Beaulieu}, {Benabed}, {Cuby}, {Elbaz}, {Fosalba}, {Gavazzi}, {Helmi}, {Hook}, {Irwin}, {Kneib}, {Kunz}, {Mannucci}, {Moscardini}, {Tao}, {Teyssier}, {Weller}, {Zamorani}, {Zapatero Osorio}, {Boulade}, {Foumond}, {Di Giorgio}, {Guttridge}, {James}, {Kemp}, {Martignac}, {Spencer}, {Walton}, {Bl{\"u}mchen}, {Bonoli},
  {Bortoletto}, {Cerna}, {Corcione}, {Fabron}, {Jahnke}, {Ligori}, {Madrid}, {Martin}, {Morgante}, {Pamplona}, {Prieto}, {Riva}, {Toledo}, {Trifoglio}, {Zerbi}, {Abdalla}, {Douspis}, {Grenet}, {Borgani}, {Bouwens}, {Courbin}, {Delouis}, {Dubath}, {Fontana}, {Frailis}, {Grazian}, {Koppenh{\"o}fer}, {Mansutti}, {Melchior}, {Mignoli}, {Mohr}, {Neissner}, {Noddle}, {Poncet}, {Scodeggio}, {Serrano}, {Shane}, {Starck}, {Surace}, {Taylor}, {Verdoes-Kleijn}, {Vuerli}, {Williams}, {Zacchei}, {Altieri}, {Escudero Sanz}, {Kohley}, {Oosterbroek}, {Astier}, {Bacon}, {Bardelli}, {Baugh}, {Bellagamba}, {Benoist}, {Bianchi}, {Biviano}, {Branchini}, {Carbone}, {Cardone}, {Clements}, {Colombi}, {Conselice}, {Cresci}, {Deacon}, {Dunlop}, {Fedeli}, {Fontanot}, {Franzetti}, {Giocoli}, {Garcia-Bellido}, {Gow}, {Heavens}, {Hewett}, {Heymans}, {Holland}, {Huang}, {Ilbert}, {Joachimi}, {Jennins}, {Kerins}, {Kiessling}, {Kirk}, {Kotak}, {Krause}, {Lahav}, {van Leeuwen}, {Lesgourgues}, {Lombardi}, {Magliocchetti}, {Maguire},
  {Majerotto}, {Maoli}, {Marulli}, {Maurogordato}, {McCracken}, {McLure}, {Melchiorri}, {Merson}, {Moresco}, {Nonino}, {Norberg}, {Peacock}, {Pello}, {Penny}, {Pettorino}, {Di Porto}, {Pozzetti}, {Quercellini}, {Radovich}, {Rassat}, {Roche}, {Ronayette}, \& {Rossetti}}]{laureijs:2011}
{Laureijs}, R., {Amiaux}, J., {Arduini}, S., {et~al.} 2011, arXiv e-prints, arXiv:1110.3193

\bibitem[{{Leauthaud} {et~al.}(2012){Leauthaud}, {George}, {Behroozi}, {Bundy}, {Tinker}, {Wechsler}, {Conroy}, {Finoguenov}, \& {Tanaka}}]{leauthaud:2012}
{Leauthaud}, A., {George}, M.~R., {Behroozi}, P.~S., {et~al.} 2012, \apj, 746, 95

\bibitem[{{Libeskind} {et~al.}(2018){Libeskind}, {van de Weygaert}, {Cautun}, {Falck}, {Tempel}, {Abel}, {Alpaslan}, {Arag{\'o}n-Calvo}, {Forero-Romero}, {Gonzalez}, {Gottl{\"o}ber}, {Hahn}, {Hellwing}, {Hoffman}, {Jones}, {Kitaura}, {Knebe}, {Manti}, {Neyrinck}, {Nuza}, {Padilla}, {Platen}, {Ramachandra}, {Robotham}, {Saar}, {Shandarin}, {Steinmetz}, {Stoica}, {Sousbie}, \& {Yepes}}]{libeskind:2018}
{Libeskind}, N.~I., {van de Weygaert}, R., {Cautun}, M., {et~al.} 2018, \mnras, 473, 1195

\bibitem[{{Lidman} {et~al.}(2012){Lidman}, {Suherli}, {Muzzin}, {Wilson}, {Demarco}, {Brough}, {Rettura}, {Cox}, {DeGroot}, {Yee}, {Gilbank}, {Hoekstra}, {Balogh}, {Ellingson}, {Hicks}, {Nantais}, {Noble}, {Lacy}, {Surace}, \& {Webb}}]{Lidman2012}
{Lidman}, C., {Suherli}, J., {Muzzin}, A., {et~al.} 2012, \mnras, 427, 550

\bibitem[{{Loh} \& {Strauss}(2006)}]{loh:2006}
{Loh}, Y.-S. \& {Strauss}, M.~A. 2006, \mnras, 366, 373

\bibitem[{{Longobardi} {et~al.}(2015){Longobardi}, {Arnaboldi}, {Gerhard}, \& {Hanuschik}}]{longobardi:2015}
{Longobardi}, A., {Arnaboldi}, M., {Gerhard}, O., \& {Hanuschik}, R. 2015, \aap, 579, A135

\bibitem[{{Lotz} {et~al.}(2021){Lotz}, {Dolag}, {Remus}, \& {Burkert}}]{lotz:2021}
{Lotz}, M., {Dolag}, K., {Remus}, R.-S., \& {Burkert}, A. 2021, \mnras, 506, 4516

\bibitem[{{Lotz} {et~al.}(2019){Lotz}, {Remus}, {Dolag}, {Biviano}, \& {Burkert}}]{lotz19}
{Lotz}, M., {Remus}, R.-S., {Dolag}, K., {Biviano}, A., \& {Burkert}, A. 2019, \mnras, 488, 5370

\bibitem[{{Mann} \& {Ebeling}(2012)}]{mann12}
{Mann}, A.~W. \& {Ebeling}, H. 2012, \mnras, 420, 2120

\bibitem[{{Mao} {et~al.}(2018){Mao}, {Wang}, {Frenk}, {Gao}, {Li}, {Wang}, {Cao}, \& {Li}}]{mao18}
{Mao}, T.-X., {Wang}, J., {Frenk}, C.~S., {et~al.} 2018, \mnras, 478, L34

\bibitem[{{Marinacci} {et~al.}(2018){Marinacci}, {Vogelsberger}, {Pakmor}, {Torrey}, {Springel}, {Hernquist}, {Nelson}, {Weinberger}, {Pillepich}, {Naiman}, \& {Genel}}]{Marinacci2018}
{Marinacci}, F., {Vogelsberger}, M., {Pakmor}, R., {et~al.} 2018, \mnras, 480, 5113

\bibitem[{{Marini} {et~al.}(2024){Marini}, {Saro}, {Borgani}, \& {Boi}}]{marini:2024}
{Marini}, I., {Saro}, A., {Borgani}, S., \& {Boi}, M. 2024, \aap, 689, A181

\bibitem[{{Martin} {et~al.}(2022){Martin}, {Bazkiaei}, {Spavone}, {Iodice}, {Mihos}, {Montes}, {Benavides}, {Brough}, {Carlin}, {Collins}, {Duc}, {G{\'o}mez}, {Galaz}, {Hern{\'a}ndez-Toledo}, {Jackson}, {Kaviraj}, {Knapen}, {Mart{\'\i}nez-Lombilla}, {McGee}, {O'Ryan}, {Prole}, {Rich}, {Rom{\'a}n}, {Shah}, {Starkenburg}, {Watkins}, {Zaritsky}, {Pichon}, {Armus}, {Bianconi}, {Buitrago}, {Bus{\'a}}, {Davis}, {Demarco}, {Desmons}, {Garc{\'\i}a}, {Graham}, {Holwerda}, {Hon}, {Khalid}, {Klehammer}, {Klutse}, {Lazar}, {Nair}, {Noakes-Kettel}, {Rutkowski}, {Saha}, {Sahu}, {Sola}, {V{\'a}zquez-Mata}, {Vera-Casanova}, \& {Yoon}}]{Martin2022}
{Martin}, G., {Bazkiaei}, A.~E., {Spavone}, M., {et~al.} 2022, \mnras, 513, 1459

\bibitem[{{Martin} {et~al.}(2024){Martin}, {Pearce}, {Hatch}, {Contreras-Santos}, {Knebe}, \& {Cui}}]{martin:2024}
{Martin}, G., {Pearce}, F.~R., {Hatch}, N.~A., {et~al.} 2024, \mnras, 535, 2375

\bibitem[{{Mart{\'\i}nez-Lombilla} {et~al.}(2023){Mart{\'\i}nez-Lombilla}, {Brough}, {Montes}, {Baena-Gall{\'e}}, {Akhlaghi}, {Infante-Sainz}, {Driver}, {Holwerda}, {Pimbblet}, \& {Robotham}}]{Martinez-Lombilla2023a}
{Mart{\'\i}nez-Lombilla}, C., {Brough}, S., {Montes}, M., {et~al.} 2023, \mnras, 518, 1195

\bibitem[{{Matthee} {et~al.}(2017){Matthee}, {Schaye}, {Crain}, {Schaller}, {Bower}, \& {Theuns}}]{matthee:2017}
{Matthee}, J., {Schaye}, J., {Crain}, R.~A., {et~al.} 2017, \mnras, 465, 2381

\bibitem[{{McBride} {et~al.}(2009){McBride}, {Fakhouri}, \& {Ma}}]{mcbride09}
{McBride}, J., {Fakhouri}, O., \& {Ma}, C.-P. 2009, \mnras, 398, 1858

\bibitem[{{McCarthy} {et~al.}(2017){McCarthy}, {Schaye}, {Bird}, \& {Le Brun}}]{mccarthy:2017}
{McCarthy}, I.~G., {Schaye}, J., {Bird}, S., \& {Le Brun}, A. M.~C. 2017, \mnras, 465, 2936

\bibitem[{{Mihos}(2004)}]{mihos:2004}
{Mihos}, J.~C. 2004, in Clusters of Galaxies: Probes of Cosmological Structure and Galaxy Evolution, ed. J.~S. {Mulchaey}, A.~{Dressler}, \& A.~{Oemler}, 277

\bibitem[{{Mihos} {et~al.}(2005){Mihos}, {Harding}, {Feldmeier}, \& {Morrison}}]{Mihos2005}
{Mihos}, J.~C., {Harding}, P., {Feldmeier}, J., \& {Morrison}, H. 2005, \apjl, 631, L41

\bibitem[{{Mihos} {et~al.}(2017){Mihos}, {Harding}, {Feldmeier}, {Rudick}, {Janowiecki}, {Morrison}, {Slater}, \& {Watkins}}]{Mihos2017}
{Mihos}, J.~C., {Harding}, P., {Feldmeier}, J.~J., {et~al.} 2017, \apj, 834, 16

\bibitem[{{Mittal} {et~al.}(2015){Mittal}, {Whelan}, \& {Combes}}]{mittal:2015}
{Mittal}, R., {Whelan}, J.~T., \& {Combes}, F. 2015, \mnras, 450, 2564

\bibitem[{{Monaco} {et~al.}(2006){Monaco}, {Murante}, {Borgani}, \& {Fontanot}}]{Monaco2006}
{Monaco}, P., {Murante}, G., {Borgani}, S., \& {Fontanot}, F. 2006, \apjl, 652, L89

\bibitem[{{Montenegro-Taborda} {et~al.}(2025){Montenegro-Taborda}, {Avila-Reese}, {Rodriguez-Gomez}, {Manuwal}, \& {Cervantes-Sodi}}]{montenegro-taborda:2025}
{Montenegro-Taborda}, D., {Avila-Reese}, V., {Rodriguez-Gomez}, V., {Manuwal}, A., \& {Cervantes-Sodi}, B. 2025, \mnras, 537, 3954

\bibitem[{{Montenegro-Taborda} {et~al.}(2023){Montenegro-Taborda}, {Rodriguez-Gomez}, {Pillepich}, {Avila-Reese}, {Sales}, {Rodr{\'\i}guez-Puebla}, \& {Hernquist}}]{montenegro:2023}
{Montenegro-Taborda}, D., {Rodriguez-Gomez}, V., {Pillepich}, A., {et~al.} 2023, \mnras, 521, 800

\bibitem[{{Montes}(2022)}]{Montes2022}
{Montes}, M. 2022, Nature Astronomy, 6, 308

\bibitem[{{Montes} {et~al.}(2023){Montes}, {Annibali}, {Bellazzini}, {Borlaff}, {Brough}, {Buitrago}, {Chamba}, {Collins}, {Dell'Antonio}, {Escala}, {Gonzalez}, {Holwerda}, {Kaviraj}, {Knapen}, {Koekemoer}, {Laine}, {Marcum}, {Martin}, {Martinez-Delgado}, {Mihos}, {Ricotti}, {Trujillo}, \& {Watkins}}]{montes:2023}
{Montes}, M., {Annibali}, F., {Bellazzini}, M., {et~al.} 2023, arXiv e-prints, arXiv:2306.09414

\bibitem[{{Montes} {et~al.}(2021){Montes}, {Brough}, {Owers}, \& {Santucci}}]{Montes2021}
{Montes}, M., {Brough}, S., {Owers}, M.~S., \& {Santucci}, G. 2021, \apj, 910, 45

\bibitem[{{Montes} \& {Trujillo}(2014)}]{MT14}
{Montes}, M. \& {Trujillo}, I. 2014, \apj, 794, 137

\bibitem[{{Montes} \& {Trujillo}(2018)}]{montes:2018}
{Montes}, M. \& {Trujillo}, I. 2018, \mnras, 474, 917

\bibitem[{{Murante} {et~al.}(2004){Murante}, {Arnaboldi}, {Gerhard}, {Borgani}, {Cheng}, {Diaferio}, {Dolag}, {Moscardini}, {Tormen}, {Tornatore}, \& {Tozzi}}]{murante:2004}
{Murante}, G., {Arnaboldi}, M., {Gerhard}, O., {et~al.} 2004, \apjl, 607, L83

\bibitem[{{Murante} {et~al.}(2007){Murante}, {Giovalli}, {Gerhard}, {Arnaboldi}, {Borgani}, \& {Dolag}}]{Murante2007}
{Murante}, G., {Giovalli}, M., {Gerhard}, O., {et~al.} 2007, \mnras, 377, 2

\bibitem[{{Naiman} {et~al.}(2018){Naiman}, {Pillepich}, {Springel}, {Ramirez-Ruiz}, {Torrey}, {Vogelsberger}, {Pakmor}, {Nelson}, {Marinacci}, {Hernquist}, {Weinberger}, \& {Genel}}]{Naiman2018}
{Naiman}, J.~P., {Pillepich}, A., {Springel}, V., {et~al.} 2018, \mnras, 477, 1206

\bibitem[{{Nelson} {et~al.}(2024){Nelson}, {Pillepich}, {Ayromlou}, {Lee}, {Lehle}, {Rohr}, \& {Truong}}]{nelson:2024}
{Nelson}, D., {Pillepich}, A., {Ayromlou}, M., {et~al.} 2024, \aap, 686, A157

\bibitem[{{Nelson} {et~al.}(2019{\natexlab{a}}){Nelson}, {Pillepich}, {Springel}, {Pakmor}, {Weinberger}, {Genel}, {Torrey}, {Vogelsberger}, {Marinacci}, \& {Hernquist}}]{Nelson2019}
{Nelson}, D., {Pillepich}, A., {Springel}, V., {et~al.} 2019{\natexlab{a}}, \mnras, 490, 3234

\bibitem[{{Nelson} {et~al.}(2018){Nelson}, {Pillepich}, {Springel}, {Weinberger}, {Hernquist}, {Pakmor}, {Genel}, {Torrey}, {Vogelsberger}, {Kauffmann}, {Marinacci}, \& {Naiman}}]{Nelson2018}
{Nelson}, D., {Pillepich}, A., {Springel}, V., {et~al.} 2018, \mnras, 475, 624

\bibitem[{{Nelson} {et~al.}(2019{\natexlab{b}}){Nelson}, {Springel}, {Pillepich}, {Rodriguez-Gomez}, {Torrey}, {Genel}, {Vogelsberger}, {Pakmor}, {Marinacci}, {Weinberger}, {Kelley}, {Lovell}, {Diemer}, \& {Hernquist}}]{nelson:2019}
{Nelson}, D., {Springel}, V., {Pillepich}, A., {et~al.} 2019{\natexlab{b}}, Computational Astrophysics and Cosmology, 6, 2

\bibitem[{{Nelson} {et~al.}(2014){Nelson}, {Lau}, {Nagai}, {Rudd}, \& {Yu}}]{nelson:2014}
{Nelson}, K., {Lau}, E.~T., {Nagai}, D., {Rudd}, D.~H., \& {Yu}, L. 2014, \apj, 782, 107

\bibitem[{{Neto} {et~al.}(2007){Neto}, {Gao}, {Bett}, {Cole}, {Navarro}, {Frenk}, {White}, {Springel}, \& {Jenkins}}]{neto07}
{Neto}, A.~F., {Gao}, L., {Bett}, P., {et~al.} 2007, \mnras, 381, 1450

\bibitem[{{Okabe} {et~al.}(2018){Okabe}, {Nishimichi}, {Oguri}, {Peirani}, {Kitayama}, {Sasaki}, \& {Suto}}]{okabe:2018}
{Okabe}, T., {Nishimichi}, T., {Oguri}, M., {et~al.} 2018, \mnras, 478, 1141

\bibitem[{{Oliva-Altamirano} {et~al.}(2015){Oliva-Altamirano}, {Brough}, {Jimmy}, {Couch}, {McDermid}, {Lidman}, {von der Linden}, \& {Sharp}}]{oliva:2015}
{Oliva-Altamirano}, P., {Brough}, S., {Jimmy}, Kim-Vy, T., {et~al.} 2015, \mnras, 449, 3347

\bibitem[{{Oliva-Altamirano} {et~al.}(2014){Oliva-Altamirano}, {Brough}, {Lidman}, {Couch}, {Hopkins}, {Colless}, {Taylor}, {Robotham}, {Gunawardhana}, {Ponman}, {Baldry}, {Bauer}, {Bland-Hawthorn}, {Cluver}, {Cameron}, {Conselice}, {Driver}, {Edge}, {Graham}, {van Kampen}, {Lara-L{\'o}pez}, {Liske}, {L{\'o}pez-S{\'a}nchez}, {Loveday}, {Mahajan}, {Peacock}, {Phillipps}, {Pimbblet}, \& {Sharp}}]{oliva-altamirano:2014}
{Oliva-Altamirano}, P., {Brough}, S., {Lidman}, C., {et~al.} 2014, \mnras, 440, 762

\bibitem[{{Owers} {et~al.}(2011){Owers}, {Randall}, {Nulsen}, {Couch}, {David}, \& {Kempner}}]{owers:2011}
{Owers}, M.~S., {Randall}, S.~W., {Nulsen}, P. E.~J., {et~al.} 2011, \apj, 728, 27

\bibitem[{{Pakmor} {et~al.}(2023){Pakmor}, {Springel}, {Coles}, {Guillet}, {Pfrommer}, {Bose}, {Barrera}, {Delgado}, {Ferlito}, {Frenk}, {Hadzhiyska}, {Hern{\'a}ndez-Aguayo}, {Hernquist}, {Kannan}, \& {White}}]{pakmor:2023}
{Pakmor}, R., {Springel}, V., {Coles}, J.~P., {et~al.} 2023, \mnras, 524, 2539

\bibitem[{{Pillepich} {et~al.}(2018{\natexlab{a}}){Pillepich}, {Nelson}, {Hernquist}, {Springel}, {Pakmor}, {Torrey}, {Weinberger}, {Genel}, {Naiman}, {Marinacci}, \& {Vogelsberger}}]{pillepich18b}
{Pillepich}, A., {Nelson}, D., {Hernquist}, L., {et~al.} 2018{\natexlab{a}}, \mnras, 475, 648

\bibitem[{{Pillepich} {et~al.}(2018{\natexlab{b}}){Pillepich}, {Springel}, {Nelson}, {Genel}, {Naiman}, {Pakmor}, {Hernquist}, {Torrey}, {Vogelsberger}, {Weinberger}, \& {Marinacci}}]{pillepich18a}
{Pillepich}, A., {Springel}, V., {Nelson}, D., {et~al.} 2018{\natexlab{b}}, \mnras, 473, 4077

\bibitem[{{Planck Collaboration} {et~al.}(2014){Planck Collaboration}, {Ade}, {Aghanim}, {Armitage-Caplan}, {Arnaud}, {Ashdown}, {Atrio-Barandela}, {Aumont}, {Baccigalupi}, {Banday}, {Barreiro}, {Bartlett}, {Battaner}, {Benabed}, {Beno{\^\i}t}, {Benoit-L{\'e}vy}, {Bernard}, {Bersanelli}, {Bielewicz}, {Bobin}, {Bock}, {Bonaldi}, {Bond}, {Borrill}, {Bouchet}, {Bridges}, {Bucher}, {Burigana}, {Butler}, {Calabrese}, {Cappellini}, {Cardoso}, {Catalano}, {Challinor}, {Chamballu}, {Chary}, {Chen}, {Chiang}, {Chiang}, {Christensen}, {Church}, {Clements}, {Colombi}, {Colombo}, {Couchot}, {Coulais}, {Crill}, {Curto}, {Cuttaia}, {Danese}, {Davies}, {Davis}, {de Bernardis}, {de Rosa}, {de Zotti}, {Delabrouille}, {Delouis}, {D{\'e}sert}, {Dickinson}, {Diego}, {Dolag}, {Dole}, {Donzelli}, {Dor{\'e}}, {Douspis}, {Dunkley}, {Dupac}, {Efstathiou}, {Elsner}, {En{\ss}lin}, {Eriksen}, {Finelli}, {Forni}, {Frailis}, {Fraisse}, {Franceschi}, {Gaier}, {Galeotta}, {Galli}, {Ganga}, {Giard}, {Giardino}, {Giraud-H{\'e}raud},
  {Gjerl{\o}w}, {Gonz{\'a}lez-Nuevo}, {G{\'o}rski}, {Gratton}, {Gregorio}, {Gruppuso}, {Gudmundsson}, {Haissinski}, {Hamann}, {Hansen}, {Hanson}, {Harrison}, {Henrot-Versill{\'e}}, {Hern{\'a}ndez-Monteagudo}, {Herranz}, {Hildebrandt}, {Hivon}, {Hobson}, {Holmes}, {Hornstrup}, {Hou}, {Hovest}, {Huffenberger}, {Jaffe}, {Jaffe}, {Jewell}, {Jones}, {Juvela}, {Keih{\"a}nen}, {Keskitalo}, {Kisner}, {Kneissl}, {Knoche}, {Knox}, {Kunz}, {Kurki-Suonio}, {Lagache}, {L{\"a}hteenm{\"a}ki}, {Lamarre}, {Lasenby}, {Lattanzi}, {Laureijs}, {Lawrence}, {Leach}, {Leahy}, {Leonardi}, {Le{\'o}n-Tavares}, {Lesgourgues}, {Lewis}, {Liguori}, {Lilje}, {Linden-V{\o}rnle}, {L{\'o}pez-Caniego}, {Lubin}, {Mac{\'\i}as-P{\'e}rez}, {Maffei}, {Maino}, {Mandolesi}, {Maris}, {Marshall}, {Martin}, {Mart{\'\i}nez-Gonz{\'a}lez}, {Masi}, {Massardi}, {Matarrese}, {Matthai}, {Mazzotta}, {Meinhold}, {Melchiorri}, {Melin}, {Mendes}, {Menegoni}, {Mennella}, {Migliaccio}, {Millea}, {Mitra}, {Miville-Desch{\^e}nes}, {Moneti}, {Montier}, {Morgante},
  {Mortlock}, {Moss}, {Munshi}, {Murphy}, {Naselsky}, {Nati}, {Natoli}, {Netterfield}, {N{\o}rgaard-Nielsen}, {Noviello}, {Novikov}, {Novikov}, {O'Dwyer}, {Osborne}, {Oxborrow}, {Paci}, {Pagano}, {Pajot}, {Paladini}, {Paoletti}, {Partridge}, {Pasian}, {Patanchon}, {Pearson}, {Pearson}, {Peiris}, {Perdereau}, {Perotto}, {Perrotta}, {Pettorino}, {Piacentini}, {Piat}, {Pierpaoli}, {Pietrobon}, {Plaszczynski}, {Platania}, {Pointecouteau}, {Polenta}, {Ponthieu}, {Popa}, {Poutanen}, {Pratt}, {Pr{\'e}zeau}, {Prunet}, {Puget}, {Rachen}, {Reach}, {Rebolo}, {Reinecke}, {Remazeilles}, {Renault}, {Ricciardi}, {Riller}, {Ristorcelli}, {Rocha}, {Rosset}, {Roudier}, {Rowan-Robinson}, {Rubi{\~n}o-Mart{\'\i}n}, {Rusholme}, {Sandri}, {Santos}, {Savelainen}, {Savini}, {Scott}, {Seiffert}, {Shellard}, {Spencer}, {Starck}, {Stolyarov}, {Stompor}, {Sudiwala}, {Sunyaev}, {Sureau}, {Sutton}, {Suur-Uski}, {Sygnet}, {Tauber}, {Tavagnacco}, {Terenzi}, {Toffolatti}, {Tomasi}, {Tristram}, {Tucci}, {Tuovinen}, {T{\"u}rler}, {Umana},
  {Valenziano}, {Valiviita}, {Van Tent}, {Vielva}, {Villa}, {Vittorio}, {Wade}, {Wandelt}, {Wehus}, {White}, {White}, {Wilkinson}, {Yvon}, {Zacchei}, \& {Zonca}}]{Planck2014}
{Planck Collaboration}, {Ade}, P.~A.~R., {Aghanim}, N., {et~al.} 2014, \aap, 571, A16

\bibitem[{{Planck Collaboration} {et~al.}(2016){Planck Collaboration}, {Ade}, {Aghanim}, {Arnaud}, {Ashdown}, {Aumont}, {Baccigalupi}, {Banday}, {Barreiro}, {Bartlett}, {Bartolo}, {Battaner}, {Battye}, {Benabed}, {Beno{\^\i}t}, {Benoit-L{\'e}vy}, {Bernard}, {Bersanelli}, {Bielewicz}, {Bock}, {Bonaldi}, {Bonavera}, {Bond}, {Borrill}, {Bouchet}, {Boulanger}, {Bucher}, {Burigana}, {Butler}, {Calabrese}, {Cardoso}, {Catalano}, {Challinor}, {Chamballu}, {Chary}, {Chiang}, {Chluba}, {Christensen}, {Church}, {Clements}, {Colombi}, {Colombo}, {Combet}, {Coulais}, {Crill}, {Curto}, {Cuttaia}, {Danese}, {Davies}, {Davis}, {de Bernardis}, {de Rosa}, {de Zotti}, {Delabrouille}, {D{\'e}sert}, {Di Valentino}, {Dickinson}, {Diego}, {Dolag}, {Dole}, {Donzelli}, {Dor{\'e}}, {Douspis}, {Ducout}, {Dunkley}, {Dupac}, {Efstathiou}, {Elsner}, {En{\ss}lin}, {Eriksen}, {Farhang}, {Fergusson}, {Finelli}, {Forni}, {Frailis}, {Fraisse}, {Franceschi}, {Frejsel}, {Galeotta}, {Galli}, {Ganga}, {Gauthier}, {Gerbino}, {Ghosh}, {Giard},
  {Giraud-H{\'e}raud}, {Giusarma}, {Gjerl{\o}w}, {Gonz{\'a}lez-Nuevo}, {G{\'o}rski}, {Gratton}, {Gregorio}, {Gruppuso}, {Gudmundsson}, {Hamann}, {Hansen}, {Hanson}, {Harrison}, {Helou}, {Henrot-Versill{\'e}}, {Hern{\'a}ndez-Monteagudo}, {Herranz}, {Hildebrandt}, {Hivon}, {Hobson}, {Holmes}, {Hornstrup}, {Hovest}, {Huang}, {Huffenberger}, {Hurier}, {Jaffe}, {Jaffe}, {Jones}, {Juvela}, {Keih{\"a}nen}, {Keskitalo}, {Kisner}, {Kneissl}, {Knoche}, {Knox}, {Kunz}, {Kurki-Suonio}, {Lagache}, {L{\"a}hteenm{\"a}ki}, {Lamarre}, {Lasenby}, {Lattanzi}, {Lawrence}, {Leahy}, {Leonardi}, {Lesgourgues}, {Levrier}, {Lewis}, {Liguori}, {Lilje}, {Linden-V{\o}rnle}, {L{\'o}pez-Caniego}, {Lubin}, {Mac{\'\i}as-P{\'e}rez}, {Maggio}, {Maino}, {Mandolesi}, {Mangilli}, {Marchini}, {Maris}, {Martin}, {Martinelli}, {Mart{\'\i}nez-Gonz{\'a}lez}, {Masi}, {Matarrese}, {McGehee}, {Meinhold}, {Melchiorri}, {Melin}, {Mendes}, {Mennella}, {Migliaccio}, {Millea}, {Mitra}, {Miville-Desch{\^e}nes}, {Moneti}, {Montier}, {Morgante}, {Mortlock},
  {Moss}, {Munshi}, {Murphy}, {Naselsky}, {Nati}, {Natoli}, {Netterfield}, {N{\o}rgaard-Nielsen}, {Noviello}, {Novikov}, {Novikov}, {Oxborrow}, {Paci}, {Pagano}, {Pajot}, {Paladini}, {Paoletti}, {Partridge}, {Pasian}, {Patanchon}, {Pearson}, {Perdereau}, {Perotto}, {Perrotta}, {Pettorino}, {Piacentini}, {Piat}, {Pierpaoli}, {Pietrobon}, {Plaszczynski}, {Pointecouteau}, {Polenta}, {Popa}, {Pratt}, {Pr{\'e}zeau}, {Prunet}, {Puget}, {Rachen}, {Reach}, {Rebolo}, {Reinecke}, {Remazeilles}, {Renault}, {Renzi}, {Ristorcelli}, {Rocha}, {Rosset}, {Rossetti}, {Roudier}, {Rouill{\'e} d'Orfeuil}, {Rowan-Robinson}, {Rubi{\~n}o-Mart{\'\i}n}, {Rusholme}, {Said}, {Salvatelli}, {Salvati}, {Sandri}, {Santos}, {Savelainen}, {Savini}, {Scott}, {Seiffert}, {Serra}, {Shellard}, {Spencer}, {Spinelli}, {Stolyarov}, {Stompor}, {Sudiwala}, {Sunyaev}, {Sutton}, {Suur-Uski}, {Sygnet}, {Tauber}, {Terenzi}, {Toffolatti}, {Tomasi}, {Tristram}, {Trombetti}, {Tucci}, {Tuovinen}, {T{\"u}rler}, {Umana}, {Valenziano}, {Valiviita}, {Van Tent},
  {Vielva}, {Villa}, {Wade}, {Wandelt}, {Wehus}, {White}, {White}, {Wilkinson}, {Yvon}, {Zacchei}, \& {Zonca}}]{Planck2016}
{Planck Collaboration}, {Ade}, P.~A.~R., {Aghanim}, N., {et~al.} 2016, \aap, 594, A13

\bibitem[{{Popesso} {et~al.}(2024){Popesso}, {Marini}, {Dolag}, {Lamer}, {Csizi}, {Biffi}, {Robothan}, {Bravo}, {Biviano}, {Vladutesku-Zopp}, {Lovisari}, {Ettori}, {Angelinelli}, {Driver}, {Toptun}, {Dev}, {Mazengo}, {Merloni}, {Zhang}, {Comparat}, {Ponti}, {Mroczkowski}, \& {Bulbul}}]{popesso:2024}
{Popesso}, P., {Marini}, I., {Dolag}, K., {et~al.} 2024, arXiv e-prints, arXiv:2411.17120

\bibitem[{{Power} {et~al.}(2012){Power}, {Knebe}, \& {Knollmann}}]{power:2012}
{Power}, C., {Knebe}, A., \& {Knollmann}, S.~R. 2012, \mnras, 419, 1576

\bibitem[{{Puchwein} {et~al.}(2010){Puchwein}, {Springel}, {Sijacki}, \& {Dolag}}]{Puchwein10}
{Puchwein}, E., {Springel}, V., {Sijacki}, D., \& {Dolag}, K. 2010, \mnras, 406, 936

\bibitem[{{Ragagnin} {et~al.}(2017){Ragagnin}, {Dolag}, {Biffi}, {Cadolle Bel}, {Hammer}, {Krukau}, {Petkova}, \& {Steinborn}}]{ragagnin17}
{Ragagnin}, A., {Dolag}, K., {Biffi}, V., {et~al.} 2017, Astronomy and Computing, 20, 52

\bibitem[{{Ragusa} {et~al.}(2023){Ragusa}, {Iodice}, {Spavone}, {Montes}, {Forbes}, {Brough}, {Mirabile}, {Cantiello}, {Paolillo}, \& {Schipani}}]{ragusa2023}
{Ragusa}, R., {Iodice}, E., {Spavone}, M., {et~al.} 2023, \aap, 670, L20

\bibitem[{{Ragusa} {et~al.}(2022){Ragusa}, {Mirabile}, {Spavone}, {Cantiello}, {Iodice}, {La Marca}, {Paolillo}, \& {Schipani}}]{Ragusa2022FrASS...952810R}
{Ragusa}, R., {Mirabile}, M., {Spavone}, M., {et~al.} 2022, Frontiers in Astronomy and Space Sciences, 9, 852810

\bibitem[{{Ragusa} {et~al.}(2021){Ragusa}, {Spavone}, {Iodice}, {Brough}, {Raj}, {Paolillo}, {Cantiello}, {Forbes}, {La Marca}, {D'Ago}, {Rampazzo}, \& {Schipani}}]{RAGUSA2021}
{Ragusa}, R., {Spavone}, M., {Iodice}, E., {et~al.} 2021, A\&A, 651, A39

\bibitem[{{Raouf} {et~al.}(2019){Raouf}, {Smith}, {Khosroshahi}, {Dariush}, {Driver}, {Ko}, \& {Hwang}}]{raouf:2019}
{Raouf}, M., {Smith}, R., {Khosroshahi}, H.~G., {et~al.} 2019, \apj, 887, 264

\bibitem[{{Reefe} {et~al.}(2025){Reefe}, {McDonald}, {Chatzikos}, {Seebeck}, {Mushotzky}, {Veilleux}, {Allen}, {Bayliss}, {Calzadilla}, {Canning}, {Donahue}, {Floyd}, {Gaspari}, {Hlavacek-Larrondo}, {McNamara}, {Russell}, {Sarkar}, {Sharon}, \& {Somboonpanyakul}}]{reefe:2025}
{Reefe}, M., {McDonald}, M., {Chatzikos}, M., {et~al.} 2025, arXiv e-prints, arXiv:2501.08527

\bibitem[{{Remus} {et~al.}(2023){Remus}, {Dolag}, \& {Dannerbauer}}]{remus:2023}
{Remus}, R.-S., {Dolag}, K., \& {Dannerbauer}, H. 2023, \apj, 950, 191

\bibitem[{{Remus} {et~al.}(2017){Remus}, {Dolag}, \& {Hoffmann}}]{remus17}
{Remus}, R.-S., {Dolag}, K., \& {Hoffmann}, T. 2017, Galaxies, 5, 49

\bibitem[{{Remus} \& {Forbes}(2022{\natexlab{a}})}]{remus:2022}
{Remus}, R.-S. \& {Forbes}, D.~A. 2022{\natexlab{a}}, \apj, 935, 37

\bibitem[{{Remus} \& {Forbes}(2022{\natexlab{b}})}]{remus22}
{Remus}, R.-S. \& {Forbes}, D.~A. 2022{\natexlab{b}}, \apj, 935, 37

\bibitem[{{Rennehan} {et~al.}(2020){Rennehan}, {Babul}, {Hayward}, {Bottrell}, {Hani}, \& {Chapman}}]{rennehan:2020}
{Rennehan}, D., {Babul}, A., {Hayward}, C.~C., {et~al.} 2020, \mnras, 493, 4607

\bibitem[{{Roberts} {et~al.}(2018){Roberts}, {Parker}, \& {Hlavacek-Larrondo}}]{roberts18}
{Roberts}, I.~D., {Parker}, L.~C., \& {Hlavacek-Larrondo}, J. 2018, \mnras, 475, 4704

\bibitem[{{Rodriguez-Gomez} {et~al.}(2016){Rodriguez-Gomez}, {Pillepich}, {Sales}, {Genel}, {Vogelsberger}, {Zhu}, {Wellons}, {Nelson}, {Torrey}, {Springel}, {Ma}, \& {Hernquist}}]{rodriguez:2016}
{Rodriguez-Gomez}, V., {Pillepich}, A., {Sales}, L.~V., {et~al.} 2016, \mnras, 458, 2371

\bibitem[{{Rudick} {et~al.}(2009){Rudick}, {Mihos}, {Frey}, \& {McBride}}]{Rudick2009}
{Rudick}, C.~S., {Mihos}, J.~C., {Frey}, L.~H., \& {McBride}, C.~K. 2009, \apj, 699, 1518

\bibitem[{{Rudick} {et~al.}(2006){Rudick}, {Mihos}, \& {McBride}}]{Rudick06}
{Rudick}, C.~S., {Mihos}, J.~C., \& {McBride}, C. 2006, \apj, 648, 936

\bibitem[{{Rudick} {et~al.}(2011){Rudick}, {Mihos}, \& {McBride}}]{Rudick2011}
{Rudick}, C.~S., {Mihos}, J.~C., \& {McBride}, C.~K. 2011, \apj, 732, 48

\bibitem[{{Sanders} {et~al.}(2013){Sanders}, {Fabian}, {Churazov}, {Schekochihin}, {Simionescu}, {Walker}, \& {Werner}}]{sanders:2013}
{Sanders}, J.~S., {Fabian}, A.~C., {Churazov}, E., {et~al.} 2013, Science, 341, 1365

\bibitem[{{Santucci} {et~al.}(2020){Santucci}, {Brough}, {Scott}, {Montes}, {Owers}, {van Sande}, {Bland-Hawthorn}, {Bryant}, {Croom}, {Ferreras}, {Lawrence}, {L{\'o}pez-S{\'a}nchez}, \& {Richards}}]{santucci:2020}
{Santucci}, G., {Brough}, S., {Scott}, N., {et~al.} 2020, \apj, 896, 75

\bibitem[{{Sartoris} {et~al.}(2020){Sartoris}, {Biviano}, {Rosati}, {Mercurio}, {Grillo}, {Ettori}, {Nonino}, {Umetsu}, {Bergamini}, {Caminha}, \& {Girardi}}]{sartoris:2020}
{Sartoris}, B., {Biviano}, A., {Rosati}, P., {et~al.} 2020, \aap, 637, A34

\bibitem[{{Schaller} {et~al.}(2015){Schaller}, {Frenk}, {Bower}, {Theuns}, {Jenkins}, {Schaye}, {Crain}, {Furlong}, {Dalla Vecchia}, \& {McCarthy}}]{schaller:2015}
{Schaller}, M., {Frenk}, C.~S., {Bower}, R.~G., {et~al.} 2015, \mnras, 451, 1247

\bibitem[{{Schaye} {et~al.}(2015){Schaye}, {Crain}, {Bower}, {Furlong}, {Schaller}, {Theuns}, {Dalla Vecchia}, {Frenk}, {McCarthy}, {Helly}, {Jenkins}, {Rosas-Guevara}, {White}, {Baes}, {Booth}, {Camps}, {Navarro}, {Qu}, {Rahmati}, {Sawala}, {Thomas}, \& {Trayford}}]{Schaye.2015}
{Schaye}, J., {Crain}, R.~A., {Bower}, R.~G., {et~al.} 2015, \mnras, 446, 521

\bibitem[{{Schaye} {et~al.}(2023){Schaye}, {Kugel}, {Schaller}, {Helly}, {Braspenning}, {Elbers}, {McCarthy}, {van Daalen}, {Vandenbroucke}, {Frenk}, {Kwan}, {Salcido}, {Bah{\'e}}, {Borrow}, {Chaikin}, {Hahn}, {Hu{\v{s}}ko}, {Jenkins}, {Lacey}, \& {Nobels}}]{schaye:2023}
{Schaye}, J., {Kugel}, R., {Schaller}, M., {et~al.} 2023, \mnras, 526, 4978

\bibitem[{{Schwinn} {et~al.}(2018){Schwinn}, {Baugh}, {Jauzac}, {Bartelmann}, \& {Eckert}}]{schwinn18}
{Schwinn}, J., {Baugh}, C.~M., {Jauzac}, M., {Bartelmann}, M., \& {Eckert}, D. 2018, \mnras, 481, 4300

\bibitem[{{Seidel} {et~al.}(2024){Seidel}, {Dolag}, {Remus}, {Sorce}, {Hern{\'a}ndez-Mart{\'\i}nez}, {Khabibullin}, \& {Aghanim}}]{seidel:2024}
{Seidel}, B.~A., {Dolag}, K., {Remus}, R.~S., {et~al.} 2024, arXiv e-prints, arXiv:2412.08708

\bibitem[{{Shaw} {et~al.}(2006){Shaw}, {Weller}, {Ostriker}, \& {Bode}}]{shaw06}
{Shaw}, L.~D., {Weller}, J., {Ostriker}, J.~P., \& {Bode}, P. 2006, \apj, 646, 815

\bibitem[{{Sorce} {et~al.}(2020){Sorce}, {Gottl{\"o}ber}, \& {Yepes}}]{sorce:2020}
{Sorce}, J.~G., {Gottl{\"o}ber}, S., \& {Yepes}, G. 2020, \mnras, 496, 5139

\bibitem[{{Spavone} {et~al.}(2017){Spavone}, {Capaccioli}, {Napolitano}, {Iodice}, {Grado}, {Limatola}, {Cooper}, {Cantiello}, {Forbes}, {Paolillo}, \& {Schipani}}]{Spavone2017b}
{Spavone}, M., {Capaccioli}, M., {Napolitano}, N.~R., {et~al.} 2017, A\&A, 603, A38

\bibitem[{{Spavone} {et~al.}(2024){Spavone}, {Iodice}, {Lohmann}, {Arnaboldi}, {Hilker}, {La Marca}, {Calvi}, {Cantiello}, {Corsini}, {D'Ago}, {Forbes}, {Mirabile}, \& {Rejkuba}}]{spavone:2024}
{Spavone}, M., {Iodice}, E., {Lohmann}, F.~S., {et~al.} 2024, \aap, 689, A306

\bibitem[{Spavone {et~al.}(2020)Spavone, Iodice, van~de Ven, Falc{\'o}n-Barroso, Raj, Hilker, Peletier, Capaccioli, Mieske, Venhola, Napolitano, Cantiello, Paolillo, \& Schipani}]{Spavone2020}
Spavone, M., Iodice, E., van~de Ven, G., {et~al.} 2020, \aap, 639, A14

\bibitem[{{Springel}(2005)}]{springel05}
{Springel}, V. 2005, \mnras, 364, 1105

\bibitem[{{Springel}(2010)}]{Springel2010}
{Springel}, V. 2010, \mnras, 401, 791

\bibitem[{{Springel} {et~al.}(2018){Springel}, {Pakmor}, {Pillepich}, {Weinberger}, {Nelson}, {Hernquist}, {Vogelsberger}, {Genel}, {Torrey}, {Marinacci}, \& {Naiman}}]{Springel2018}
{Springel}, V., {Pakmor}, R., {Pillepich}, A., {et~al.} 2018, \mnras, 475, 676

\bibitem[{{Springel} {et~al.}(2001){Springel}, {Yoshida}, \& {White}}]{Springel_et_al_2001}
{Springel}, V., {Yoshida}, N., \& {White}, S. D.~M. 2001, \na, 6, 79

\bibitem[{{Stevens} {et~al.}(2019){Stevens}, {Diemer}, {Lagos}, {Nelson}, {Pillepich}, {Brown}, {Catinella}, {Hernquist}, {Weinberger}, {Vogelsberger}, \& {Marinacci}}]{stevens:2018}
{Stevens}, A. R.~H., {Diemer}, B., {Lagos}, C. d.~P., {et~al.} 2019, \mnras, 483, 5334

\bibitem[{{Swindle} {et~al.}(2011){Swindle}, {Gal}, {La Barbera}, \& {de Carvalho}}]{swindle:2011}
{Swindle}, R., {Gal}, R.~R., {La Barbera}, F., \& {de Carvalho}, R.~R. 2011, \aj, 142, 118

\bibitem[{{Teklu} {et~al.}(2015){Teklu}, {Remus}, {Dolag}, {Beck}, {Burkert}, {Schmidt}, {Schulze}, \& {Steinborn}}]{teklu15}
{Teklu}, A.~F., {Remus}, R.-S., {Dolag}, K., {et~al.} 2015, \apj, 812, 29

\bibitem[{{Teyssier}(2002)}]{Teyssier.2002}
{Teyssier}, R. 2002, \aap, 385, 337

\bibitem[{{Tremaine} \& {Richstone}(1977)}]{tremaine:1977}
{Tremaine}, S.~D. \& {Richstone}, D.~O. 1977, \apj, 212, 311

\bibitem[{{Tremblay} {et~al.}(2015){Tremblay}, {O'Dea}, {Baum}, {Mittal}, {McDonald}, {Combes}, {Li}, {McNamara}, {Bremer}, {Clarke}, {Donahue}, {Edge}, {Fabian}, {Hamer}, {Hogan}, {Oonk}, {Quillen}, {Sanders}, {Salom{\'e}}, \& {Voit}}]{tremblay:2015}
{Tremblay}, G.~R., {O'Dea}, C.~P., {Baum}, S.~A., {et~al.} 2015, \mnras, 451, 3768

\bibitem[{{Tremmel} {et~al.}(2019){Tremmel}, {Quinn}, {Ricarte}, {Babul}, {Chadayammuri}, {Natarajan}, {Nagai}, {Pontzen}, \& {Volonteri}}]{tremmel:2019}
{Tremmel}, M., {Quinn}, T.~R., {Ricarte}, A., {et~al.} 2019, \mnras, 483, 3336

\bibitem[{{Turner} {et~al.}(2025){Turner}, {Tacchella}, {D'Eugenio}, {Carniani}, {Curti}, {Glazebrook}, {Johnson}, {Lim}, {Looser}, {Maiolino}, {Nanayakkara}, \& {Wan}}]{turner:2025}
{Turner}, C., {Tacchella}, S., {D'Eugenio}, F., {et~al.} 2025, \mnras, 537, 1826

\bibitem[{{Tweed} {et~al.}(2009){Tweed}, {Devriendt}, {Blaizot}, {Colombi}, \& {Slyz}}]{Tweed2009}
{Tweed}, D., {Devriendt}, J., {Blaizot}, J., {Colombi}, S., \& {Slyz}, A. 2009, \aap, 506, 647

\bibitem[{{Valentino} {et~al.}(2023){Valentino}, {Brammer}, {Gould}, {Kokorev}, {Fujimoto}, {Jespersen}, {Vijayan}, {Weaver}, {Ito}, {Tanaka}, {Ilbert}, {Magdis}, {Whitaker}, {Faisst}, {Gallazzi}, {Gillman}, {Gim{\'e}nez-Arteaga}, {G{\'o}mez-Guijarro}, {Kubo}, {Heintz}, {Hirschmann}, {Oesch}, {Onodera}, {Rizzo}, {Lee}, {Strait}, \& {Toft}}]{valentino:2023}
{Valentino}, F., {Brammer}, G., {Gould}, K. M.~L., {et~al.} 2023, \apj, 947, 20

\bibitem[{{V{\'e}liz Astudillo} {et~al.}(2024){V{\'e}liz Astudillo}, {Carrasco}, {Nilo Castell{\'o}n}, {Zenteno}, \& {Cuevas}}]{veliz24}
{V{\'e}liz Astudillo}, S., {Carrasco}, E.~R., {Nilo Castell{\'o}n}, J.~L., {Zenteno}, A., \& {Cuevas}, H. 2024, arXiv e-prints, arXiv:2408.02519

\bibitem[{{Yoo} {et~al.}(2024){Yoo}, {Park}, {Sabiu}, {Singh}, {Ko}, {Lee}, {Pichon}, {Jee}, {Gibson}, {Snaith}, {Kim}, {Shin}, {Kim}, \& {Kim}}]{yoo:2024}
{Yoo}, J., {Park}, C., {Sabiu}, C.~G., {et~al.} 2024, \apj, 965, 145

\bibitem[{{Zenteno} {et~al.}(2020){Zenteno}, {Hern{\'a}ndez-Lang}, {Klein}, {Vergara Cervantes}, {Hollowood}, {Bhargava}, {Palmese}, {Strazzullo}, {Romer}, {Mohr}, {Jeltema}, {Saro}, {Lidman}, {Gruen}, {Ojeda}, {Katzenberger}, {Aguena}, {Allam}, {Avila}, {Bayliss}, {Bertin}, {Brooks}, {Buckley-Geer}, {Burke}, {Capasso}, {Carnero Rosell}, {Carrasco Kind}, {Carretero}, {Castander}, {Costanzi}, {da Costa}, {De Vicente}, {Desai}, {Diehl}, {Doel}, {Eifler}, {Evrard}, {Flaugher}, {Floyd}, {Fosalba}, {Frieman}, {Garc{\'\i}a-Bellido}, {Gerdes}, {Gonzalez}, {Gruendl}, {Gschwend}, {Gutierrez}, {Hartley}, {Hinton}, {Honscheid}, {James}, {Kuehn}, {Lahav}, {Lima}, {McDonald}, {Maia}, {March}, {Melchior}, {Menanteau}, {Miquel}, {Ogando}, {Paz-Chinch{\'o}n}, {Plazas}, {Roodman}, {Rykoff}, {Sanchez}, {Scarpine}, {Schubnell}, {Serrano}, {Sevilla-Noarbe}, {Smith}, {Soares-Santos}, {Suchyta}, {Swanson}, {Tarle}, {Thomas}, {Varga}, {Walker}, {Wilkinson}, \& {DES Collaboration}}]{zenteno20}
{Zenteno}, A., {Hern{\'a}ndez-Lang}, D., {Klein}, M., {et~al.} 2020, \mnras, 495, 705

\bibitem[{{Zhang} {et~al.}(2024){Zhang}, {Golden-Marx}, {Ogando}, {Yanny}, {Rykoff}, {Allam}, {Aguena}, {Bacon}, {Bocquet}, {Brooks}, {Carnero Rosell}, {Carretero}, {Cheng}, {Conselice}, {Costanzi}, {da Costa}, {Pereira}, {Davis}, {Desai}, {Diehl}, {Doel}, {Ferrero}, {Flaugher}, {Frieman}, {Gruen}, {Gruendl}, {Hinton}, {Hollowood}, {Honscheid}, {James}, {Jeltema}, {Kuehn}, {Kuropatkin}, {Lahav}, {Lee}, {Lima}, {Mena-Fern{\'a}ndez}, {Miquel}, {Palmese}, {Pieres}, {Plazas Malag{\'o}n}, {Romer}, {Sanchez}, {Smith}, {Suchyta}, {Tarle}, {To}, {Tucker}, {Weaverdyck}, \& {DES Collaboration}}]{zhang:2024}
{Zhang}, Y., {Golden-Marx}, J.~B., {Ogando}, R. L.~C., {et~al.} 2024, \mnras, 531, 510

\bibitem[{{Zibetti} {et~al.}(2005){Zibetti}, {White}, {Schneider}, \& {Brinkmann}}]{Zibetti2005}
{Zibetti}, S., {White}, S.~D.~M., {Schneider}, D.~P., \& {Brinkmann}, J. 2005, \mnras, 358, 949

\end{thebibliography}





\begin{appendix}
\label{app}

\section{Partial Correlations and Fitting}
\label{app:fitting}

\begin{figure}
    \includegraphics[width=\columnwidth]{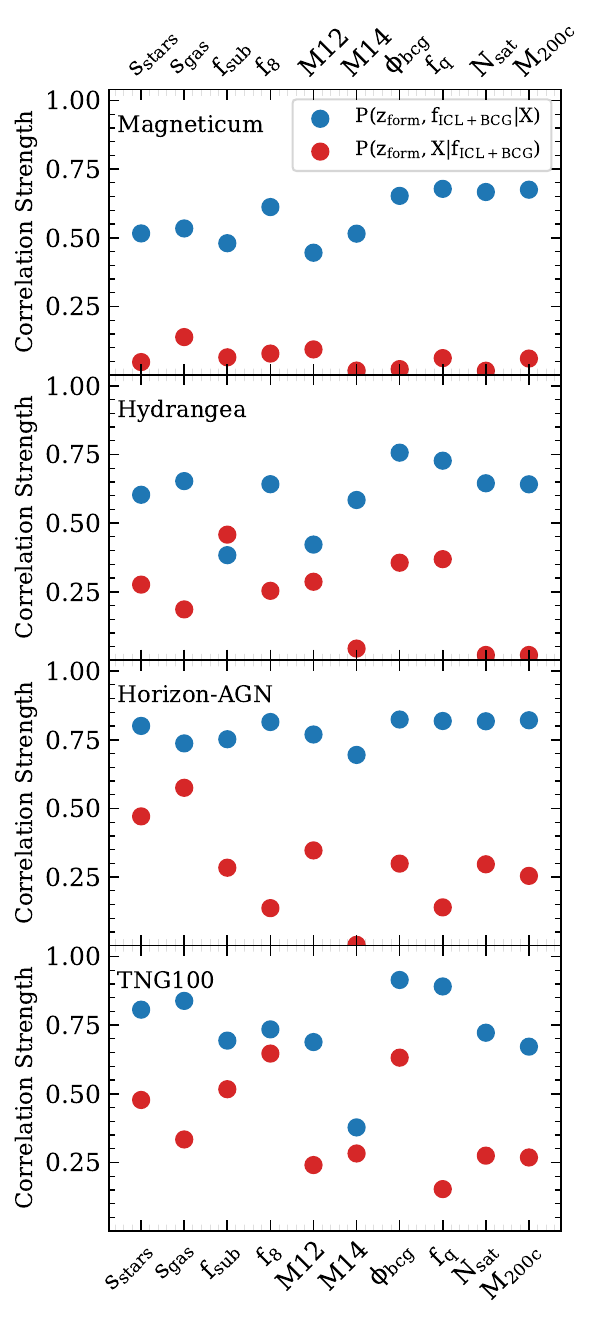}
    \caption{Partial correlation of $z_\mathrm{form}$ with $f_\mathrm{ICL+BCG}$, accounting for other parameters X (blue), or between $z_\mathrm{form}$ and X, accounting for $f_\mathrm{ICL+BCG}$ (red), for all four simulations (panels).
    }
    \label{fig:parcor}
\end{figure}

Given that in Fig.~\ref{fig:dyncor} we find $f_\mathrm{ICL+BCG}$ to tightly trace with the cluster formation redshift $z_\mathrm{form}$, it is of relevance if this correlation is driven by some other parameter that provides the link. To test the true driver of the relation to $z_\mathrm{form}$, we calculate the partial correlation to $f_\mathrm{ICL+BCG}$ while accounting for all other parameters considered in this study individually. The partial correlation between two parameters x and y, accounting for z, is determined by fitting two linear regressions: between x and z as well as y and z. The residuals in x and y of these fits are thus corrected for any possible correlation with z, and the Pearson correlation coefficient of these residuals is defined as the partial correlation coefficient, which we plot in Fig.~\ref{fig:parcor}. In all cases and for all simulations, the correlation between $z_\mathrm{form}$ and $f_\mathrm{ICL+BCG}$ remains significantly in place. Indeed, aside from $f_\mathrm{sub}$ for Hydrangea, the correlation of any parameter (from magnitude gap to center shift to halo mass) with $z_\mathrm{form}$ is less significant once accounting for that parameters correlation with $f_\mathrm{ICL+BCG}$. While it is possible that there is another underlying driver which best determines a galaxy cluster's formation redshift, of all the parameters discussed in this study, $f_\mathrm{ICL+BCG}$ is the strongest.

\begin{figure*}
    \includegraphics[width=\textwidth]{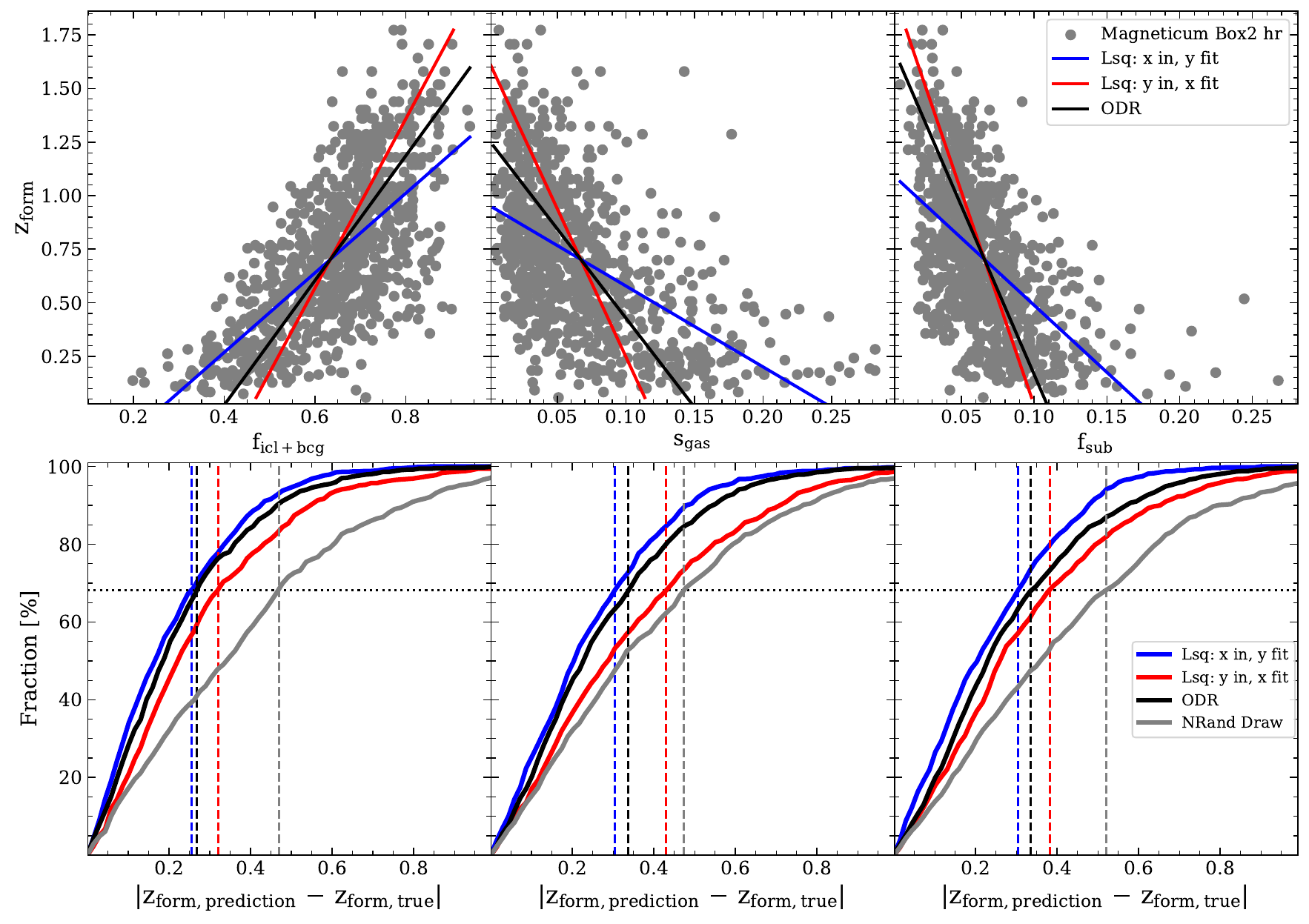}
    \caption{\textit{Top row:} The Magneticum data as in Fig.~\ref{fig:dyncor}, with the axes swapped. Solid lines show the best fits using either the least-squares method, fitting toward x (red) or y (blue), as well as the best fit using orthogonal distance regression (black). 
    \textit{Bottom row:} The cumulative histogram of the absolute offset between the predicted versus actual formation redshifts for the three best fit lines, as well as when drawing randomly from a normal distribution of formation redshifts with the mean and standard deviation calculated from the total sample of clusters (gray). The $1\sigma$-bounds ($68.27\%$) are shown as colored vertical lines. 
    }
    \label{fig:diff_fit}
\end{figure*}

This correlation allows us to define a fitting function, using $f_\mathrm{ICL+BCG}$ to predict $z_\mathrm{form}$, as discussed for Fig.~\ref{fig:dyncor}. In the top row of Fig.~\ref{fig:diff_fit} we show the best fit lines between $z_\mathrm{form}$ and $f_\mathrm{ICL+BCG}$, as well as to $\mathrm{s}_\mathrm{gas}$ and $f_\mathrm{sub}$, by using either the least-squares method, reducing the errors in either the x (red) or y-direction (blue), as well as using orthogonal distance regression (black, ODR), where both axes are weighted equally over the full range of their data. Using these best fit lines, we calculate the predicted formation redshift $z_\mathrm{form,prediction}$ and plot the cumulative distribution in absolute offset between the prediction and true $z_\mathrm{form}$ in the bottom row of Fig.~\ref{fig:diff_fit}. The average 1$\sigma$-error is given by the vertical colored lines. 

We find that the least-squares fit in the direction of x to y produces the overall tightest predictions, as expected because it minimizes precisely this error. ODR performs nearly as well in all cases, and additionally accounts for errors in the x-direction. As such, we chose ODR when calculating the best fit parameters defined in Table~\ref{tab:sims3}. We note that of all three dynamical tracers, $f_\mathrm{ICL+BCG}$ performs the best in predicting $z_\mathrm{form}$, with a 1$\sigma$-error of around $\pm0.25$. While this is not insignificant, it is still remarkable that the time when a galaxy cluster had first assembled half of its present day total mass is imprinted into the current stellar distribution of mass to the point of being able to predict that time with just a $30$\%~error on average (given the median $z_\mathrm{form}=0.67$). The greatest deviations occur for galaxy clusters which assembled very recently (low $z_\mathrm{form}$) but which already have merged most of that stellar mass into the BCG or ICL (high $f_\mathrm{ICL+BCG}$). For these cases, we find that it can be useful to consider other dynamical tracers as well, as often the center shift in the gas or stars still shows that there was a recent merger, even if the overall stars have already become bound to the central halo.

\section{Definitions for Center Shift and BCG vs ICL}
\label{app:center}

\begin{figure*}
    \includegraphics[width=\textwidth]{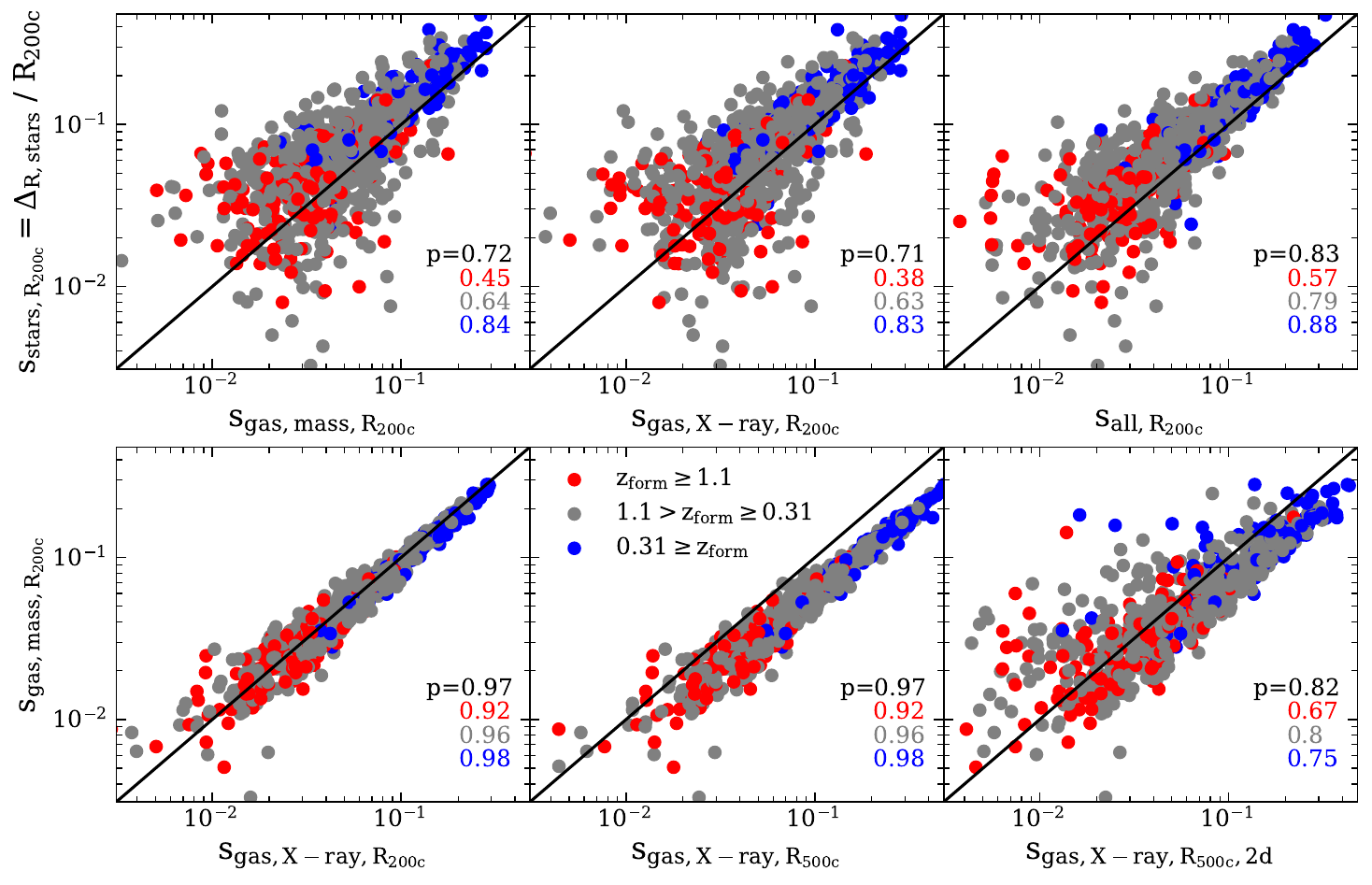}
    \caption{Comparison of various center shift definitions for the Magneticum galaxy clusters. Coloring indicates whether the galaxy cluster has a formation redshift in the upper (red), middle (gray), or lower third (blue) of the total sample, as given in the legend.
    }
    \label{fig:diff_cent}
\end{figure*}

There are different options when defining a center shift, the offset between what represents the deepest point of the potential, traced either via the BCG or peak in the X-ray surface brightness, and the larger scale barycenter, either of the stellar mass, light, gas mass or X-ray luminosity. We chose for our comparison the BCG as the first point, and then determined the offset relative to the three-dimensional barycenter of the stellar (s$_\mathrm{stars}$) or gas mass (s$_\mathrm{gas}$). Here we consider the impact of alternative definitions, plotted in Fig.~\ref{fig:diff_cent}. To distinguish the different center shifts, we add additional quantifiers in particular to the gas, such as the radius considered ($R_\mathrm{200c}$ or $R_\mathrm{500c}$), whether we take the barycenter of the mass or X-ray luminosity, and finally if we determine it in projection (2d). We calculate the X-ray luminosity of individual gas particles via their mass~M and temperature~T, and approximate the electron density by assuming hydrogen and helium to be fully ionized ($n_\mathrm{e} = n_\mathrm{e,H} + n_\mathrm{e,He}$), to get the luminosity in the range of $E_1=0.5$~to~$E_2=2$~keV via:
\begin{equation*}
    L \propto n_\mathrm{e}\cdot M \cdot \sqrt{T} \cdot\left(\exp\left(\frac{-E_2}{k_B\cdot T}\right)-\exp\left(\frac{-E_2}{k_B\cdot T}\right)\right).
\end{equation*}

The top row of Fig.~\ref{fig:diff_cent} shows the stellar center shift s$_\mathrm{stars}$ as a function of the gas mass center shift s$_\mathrm{gas,m,R200c}$, the X-ray luminosity s$_\mathrm{gas,X-ray,r200c}$, and the total barycenter s$_\mathrm{all,R200c}$. We find the agreement between the center shifts to be very tight, with Pearson correlation coefficients in excess of $p=0.7$. The stellar center shift apparently best correlated with the total mass center shift, which is likely because the mass is dominated by dark matter which, like the stellar particles, is dispersionless. Additionally, we color the points by the galaxy cluster formation redshift, with the upper third of the earliest forming clusters in red, the middle third in gray, and finally those clusters most recently formed in blue (with the formation redshift cuts used given in the legend). The colored values are the Pearson correlation coefficients for their respective third, where we find that the agreement among the different types of center shift is always by far tightest for the most recently formed clusters. We attribute this to the fact that differences in the smaller center shift of a very relaxed, early formed cluster are inherently more spurious.

The second row of Fig.~\ref{fig:diff_cent} plots the gas mass center shift s$_\mathrm{gas,m,r200c}$ versus the X-ray center shift, determined in the same $r_\mathrm{200c}$ (left, s$_\mathrm{gas,X-ray,r200c}$), or within $r_\mathrm{500c}$ either in three-dimensions (center, s$_\mathrm{gas,X-ray,r500c}$) or projected along the z-axis (left, s$_\mathrm{gas,X-ray,r500c,2d}$). We find the gas mass center shift to trace that of the X-ray luminosity remarkably well, even when considering regions farther in or in projection. This is surprisingly true even for the earliest forming clusters (red). 

\begin{figure*}
    \includegraphics[width=\textwidth]{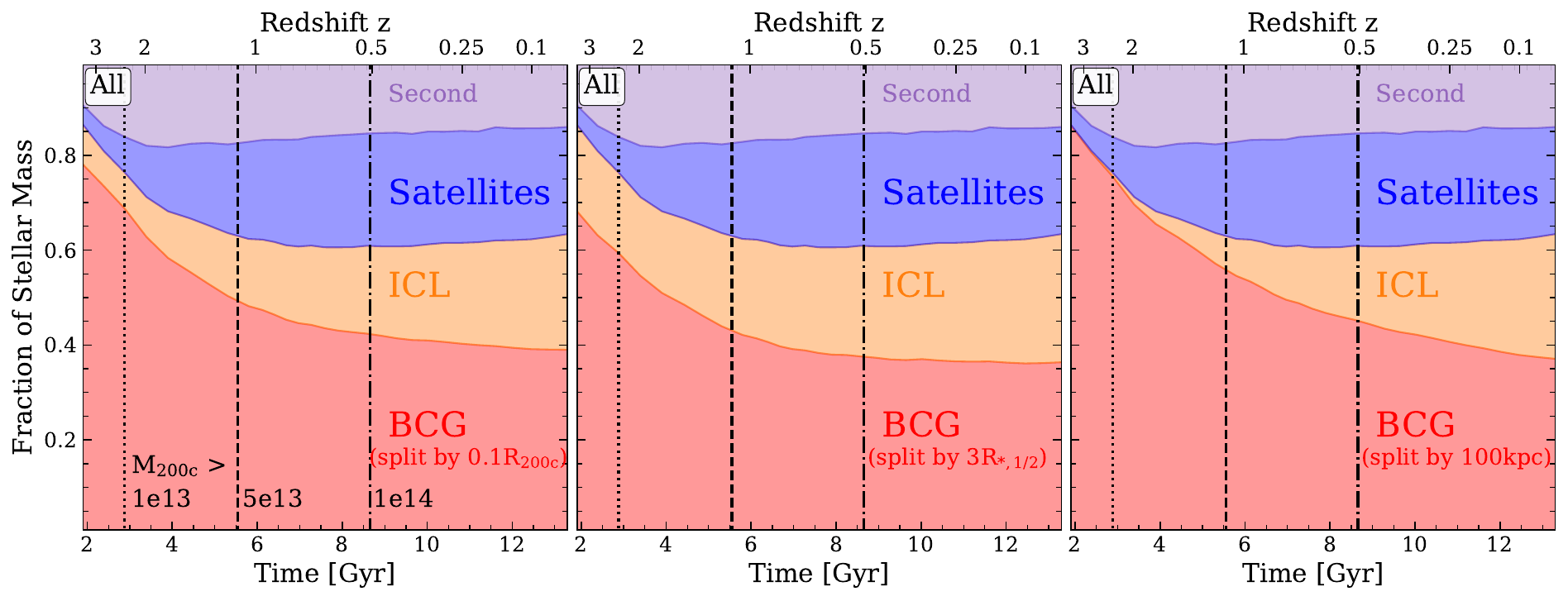}
    \caption{Same as Fig.~\ref{fig:evo} for the Magneticum galaxy clusters using three different definitions of the split between ICL and BCG. \textit{Left panel:} Same as the left panel of Fig.~\ref{fig:evo}, with a split between BCG and ICL at $0.1\cdot R_\mathrm{200,cri}$.
    \textit{Central panel:} BCG and ICL split at $3\cdot R_\mathrm{*,1/2}$, with $R_\mathrm{*,1/2}$ the stellar half mass radius.
    \textit{Right panel:} BCG and ICL split at a fixed aperture of 100kpc.
    }
    \label{fig:diff_def}
\end{figure*}

Finally, we also consider differing definitions of where to split the BCG and ICL components used to trace their evolution in Fig.~\ref{fig:diff_def}. We show three different definitions for direct comparison. In the left panel, $r_\mathrm{cut}=0.1\cdot R_\mathrm{200c}$ is used as in Fig.~\ref{fig:evo}, which cuts based on the size of the overall halo. The middle panel separates the BCG and the ICL based on the distribution of the central galaxy stars, namely $r_\mathrm{cut}=3\cdot R_\mathrm{*,1/2}$, where $R_\mathrm{*,1/2}$ is the radius which contains 50\% of the central stellar component within $10\%R_\mathrm{vir}$. This is particularly interesting when comparing to recent observations by \citet{joo:2023_icl} who report little evolution in the ICL fraction with redshift. They find $f_\mathrm{ICL}\approx14\%$ already at $z>1$, which matches the values we find for the split at $r_\mathrm{cut}=3\cdot R_\mathrm{*,1/2}$ at that redshift. This raises the question of whether the radial luminosity profiles at higher redshifts result in a split radius between BCG and ICL that moves increasingly inwards, a process which may be driven by increasingly centrally compact objects at these redshifts \citep{kimmig:2023q,ito:2024}, consequently resulting in a flat evolution in $f_\mathrm{ICL}$. If instead we split by a fixed aperture, as shown in the right panel of Fig.~\ref{fig:diff_def}, we find that at high redshifts practically all of the non-satellite stellar mass in the cluster would be considered part of the BCG. Overall, the fixed aperture and halo based cuts are most similar in terms of the later evolution of galaxy clusters. 


\end{appendix}


\label{lastpage}
\end{document}